\documentclass{aa} 
\usepackage{natbib}
\bibpunct{(}{)}{;}{a}{}{,}

\usepackage{graphicx}
\usepackage{lscape}
\usepackage{subfigure}
\usepackage{txfonts}
\usepackage{xcolor}

\newcommand{\Ha}{\mbox{H$\alpha$}}
\newcommand{\Hb}{\mbox{H$\beta$}}

\begin{document}

\title{Warm ionized gas in the blue compact galaxy Haro~14 viewed by MUSE\thanks{Based on observations made with ESO Telescopes at Paranal Observatory under programme 
ID 60.A-9186(A)}}

\subtitle{The diverse ionization mechanisms acting in low-mass starbursts}

\author{L. M. Cair\'os
    \inst{1}
    \and
    J.N. Gonz\'alez-P\'erez
    \inst{2}
    \and
    P. M. Weilbacher
    \inst{3}
    \and
    R. Manso Sainz
    \inst{4}
    }
\institute{Institut f{\"u}r Astrophysik, Georg-August-Universit{\"a}t,
Friedrich-Hund-Platz 1, D-37077 G{\"o}ttingen, Germany \\
           \email{luzma@astro.physik.uni-goettingen.de}
           \and Hamburger Sternwarte,
            Gojenbergsweg 112,
21029 Hamburg, Germany
\and Leibniz-Institut f{\"u}r Astrophysik (AIP),
An der Sternwarte 16, 14482 Potsdam, Germany 
\and Max Planck Institute for Solar System Research,
      Justus-von-Liebig-Weg 3,
       D-37077 G{\"o}ttingen, Germany 
     }

\date{January 2022}

\abstract{We investigate the warm ionized  gas in the blue compact galaxy (BCG) Haro~14 by means of  integral field spectroscopic observations taken with the Multi Unit Spectroscopic Explorer (MUSE) at the Very Large Telescope.  The large field of view of MUSE   and its unprecedented sensitivity enable observations of the galaxy nebular emission up to large galactocentric distances, 
even in the important but very faint  [\ion{O}{i}] $\lambda$6300 diagnostic line. 
This allowed us to trace the ionized gas morphology and ionization structure of Haro~14 up to kiloparsec scales and, for the first time, to accurately investigate the excitation mechanism operating in the outskirts of a typical BCG. The intensity and diagnostic maps reveal at least two highly distinct components of ionized gas: the bright central regions, mostly made of individual clumps, and a faint component which extends up to kiloparsec scales and consists of  widespread diffuse emission,  well-delineated filamentary structures, and faint knots. Noteworthy are  the two curvilinear filaments extending up to 2 and 2.3 kpc southwest, which likely trace the edges of supergiant expanding bubbles driven by galactic outflows. 
We find that while the central clumps in Haro 14 are \ion{H}{ii}-region complexes, 
the morphology and line ratios of the whole low-surface-brightness component are not compatible with star formation photoionization.
In the spatially resolved emission-line-ratio diagnostic diagrams, spaxels above the maximum starburst line form the majority ($\sim$75\% and $\sim$50\% in the diagnostic diagrams involving [\ion{O}{i}] and [\ion{S}{ii}] respectively). 
Moreover,  our findings suggest that more than one alternative mechanism is ionizing the outer galaxy regions. The properties of the diffuse component are consistent 
with ionization by diluted radiation and the large filaments and shells are most probably shocked areas at the edge of bubbles. The mechanism responsible for the ionization of the faint individual clumps observed in the galaxy periphery is more difficult to assess. These clumps could be the shocked debris of fragmented shells or regions where star formation is proceeding under extreme conditions. 
}


\maketitle
%

\section{Introduction}

Blue compact galaxies (BCGs) are low-mass, metal-poor, gas-rich objects that form stars at extremely high rates \citep{ThuanMartin1981,Thuan1999,Cairos2001a,GildePaz2003,Shi2005}. These unusual characteristics mean that they are key targets in extragalactic  astronomy. In particular,  research into BCGs is fundamental for improving our understanding of large-scale (galactic) star formation (SF) and the impact of massive stellar feedback on galaxy formation and evolution \citep{Marlowe1995,Marlowe1997,Martin1998,Cairos2017a,Cairos2017b,Cairos2020,McQuinn2018,McQuinn2019}.  

\smallskip

BCGs cannot develop spiral density waves or strong shear forces, and therefore enable us to probe  the process of SF in a relatively simple environment  \citep{Hunter1986,Hunter1997}. In addition, while in larger galaxies the collapse of gas clouds is explained (at least in part) as being due to the effect of density waves \citep{Shu1972, Seigar2002}, the mechanism responsible for the SF ignition in low-mass systems is still unclear \citep{Pustilnik2001,Brosch2004,HunterElmegreen2006, Elmegreen2012}. Several  internal processes have been proposed (e.g., stochastic self-propagating SF (\citealp{Gerola1980}), or cyclic gas re-processing (\citealp{Salzer1999})), but recent findings  point to gravitational interactions as the SF trigger in BCGs \citep{Ostlin2001,Bekki2008,Stierwalt2015,Pearson2016,Watts2016}. In particular, observational evidence for dwarf--dwarf mergers has considerably increased in the last decade \citep{MartinezDelgado2012,Paudel2015,Paudel2018,Privon2017,Zhang2020a,Zhang2020b}. Observations of mergers in nearby, metal-poor,  and gas-rich objects create a possibility to  probe the hierarchical scenario in conditions very similar to those found in the early Universe  \citep{Thuan2008}.

\smallskip 

Massive stars release a huge amount of momentum and energy into the surrounding interstellar medium (ISM), mostly in the form of ionizing photons, stellar winds, and supernova (SN) explosions. This stellar feedback  
strongly  impacts further SF in the galaxy: in extreme cases it disrupts the natal cloud  and  halts the SF \citep{Walch2012,Krumholz2014,Federrath2015} or, alternatively,  triggers a  subsequent  second episode of SF \citep{Elmegreen1977,  McCray1987,Whitworth1994,Oey2005}.  Stellar winds and SN explosions produce  large  bubbles in the ISM that  
may break out of the galaxy disk and expel metal-enriched matter into the interstellar or even the intergalactic medium \citep{Heckman1990,DeYoung1994,Veilleux2005}---these mass losses  being particularly dramatic within the  shallow gravitational potential of a dwarf  galaxy  \citep{Larson1974,Dekel1986,MacLow1999}. In spite of its unquestionable importance,
stellar feedback  remains  a   
poorly understood mechanism, as the complexity of the physics involved make the problem quite difficult to resolve, both analytically and numerically
  \citep{Mckee2007,Somerville2015,Naab2017}.

\smallskip

 The study of the warm ionized gas in star forming dwarf galaxies is
crucial for characterizing their ongoing SF episode  and 
the impact of massive stellar feedback  into their ISM, as the bright  optical emission lines provide information  on the star formation rate (SFR) and SF pattern, the chemical abundances and physical parameters of the gas, and the dust distribution, as well as on the power source(s) heating the nebula  \citep{Marlowe1995,Marlowe1997, Martin1997, Martin1998,HunterElmegreen2004}. 
Such investigations, using traditional
observing techniques, are nevertheless prohibitively costly, 
as proper characterization of such highly irregular galaxies requires accurate mapping of the gas distribution \citep{MacKenty2000,Calzetti2004,Buckalew2006,James2016}. 
Fortunately, integral field spectroscopy and, in particular, the advent of a new generation of wide-field integral field  spectrographs (IFS) working in 8-m class telescopes has considerably widened the scope of this research area. Specifically, the Multi Unit Spectroscopic Explorer (MUSE; \citealp{Bacon2010}), with its unique combination of high spatial and spectral resolution, large field of view
(FoV), and extended wavelength range,  has proved to be a powerful tool in BCG research   \citep{Bik2015,Bik2018,Cresci2017,Herenz2017,Kehrig2018,Menacho2019,Menacho2021,Cairos2021}.

\begin{table}
\small
\caption{Basic data for Haro\,14
\label{tab:data}}
\begin{center}
\begin{tabular}{lcc}
\hline
Parameter   & Value & Reference \\ 
\hline
Other names & NGC 0244, UGCA 10 & \\
& VV~728, PGC~2675                                  & \\
RA (J2000)                                          &  00$^h$45$^m$46$\fs$4 &    \\
Dec (J2000)                                         &   -15$\degr$35$\arcmin$49$\arcsec$  &     \\
 Distance                                                   &  13.0$\pm$0.1     Mpc                                             &     \\
 Spatial scale                                              & 63~pc arcsec$^{-1}$ &                                     \\
 R$_{25}$                                           & 2.32~kpc                                            &    RC3 \\
 m$_{V}$                                            & 13.24                                     &  Paper~I \\
 M$_{V}$                                            & -17.33                                    & Paper~I \\
 V-R                                                       & 0.34                            & Paper~I \\
 V-I                                                       & 0.71                           & Paper~I \\
 M$_{\rm H I}$         & 3.2$\times$10$^{8}$M$\odot$    & TM81  \\
 M$_{T}$          & 3.8$\times$10$^{8}$M$\odot$ & TM81  \\ 
 12+log(O/H)      & 8.2$\pm$0.1     & CG17      \\
Morphology                       & SOpec, BCG, iE BCD                                                   & \\
\hline
\end{tabular}
\end{center}
\small Notes:
RA, DEC, distance, and apparent major isophotal radius R$_{25}$ measured
at a surface brightness level of 25.0 mag~arcsec$^{-2}$ are
taken from NED ({\tt http://nedwww.ipac.caltech.edu/}).  References.- CG17: \cite{Cairos2017a}; RC3: \cite{deVaucouleurs1991}; 
TM81:\cite{ThuanMartin1981}.  
\end{table}

\smallskip

\begin{figure}
\centering
\includegraphics[angle=0, width=0.9\linewidth]{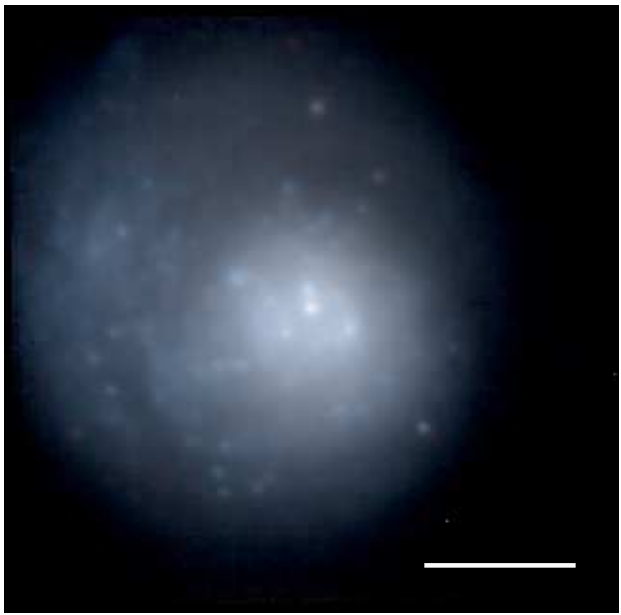}
\caption{Color image of Haro\,14  from the MUSE data (logarithmic intensity scale). A scale bar of 1~kpc  (15.9~arcsec) in length is shown to the bottom right. North is at the top and east to the left, as is the case for all the maps presented hereafter.}
\label{Figure:Haro14_rgb} 
\end{figure}

This paper is the second  of  a series presenting results from 
MUSE observations of the BCG Haro~14 (Table~1; Figure~\ref{Figure:Haro14_rgb}).
In the first paper (Cair\'os et al. 2021, hereafter Paper~I),  we introduced the observations and data analysis and performed an exhaustive investigation of the morphology, structure, and stellar populations of the galaxy. 
As in the great majority of BCGs, Haro~14 is made of an irregular high-surface-brightness (HSB) 
region placed atop a smooth low-surface-brightness (LSB) host.
The large MUSE FoV ($1\arcmin\times1\arcmin$=$3.8\times3.8$~kpc$^{2}$ at the adopted distance of 13~Mpc)  allowed  simultaneous observations of both the starburst  and the  older stellar component. This represented a huge improvement with respect to most previous integral field spectroscopic studies of BCGs, which, due to their limited FoV,
could only cover the central HSB area  \citep{Vanzi2008,Vanzi2011,GarciaLorenzo2008,Cairos2009a,Cairos2009b,Cairos2017a,Cairos2017b,Cairos2020,James2010,James2013a,James2013b,Kumari2017,Kumari2019}. Haro~14 presents a  markedly asymmetric stellar distribution, with the  peak in continuum, a bright stellar cluster with M$_{V}=-12.8$ (most probably a super stellar cluster), being displaced by  about 500~pc southwest, and  a highly
distorted, blue, but nonionizing stellar component occupying almost the whole eastern part of the galaxy.  
We find evidence of (at least) three different stellar populations: a very young starburst (ages $\leq$6~Myr), an intermediate-age stellar component, with ages of between 30 and a few hundred million years, and a smooth stellar population with ages of several gigayears. Although it is not clear how these different stellar components are related to each other, the distorted galaxy morphology, the pronounced lopsidedness in the color map, and the presence of numerous young massive stellar clusters point to previous episodes  of mergers or interactions.

\smallskip 

In this second paper, we study the ionized gas emission in Haro~14. The wide FoV of MUSE and its unprecedented sensitivity allow observations of a large set  of diagnostic emission lines, including the faint [\ion{O}{i}]$\lambda$6300,  out to large galactocentric distances (up to 2.7~kpc) and very low surface brightness levels (about 10$^{-18}$erg~s$^{-1}$~cm$^{-2}$~arcsec$^{-2}$). This dataset  enables us to perform,  for the first time,  an exhaustive investigation of the nebular morphology and the ionization structure in the very faint outskirts of a BCG.

\smallskip

The paper is structured as follows: in Sect.~\ref{Section:observations} we briefly summarize the observations and main steps in the data process and introduce
the method employed to derive the emission line maps.  In Sect.~\ref{Section:ionizedgas} we present the main outcomes from our data analysis, which include a thorough discussion of the  Haro~14 spatial pattern in emission lines and in diagnostic line ratios, spatially resolved diagnostic diagrams, and integrated spectroscopy of the galaxy and the most interesting galaxy regions. 
Finally, the main findings of the work  are discussed
in Sect.\ref{discussion} and are summarized in Sect.~\ref{conclusions}.

\section{Observations and data processing}
\label{Section:observations}

\begin{figure*}
\centering
\includegraphics[angle=0, width=\linewidth]{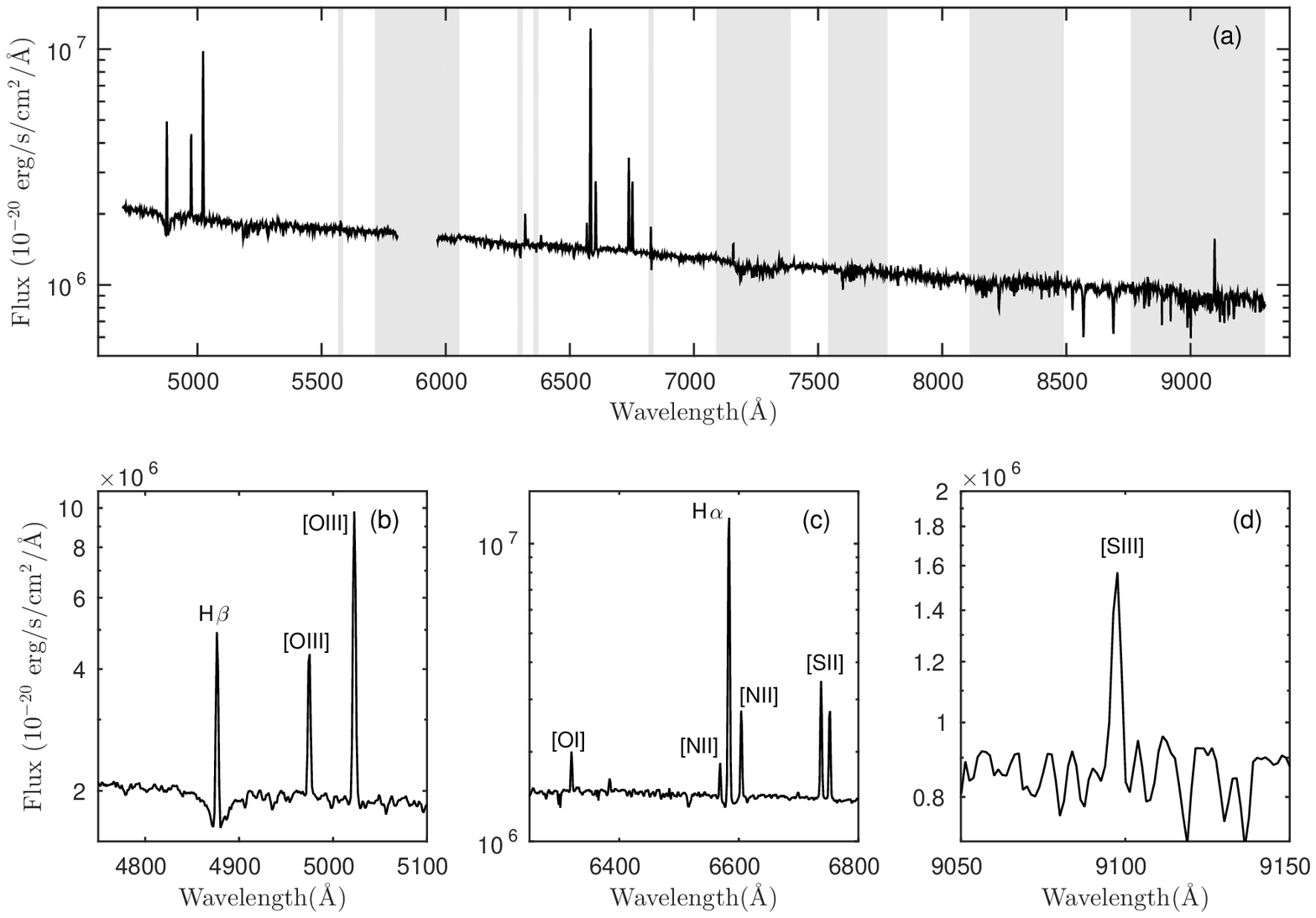}
\caption{Flux-calibrated integrated spectrum of the BCG Haro\,14. (a) Spectrum in the whole observed wavelength range (4750-9300 \AA).  
Shaded bars mark spectral windows heavily contaminated by telluric lines which were excluded from the fitting of the stellar component. 
The 5800-5950~\AA\ gap corresponds to the spectral area masked out due to the AO lasers. 
Bottom panels show spectral regions of particular interest, with prominent lines  labeled. (b) Zoom into the 4750-5100~\AA\ wavelength region, with the pronounced absorption wings around H$\beta$ clearly resolved. (c) Zoom into the 6250-6800~\AA\ spectral region. (d) Zoom into the region of 9050-9150~\AA\ .}
\label{Figure:Haro14_spec} 
\end{figure*}

Data on Haro\,14 were collected with 
MUSE \citep{Bacon2010} at the Very Large Telescope
(VLT; ESO Paranal Observatory, Chile). MUSE, working in wide field mode, provides  a FoV of $1\arcmin\times1\arcmin$ with a  
spatial sampling of 0\farcs2. We observed in the 
wavelength range 4750-9300~\AA, with a spectral sampling of 1.25~\AA\,pix$^{-1}$ in  the
dispersion direction, and an average  resolving power of R$\sim$3000. 
The data were acquired in September 2017 during 
the science verification of the MUSE Adaptive Optics Facility (AOF; 
\citealp{Leibundgut2017}).   
There is a gap between $\sim$5800 and 5950~\AA,\  because the  AOF works with four sodium lasers and otherwise the detector would saturate  in that range. 

\smallskip 

We took
four exposures of Haro\,14, each of 1370\,s, with a pattern of
90$^{\circ}$ rotations between exposures, giving a total of 5480\,s on source. Because the target fills the MUSE FoV, we took separate sky fields of 120\,s  between the science exposures.
The data were processed using the standard MUSE pipeline
\citep{Weilbacher2016,Weilbacher2020} working within the
{\sc esorex} environment with the default set of calibrations. The complete description of the data reduction and sky subtraction is provided in 
Paper~I. 

\subsection{Emission line fitting}
\label{Section:fittinglines}

The next step in the IFS data process is the measurement of the parameters of the emission lines (fluxes, radial velocities, and full width at half maximum (FWHM))  required to build the maps. To derive accurate fluxes  of emission lines in star forming galaxies is not straightforward, because in every element of spatial resolution the observed spectrum is the sum of the light generated in the nebula and the light emitted by the stars.  Particularly problematic is the measurement of  the fluxes of the hydrogen Balmer series in emission, because the  strength of these lines can be severely affected by the absorption of the underlying population of stars  \citep{McCall1985,Olofsson1995,GonzalezDelgado1999a,Levesque2013}.  Single stellar population (SSP) models predict equivalent width values of between 2 and 16 \AA\  during the first  1~Gyr of a solar metallicity burst, with the equivalent widths reaching their maximum  at about a few hundred million years, when the stellar population is dominated by A-type stars  \citep{GonzalezDelgado1999b,GonzalezDelgado2005}.

\smallskip 

\begin{figure*}
\centering
\includegraphics[angle=0, width=\linewidth]{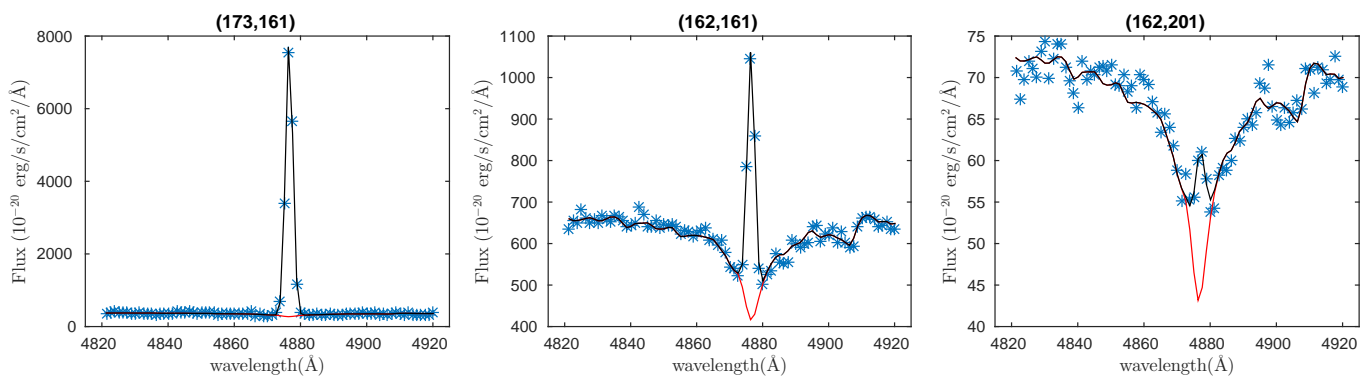}
\caption{Example fits to the observed spectrum (asterisks) in the vicinity of the H$\beta$ line at three  
positions in the map (spaxel coordinates on top), illustrating where the emission line is dominant
(left panel), severely affected by underlying stellar absorption (central panel), and
subdominant (right panel). Red line: stellar component. Black line: stellar 
plus emission (Gaussian) component.}
\label{Figure:Haro14_spechbeta} 
\end{figure*}

The influence of the underlying stellar absorption in the integrated spectrum of Haro~14 is already clear from Figure~\ref{Figure:Haro14_spec}. Sharp absorption wings are visible around H$\beta$ in emission, and   more moderate wings are also visible around H$\alpha$. An inspection of the spectra  of the individual  spatial pixels (spaxels) shows that the strengths of the absorption strongly depends on the position across the galaxy (Figure~\ref{Figure:Haro14_spechbeta}). Therefore, the determination of accurate Balmer fluxes in emission requires modelling the spectral energy distribution (SED) of the stellar component for every  spaxel. 

\smallskip

Before performing  the spectral fit of the stellar component,  we  grouped the  104006 spectra of the MUSE data cube into bins of  similar signal-to-noise ratio (S/N) using
the  Voronoi adaptive binning algorithm  \citep{Cappellari2003}. We required a S/N on the stellar continuum $\geq$ 50 per wavelength channel in each bin in the spectral range 5400-5600\AA\ and 
discarded spaxels with S/N$\leq$5.  This resulted in a total of 1793 spectra after the binning process.

\smallskip

Next we modeled the SED of these 1793 individual spectra with the Galaxy IFU Spectroscopy Tool (GIST\footnote{GIST is a modular pipeline, entirely written in Python~3, developed for the scientific analysis of reduced integral field spectroscopic data. {\tt http://ascl.net/1907.025}}; \citealp{Bittner2019}).  In a first step,   we determined the 
line-of-sight velocity distribution  
by means of the  
penalized pixel-fitting technique ({\tt ppxf}; \citealp{Cappellari2004, Cappellari2017}).
This method allows the user to simultaneously fit the
optimal linear combination of stellar templates to the observed
spectrum and to derive the stellar kinematics, using a maximum
likelihood approach to suppress noise solutions. 
The code convolves the stellar templates with Gauss-Hermite functions
to reproduce the shape of the galaxy absorption lines; we considered here only the first four  Gauss-Hermite moments (velocity, sigma, h$_{3}$ and h$_{4}$).
The {\tt ppxf}  technique can work with empirical libraries or with theoretical spectra generated by population synthesis models. We opt for the latter, and use synthesis model templates because they enable more flexibility concerning the spectral range and spectral resolution. 
We used the  Binary Population
and Spectral Synthesis models (BPASS, v.2.2\footnote{{\tt https://bpass.auckland.ac.nz/}}; \citealp{Eldridge2009,Eldridge2017,Stanway2018}), a relatively new set of  population synthesis models that include the effects of binary evolution, conferring a major advantage. The models 
predict the properties of SSPs at ages from 1~Myr to 100~Gyr in 51 steps (we incorporated  43 steps in our calculations, from 1 to 16~Gyr) and 14 metallicities, from  5$\times$10$^{-5}$ to  2~Z$_\odot$. 
We adopted an initial mass function (IMF) with a maximum stellar mass of  100$M_\odot$ and a broken power law with
a slope of  -2.0 above 0.5~M$_\odot$ and -1.3 below.

\smallskip

 Once the kinematical parameters are known, GIST provides a guess of the spectrum of the underlying stellar population  in each bin.  After removing this stellar spectrum from the observed one, the emission line parameters are computed 
making use of  the {\em Gas and Absorption Line Fitting} software (GandALF; \citealp{Sarzi2006, FalconBarroso2006}).  The emission line spectrum  is then removed from the observed one and the {\tt ppxf} algorithm is run again to obtain the final stellar spectrum of the bin.
Spectral regions heavily affected by telluric lines were excluded from the fitting 
(see Figure~\ref{Figure:Haro14_spec}).

\smallskip

Finally,  the emission-line parameters are calculated on every spaxel by fitting Gaussians plus the estimated stellar spectrum of the corresponding bin. 
The fit
was carried out with the {\em Trust-region} algorithm for nonlinear least squares,
using the function {\tt fit} of {\sc Matlab}. We run an automatic procedure,
which fits a series of lines for every spaxel, namely, H$\beta$, [\ion{O}{iii}] $\lambda\lambda 4959, 5007$, [\ion{O}{i}] $\lambda\lambda 6300, 6363$,
H$\alpha$, [\ion{N}{ii}] $\lambda\lambda 6548, 6583$, [\ion{S}{ii}] $\lambda\lambda 6716, 6731$, [\ion{Ar}{iii}] $\lambda7135,$ and [\ion{S}{iii}] $\lambda9069$. The fit algorithm provides the 
relevant parameters of the emission lines (line flux, centroid position, line
width)  as well as their associated errors.

\begin{table}
\small
\begin{center}
\caption{Emission lines of Haro 14 in the MUSE observed range for which line maps have been generated. \label{tab:lines}}
\begin{tabular}{lccccc}
\hline\hline 
Line                            & Ion                           & IP                                    &           $F_{\rm \lambda}^{obs}$                  &    $F_{\rm \lambda}^{cor}$  &  $F_{\rm \lambda}^{lim}$     \\
(\AA)                   &                                       &  (eV)                          &  ($10^{-14}$ )                                                    &    ($10^{-14}$ )                &   ($10^{-19}$ )  \\                                      
 \hline
4861 (H$\beta$) & \ion{H}{i}                    & 13.60                                 & 11.0                                                                & 15.9                                         & 27\\
5007                    &  $[\ion{O}{iii}]$             & {\em 35.12}   &  23.5                                                              & 33.5                                          &  12 \\
6300                    & $[$O {\sc i}$]$               & 13.62                 & 1.6                                                                        & 2.1                                            &   11 \\
6563 (H$\alpha$)        & \ion{H}{i}                    & 13.60                                 & 35.4                                                              & 45.6                                            &  9 \\                      
6583                    & $[$N {\sc ii}$]$              & {\em 14.53}           & 4.1                                                             & 5.3                                      &  8\\
6716, 6731              & $[$S {\sc ii}$]$              & {\em 10.36}           & 10.8                                                            & 13.9                                             &  10 \\
7135             & $[$Ar {\sc iii}$]$           & {\em 27.63}           & 1.0                                                               & 1.3                                             &   8\\
9069            & $[$S {\sc iii}$]$                      & {\em 23.34}          & 2.2                                                               & 2.5                                             &  15 \\
\hline\hline 
\end{tabular}
\end{center}
Notes.- IP: ionization potential of the ionization stage of the line, for recombination (\ion{H}{i})
and charge exchange (\ion{O}{i}) dominated lines; of the previous (lower) ionization stage for the rest of the
forbidden lines (in italics). $F_\lambda^{obs}$  and $F_\lambda^{cor}$ are the observed and interstellar-extinction-corrected 
fluxes as measured in the integrated spectrum (units are $10^{-14}$ erg s$^{-1}$ cm$^{-2}$).  $F_{\rm \lambda}^{lim}$ is the limit 
surface brightness in units of $10^{-19}$ erg s$^{-1}$ cm$^{-2}$ arcsec$^{-2}$.
\end{table}

\subsection{Generating the galaxy maps}
\label{Section:themaps}

The parameters derived from the fit (line fluxes, centroid position, line width, and continuum) were then used to build the bidimensional maps of the galaxy. The full wavelength range of MUSE covers a good number of diagnostic emission lines: recombination lines of hydrogen and helium\footnote{Unfortunately, the important \ion{He}{i}~$\lambda5876$ line falls in the spectral gap used by the AO system.},  high and low ionization, collisionally excited forbidden lines of oxygen and sulfur, and low ionization collisional forbidden lines of nitrogen (see Table~\ref{tab:lines}, and Figure~\ref{Figure:Haro14_spec}).  We built emission line maps of the brighter lines, that is, H$\beta$,  [\ion{O}{iii}],  [\ion{O}{i}], H$\alpha$,  [\ion{N}{ii}],  [\ion{S}{ii}],  [\ion{Ar}{iii}], and  [\ion{S}{iii}]. 

\smallskip

Line ratio maps were generated by simply dividing the corresponding flux maps. Because the seeing depends on the spectral region observed, when two lines are widely separated (as is the case of the \Ha/\Hb{} and  [\ion{O}{iii}]/\Ha{} maps; see Sections~\ref{Section:extinctionmap} and~\ref{Section:lineratiomaps}) we have corrected for the different seeing. The H$\alpha$ map was degraded by convolving it with a bi-dimensional Gaussian 
($\sigma$=1.2 pix) to match both point spread functions (PSFs). 
When computing line ratio maps, we included only those spaxels containing values higher than the 3$\sigma$ level. The maps have not been corrected for interstellar extinction  (see Section~\ref{Section:extinctionmap}) .

\section{Results}
\label{Section:ionizedgas}

\subsection{Morphology of the warm ionized gas}
\label{Section:emissionlinemaps}

\begin{figure*}
\centering
\begin{subfigure}{}
\includegraphics[width=8cm]{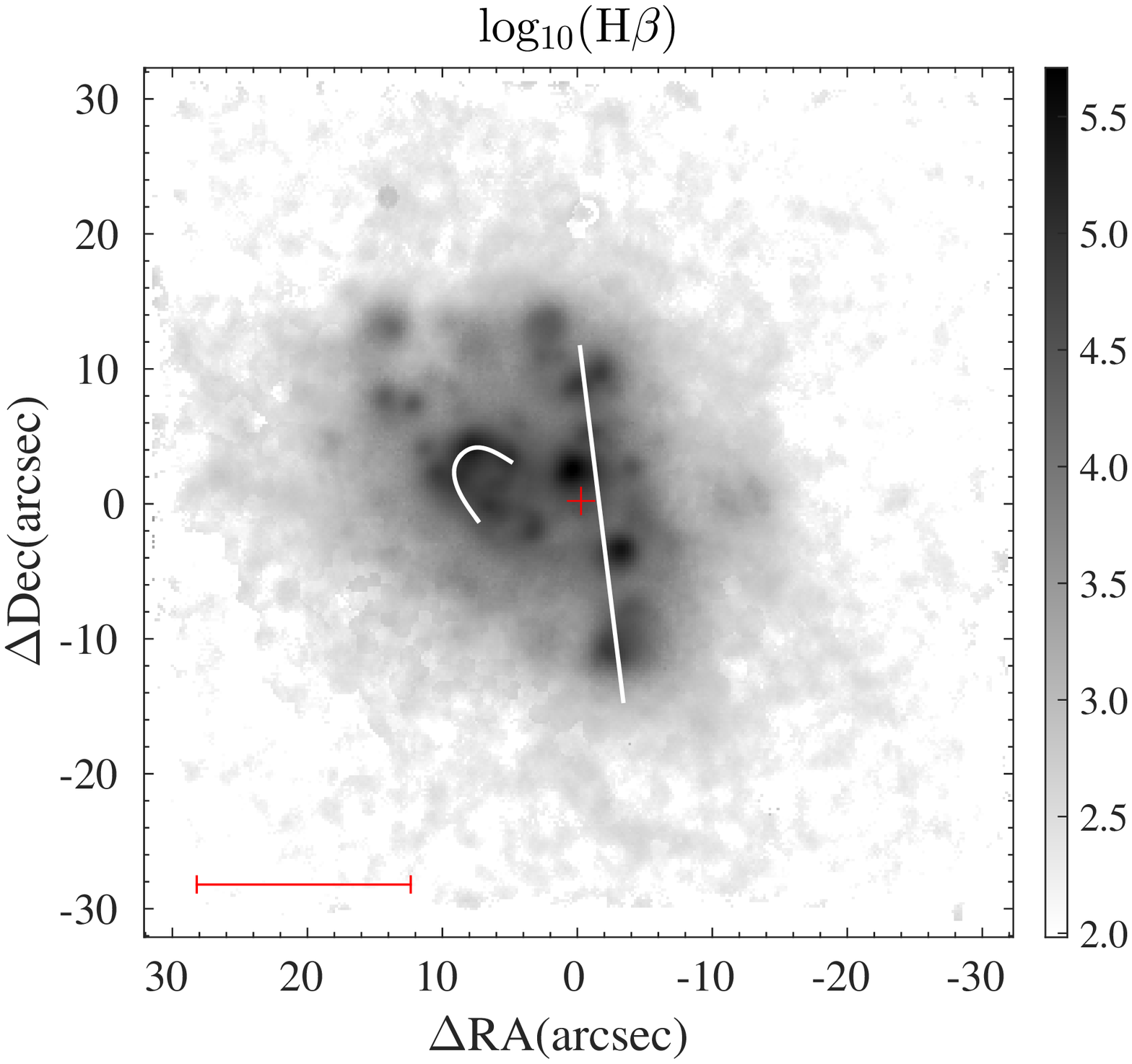}
\end{subfigure}
\begin{subfigure}{}
\includegraphics[width=8cm]{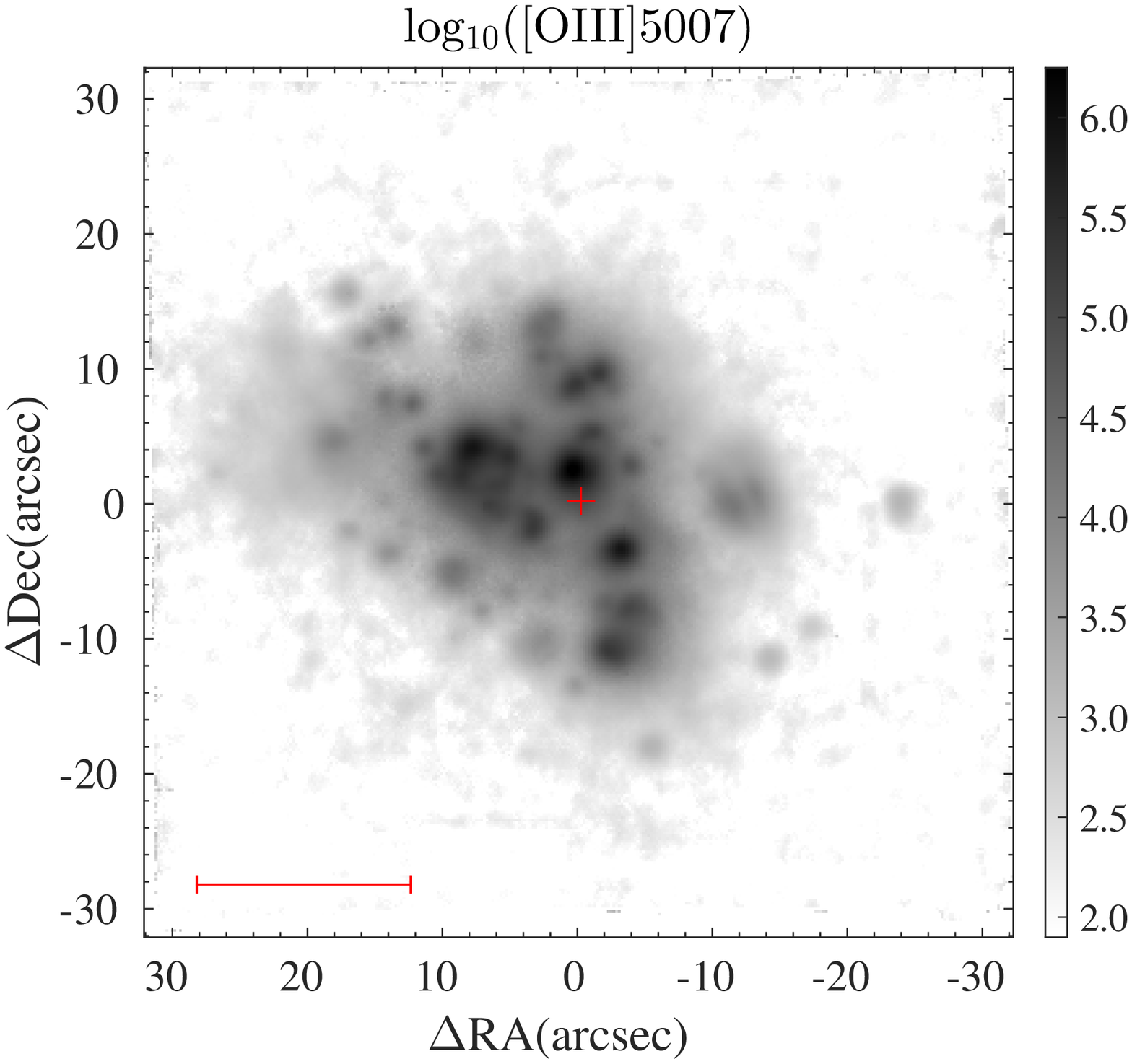}
\end{subfigure}
\begin{subfigure}{}
\includegraphics[width=8cm]{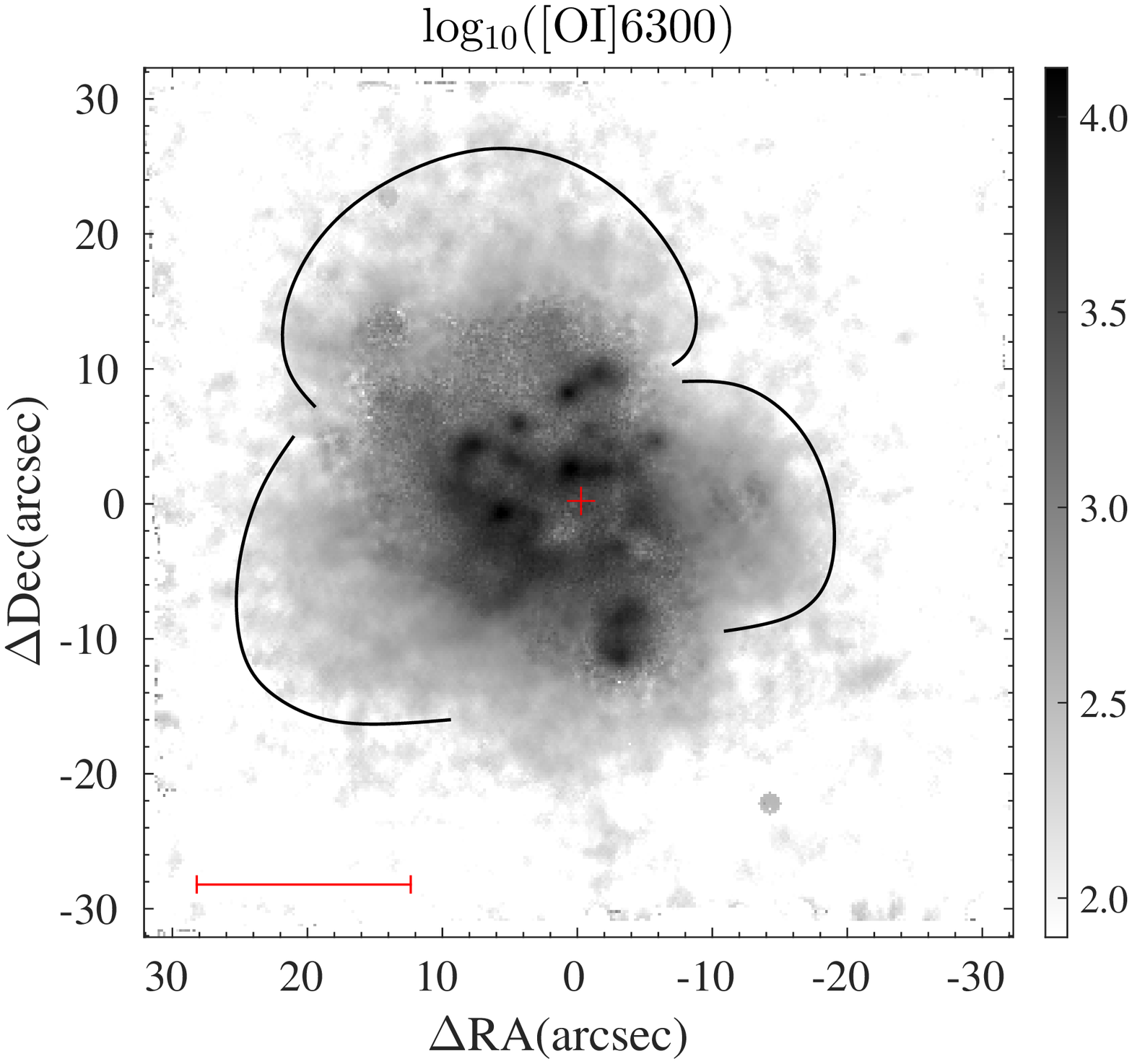}
\end{subfigure}
\begin{subfigure}{}
\includegraphics[width=8cm]{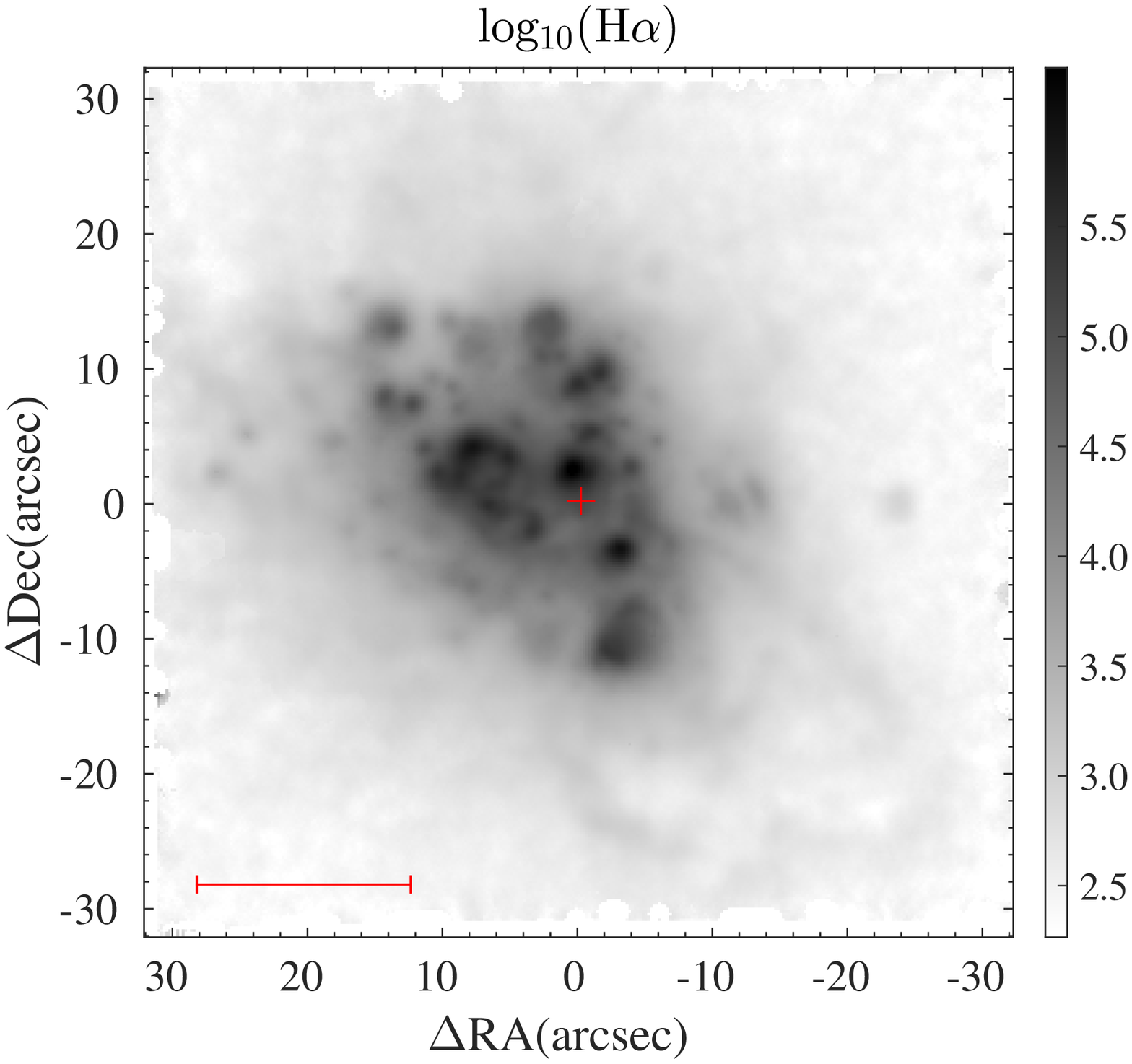}
\end{subfigure}
\caption{Emission-line flux maps of Haro\,14. {\em Top-left}:  H$\beta$ map, with the central bar-like  and  curvilinear structures indicated with white lines.  {\em Top-right}: [\ion{O}{iii}]~$\lambda5007$ map. {\em Bottom-left}: [\ion{O}{i}]~$\lambda6300$ map; the black  lines indicate lobes of enhanced emission  in the galaxy periphery.  {\em Bottom-right}: H$\alpha$  map. 
The maps are in logarithmic scale and flux units are 10$^{-20}$ erg$\,$s$^{-1}\,$cm$^{-2}$\,arcsec$^{-2}$. The red cross indicates the position of the continuum peak and the red line (bottom left) corresponds to 1~kpc. }
\label{Figure:emissionmaps1} 
\end{figure*}

\begin{figure*}
\centering
\begin{subfigure}{}
\includegraphics[width=8cm]{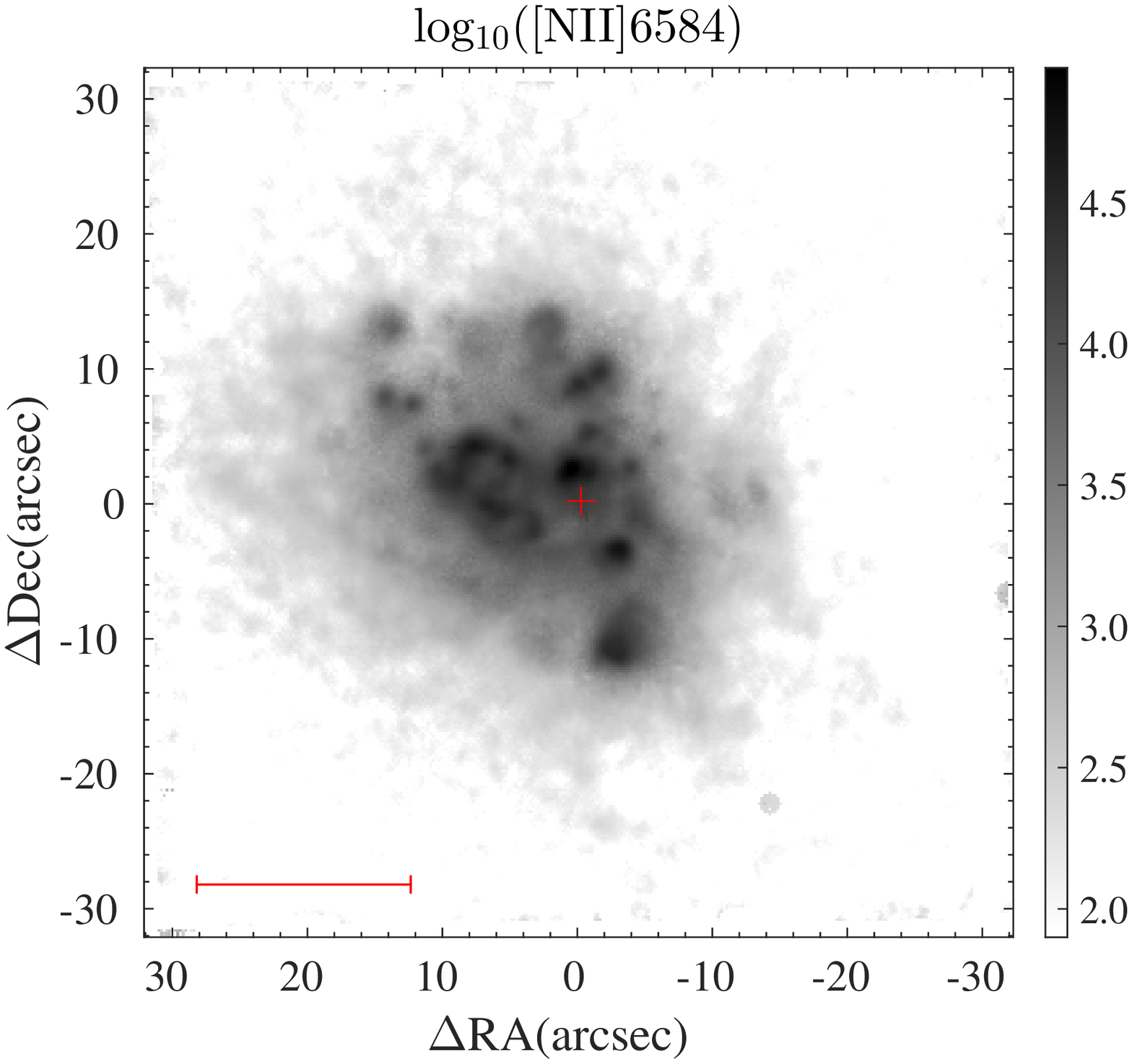}
\end{subfigure}
\begin{subfigure}{}
\includegraphics[width=8cm]{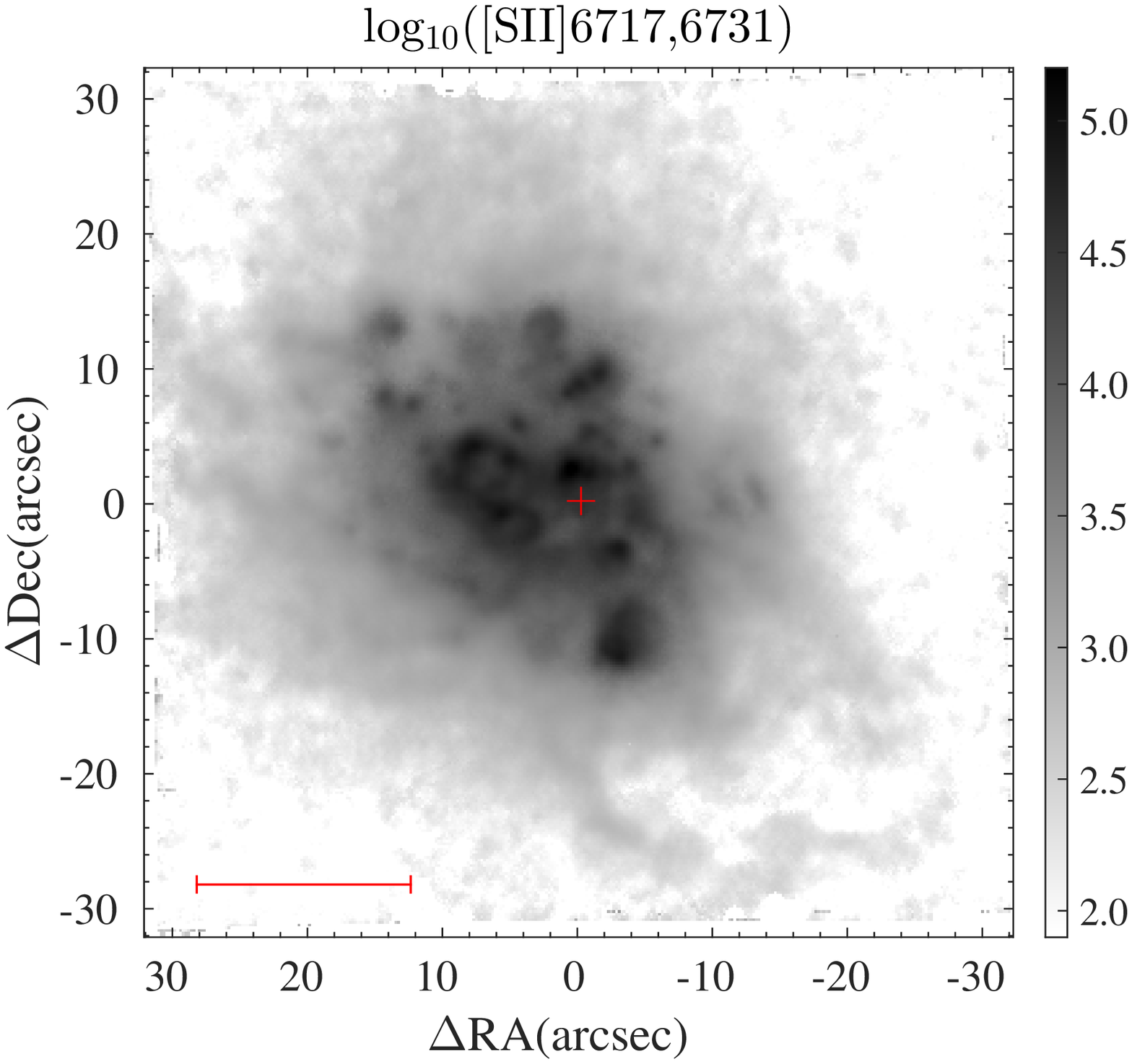}
\end{subfigure}
\begin{subfigure}{}
\includegraphics[width=8cm]{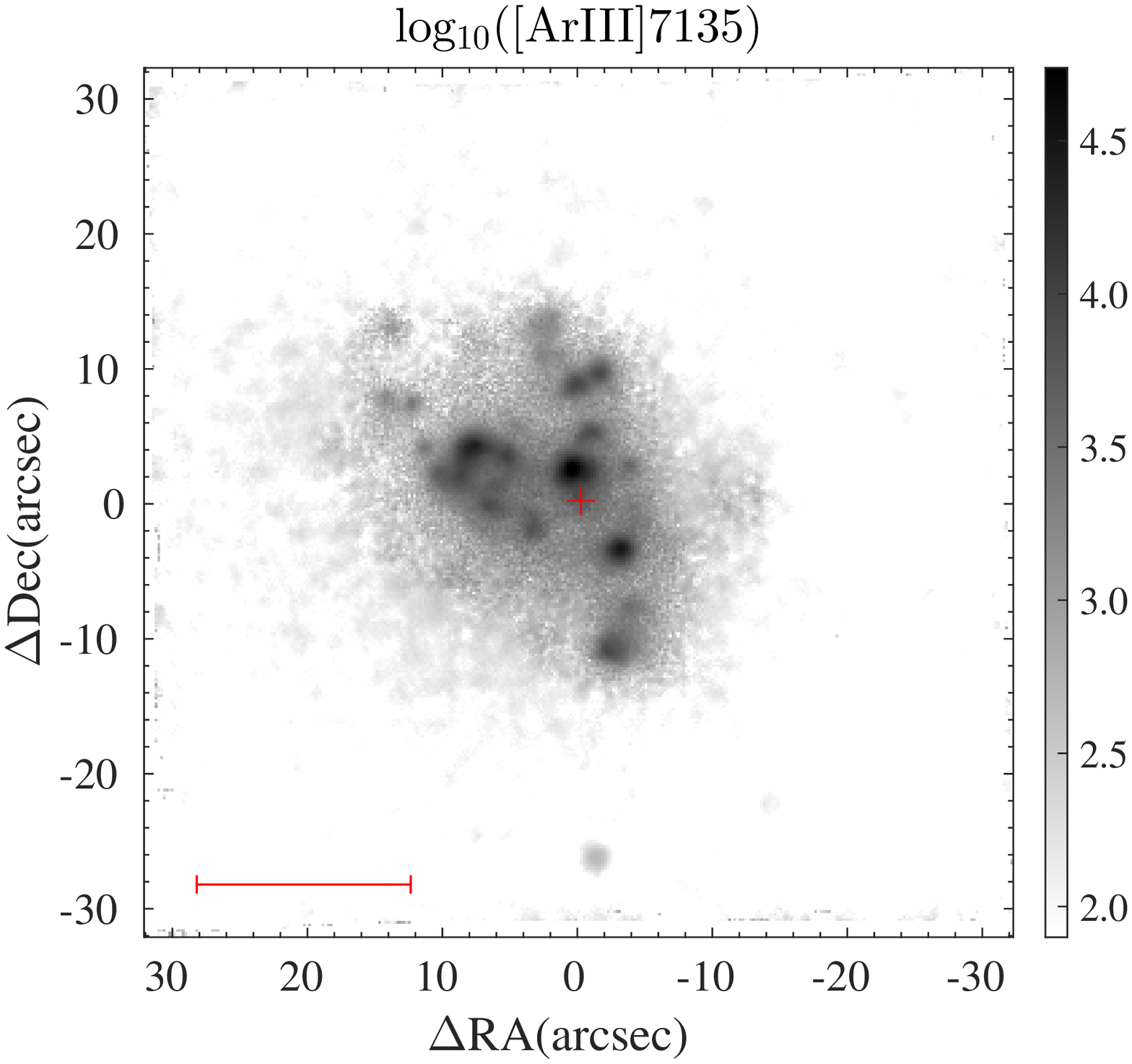}
\end{subfigure}
\begin{subfigure}{}
\includegraphics[width=8cm]{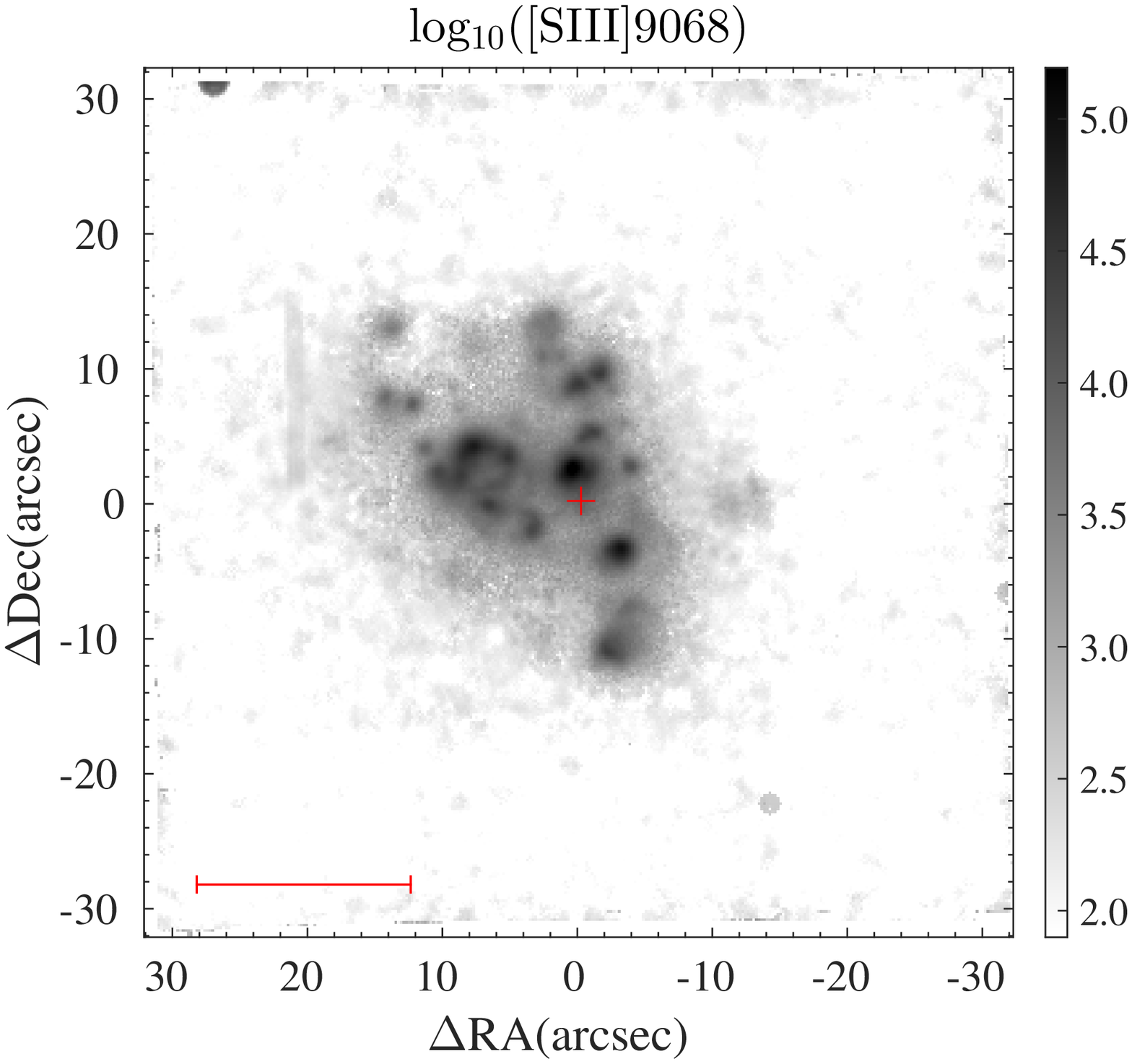}
\end{subfigure}
\caption{Emission-line flux maps of Haro\,14. {\em Top left:} [\ion{N}{ii}]~$\lambda6584$. {\em Top right:} [\ion{S}{ii}]~$\lambda\lambda6717,\,6731$. {\em Bottom left:} [\ion{Ar}{iii}]~$\lambda7135$. 
{\em Bottom right:} [\ion{S}{iii}]~$\lambda9068$. The maps are in logarithmic scale and flux units are 10$^{-20}$ erg$\,$s$^{-1}\,$cm$^{-2}$\,arcsec$^{-2}$. The red cross indicates the position of the continuum peak and the red line (bottom left) corresponds to 1~kpc.}
\label{Figure:emissionmaps2} 
\end{figure*}

\begin{figure*}
\centering
\begin{subfigure}{}
\includegraphics[width=8cm]{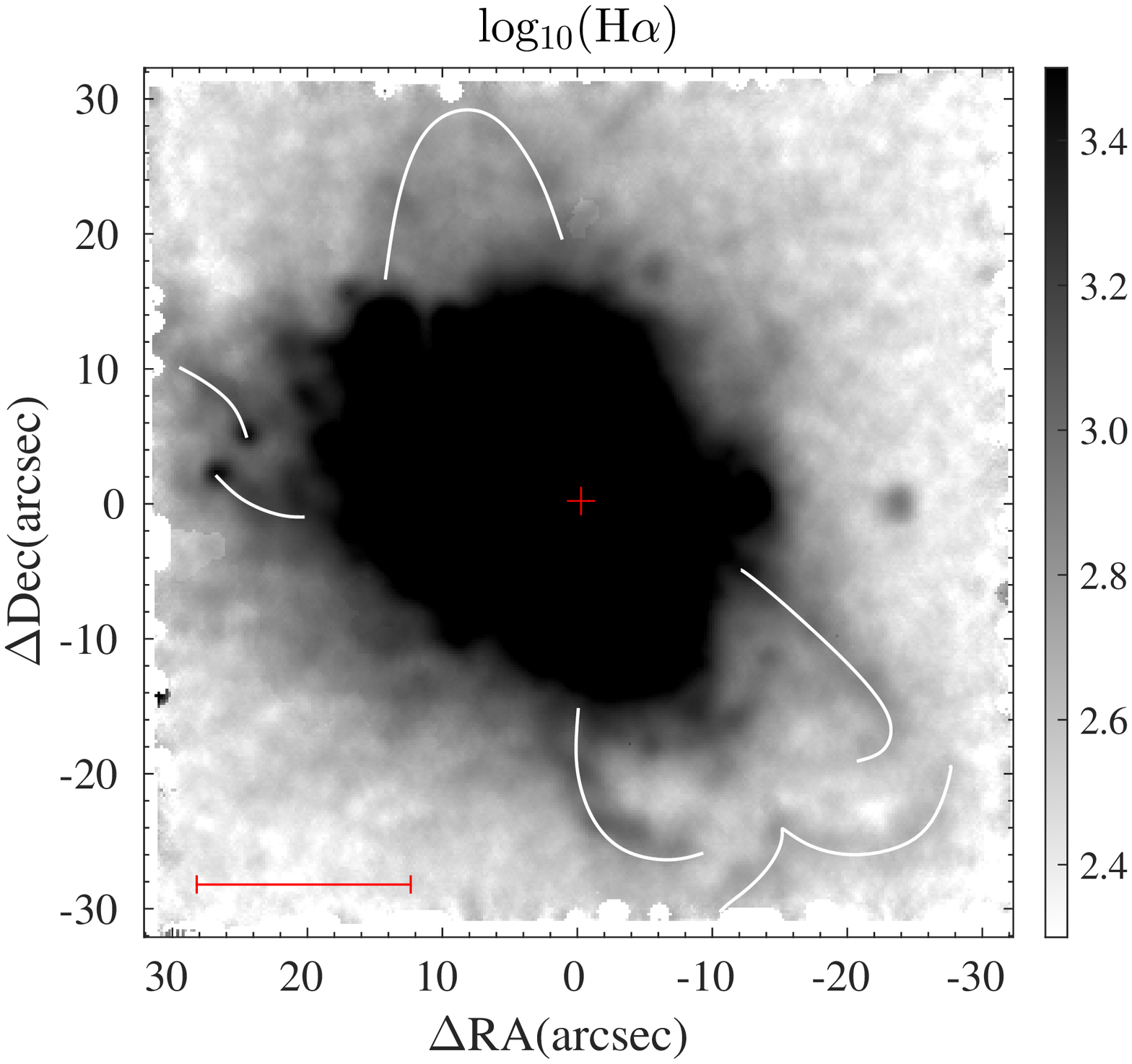}
\end{subfigure}
\begin{subfigure}{}
\includegraphics[width=8cm]{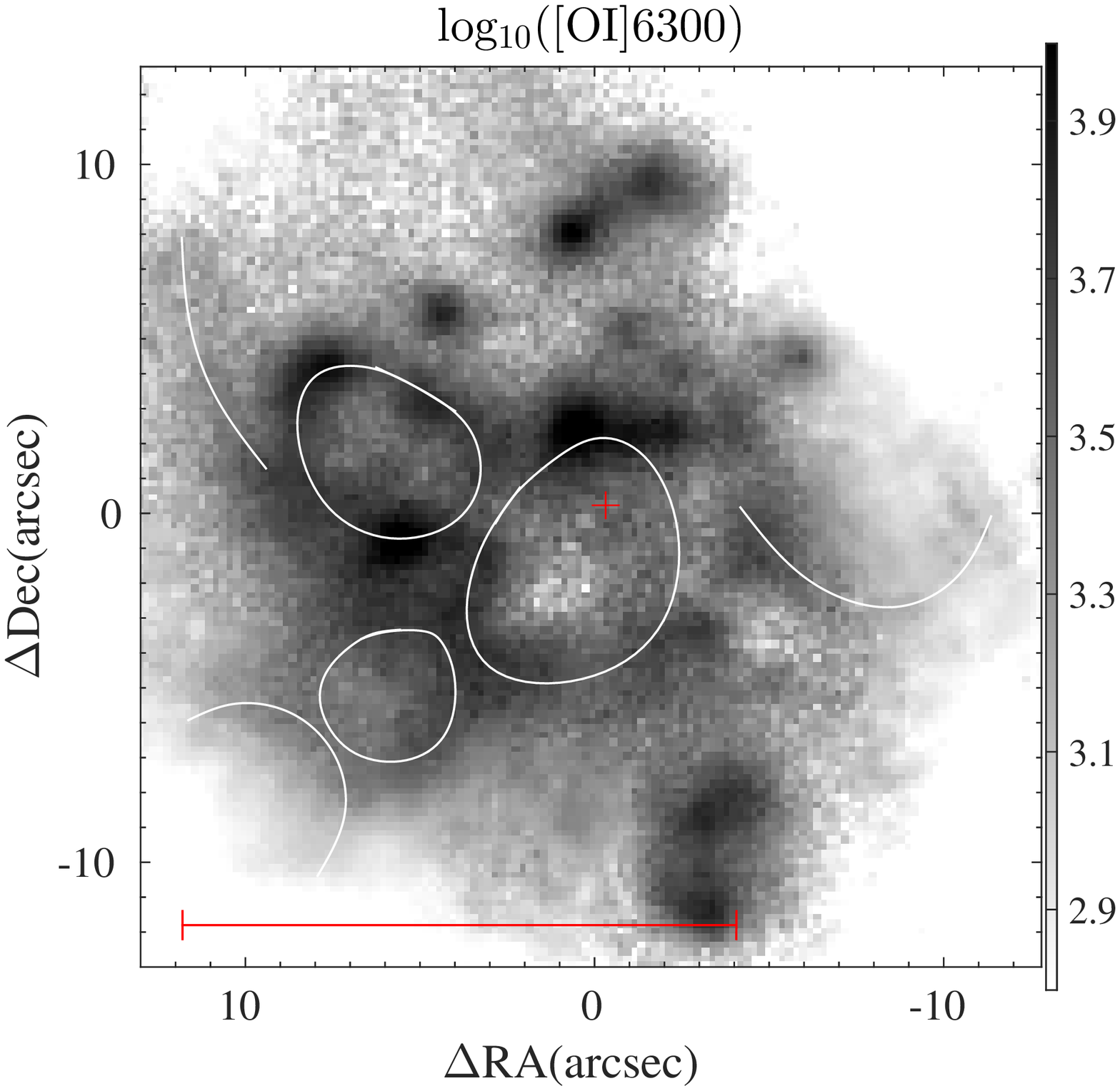}
\end{subfigure}
\caption{Intensity maps of Haro\,14 in H$\alpha$ (left) and [\ion{O}{i}] (right). In the H$\alpha$ map, the central regions are saturated in order to enhance the faint surface brightness features in the galaxy outskirts; the white
lines indicate the most prominent shells and filaments in the galaxy periphery. In the [\ion{O}{i}]  map, we have zoomed into the HSB area in order to better distinguish its structure; major bubbles and filaments are indicated with white lines. Both maps are in logarithmic scale and flux units are 10$^{-20}$ erg$\,$s$^{-1}\,$cm$^{-2}$\,arcsec$^{-2}$.}
\label{Figure:emissionmapssat} 
\end{figure*}

The intensity maps of Haro~14 in the brightest emission lines are shown in Figures~\ref{Figure:emissionmaps1},  \ref{Figure:emissionmaps2}, and \ref{Figure:emissionmapssat}.
Maps in H$\alpha$ and [\ion{O}{iii}]~$\lambda5007$ were already presented and briefly discussed in Cair\'os et al. (2021). The galaxy shows a clumpy morphology, with the brightest knots located  in the central regions forming a bar-like arrangement and a horseshoe-like curvilinear feature. These clumps are embedded in an extended halo of ionized gas, from which curvilinear structures such as loops, shells, and filaments emerge. Despite the globally similar appearance, some important differences are evident among the maps in different emission lines, in particular in the dimmer regions of the galaxy periphery.

\smallskip 

The maps  in H$\alpha$ and H$\beta$  display the same pattern, but in H$\alpha$,  which is significantly brighter, the intricate distribution of shells, filaments, holes, and bubbles appears much more   clearly.
Particularly noteworthy are  two curvilinear filaments extending up to 2 and 2.3~kpc southwest, and a shell  with a diameter of about 800~pc  visible to the north (Figure~\ref{Figure:emissionmaps1}, bottom-right and Figure~\ref{Figure:emissionmapssat}, left).

\smallskip 

The morphology in the high-ionization [\ion{O}{iii}] line is roughly similar to H$\alpha$ in the central area but notable differences appear in the galaxy outskirts. As a general rule, the knots  situated in the peripheral zone appear more  compact and round in [\ion{O}{iii}] than in H$\alpha$ and, although 
the bulk of the knots are detected in both maps, some of the faintest ones are visible only in H$\alpha$ or only in [\ion{O}{iii}]. Remarkably, the large filaments of ionized gas extending southwest and the shell  at the north, which are highly visible in H$\alpha$,  are not detected in 
[\ion{O}{iii}] (cf upper and lower right panels in Figure~\ref{Figure:emissionmaps1}).  The maps in the other two high-ionization  lines, namely [\ion{Ar}{iii}]~$\lambda7135$ and [\ion{S}{iii}]~$\lambda9069$, show the same morphology as  [\ion{O}{iii}]  but a poorer S/N, and mostly delineate the HSB central region of the galaxy.

\smallskip

The  map in the  low-ionization [\ion{O}{i}] $\lambda$6300 line presents some distinctive features. We see fewer and smaller blobs  than in the other line maps. In the central galaxy regions,  we discern several bubbles and filaments which are barely detected (if at all) in H$\alpha$   (Figure~\ref{Figure:emissionmaps1}, bottom and Figure~\ref{Figure:emissionmapssat}, right). 
In the galaxy periphery, the emission is considerably enhanced: 
we see extended [\ion{O}{i}] emission spatially coincident with the southern filaments and northern bubble detected in H$\alpha$, but also two large  lobes of diffuse emission departing  southeast and west, and extending about 1.3 and 1~kpc, respectively (Figure~\ref{Figure:emissionmaps1}, lower left). 

\smallskip 

The maps in the low-ionization  lines  [\ion{N}{ii}]~$\lambda6584$ and [\ion{S}{ii}]~$\lambda\lambda6717,\,6731$ are almost indistinguishable in the central clumpy area, but the brighter sulfur lines  delineate  the galaxy outskirts much more clearly; both lines  
reproduce the morphology in H$\alpha$  very well (Figure~\ref{Figure:emissionmaps2}, top panels).

\smallskip

The strength and spatial distribution of the different emission lines provide an initial view of the physical processes taking place in a galaxy.   While \ion{H}{ii}-regions are characterized by strong high-ionization emission lines  (such as  [\ion{O}{iii}]) and weak low-ionization forbidden lines (such as [\ion{N}{ii}], [\ion{S}{ii}], or [\ion{O}{i}]),  spectra generated by an active galactic nucleus (AGN)  or shock heating show a much wider range of ionization, with enhanced emission from both  high-  and low-ionization forbidden lines \citep{Aller1984,Veilleux1987,Dopita2003,Osterbrock2006}. 

\smallskip

In Haro~14, the morphology of the central HSB regions is consistent with SF photoionization: the emission is mainly concentrated in large and bright clumps, which appear very well delineated in recombination (H$\alpha$, H$\beta$) and 
high-ionization ([\ion{O}{iii}], [\ion{Ar}{iii}], [\ion{S}{iii}]) lines, but 
are not as clearly defined 
in the maps of forbidden lines with a low ionization potential ([\ion{O}{i}], [\ion{N}{ii}], [\ion{S}{ii}]). 
However, the mechanism responsible for the gas ionization in the galaxy periphery is difficult to establish. 
The diffuse emission in the H$\alpha$ and [\ion{S}{ii}] maps fills almost the whole FoV,  extending up to galactocentric distances of $\sim$2.7~kpc---
different scenarios have been proposed to explain how could Lyman continuum photons leaking from \ion{H}{ii}-regions travel such large distances without being absorbed,  but this  is still unclear
\citep[e.g.,][]{Miller1993, Dove1994, Dove2000, Hoopes2003,Giammanco2005}.
In addition, kiloparsec-scale (possibly expanding) structures, such as bubbles and filaments, are highly visible in our [\ion{O}{i}], H$\alpha,$ and [\ion{S}{ii}] maps. The existence of such well-traced structures, which is particularly difficult to reconcile with a scenario of ionization by hot stars,  is often interpreted in terms of shocks \citep{Rand1998,Collins2001}. This is further explored below  by means of diagnostic line ratio maps and diagnostic diagrams.

\begin{figure*}
\centering
\begin{subfigure}{}
\includegraphics[width=8cm]{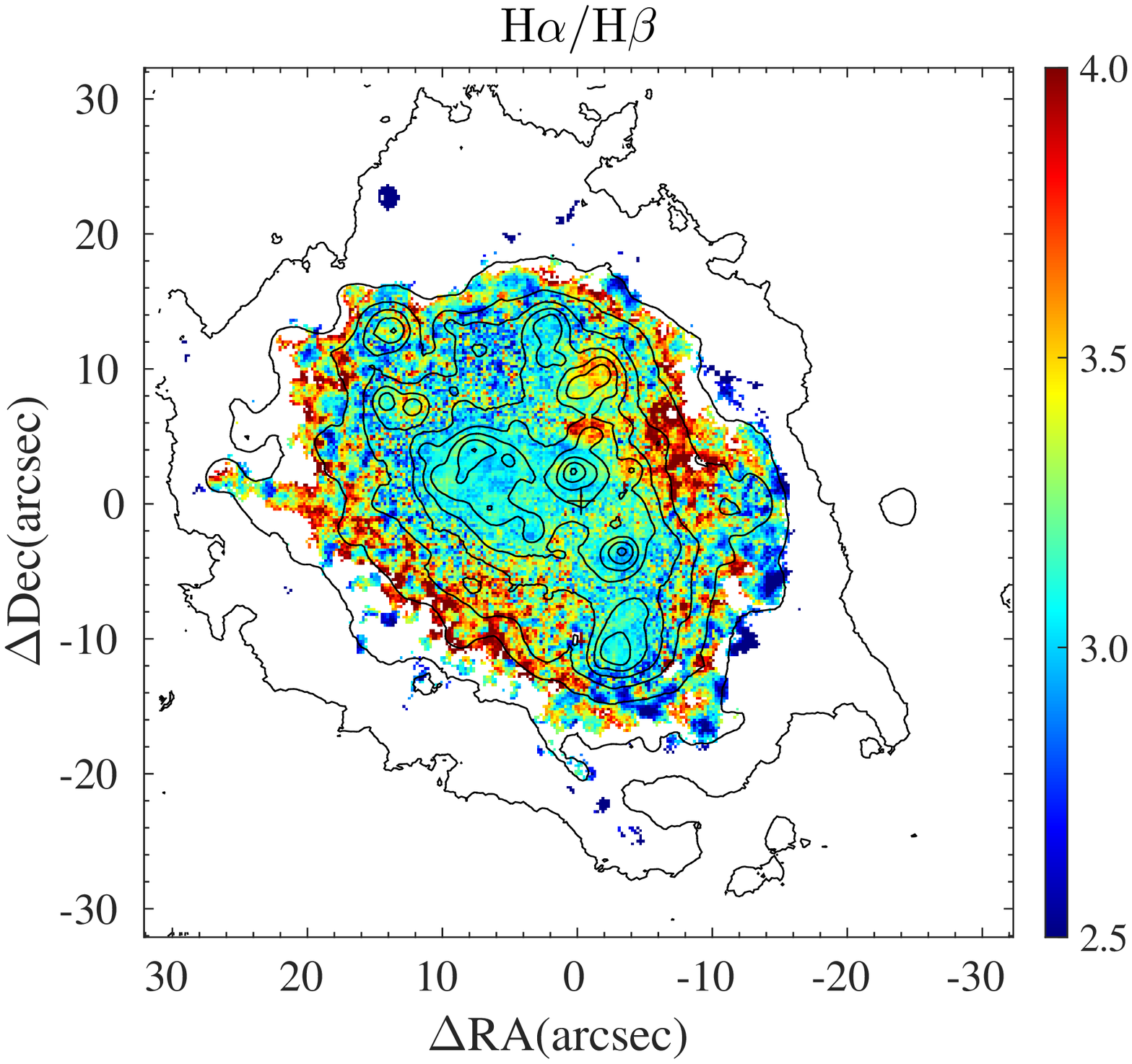}
\end{subfigure}
\begin{subfigure}{}
\includegraphics[width=8cm]{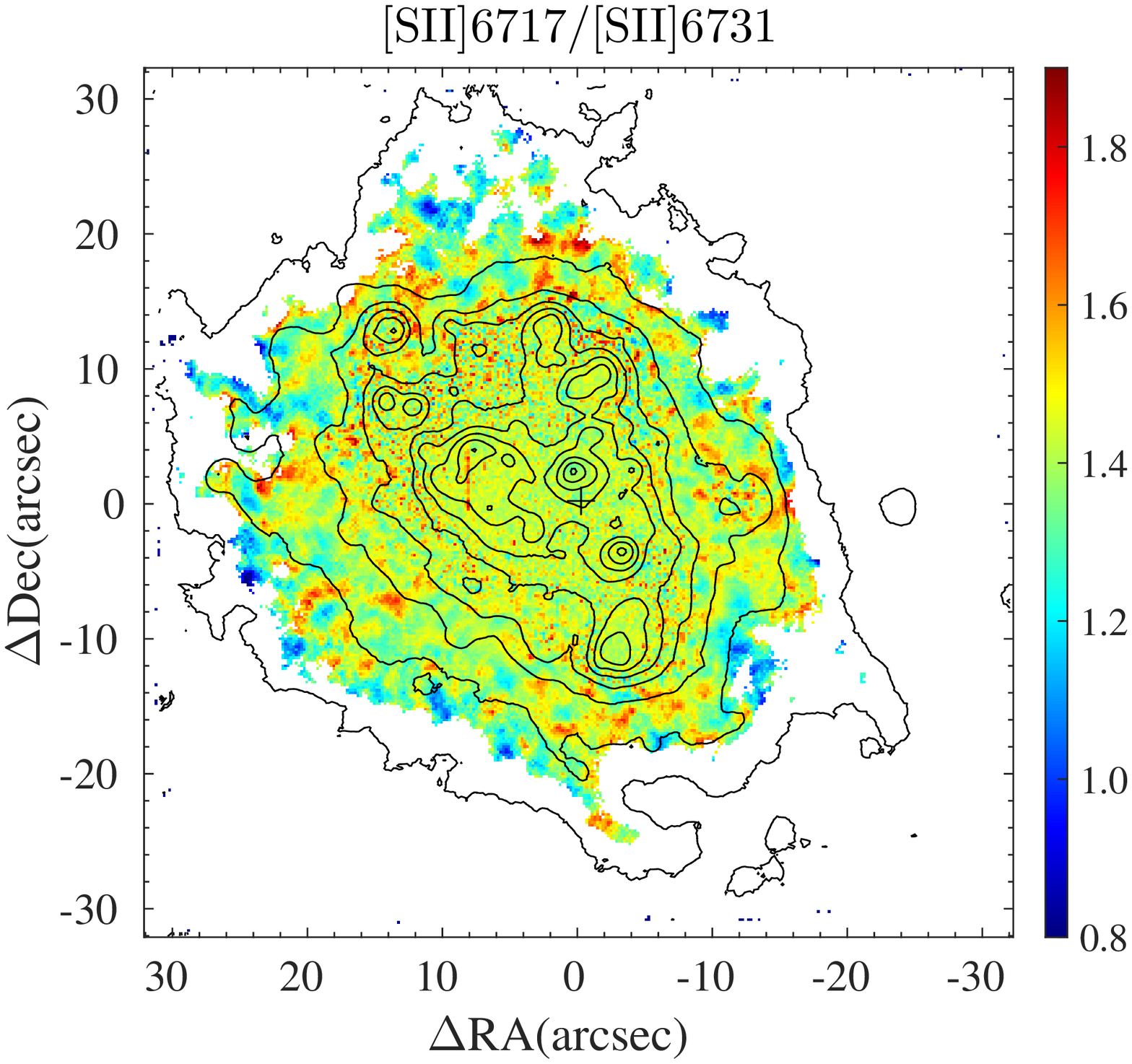}
\end{subfigure}
\begin{subfigure}{}
\includegraphics[width=8cm]{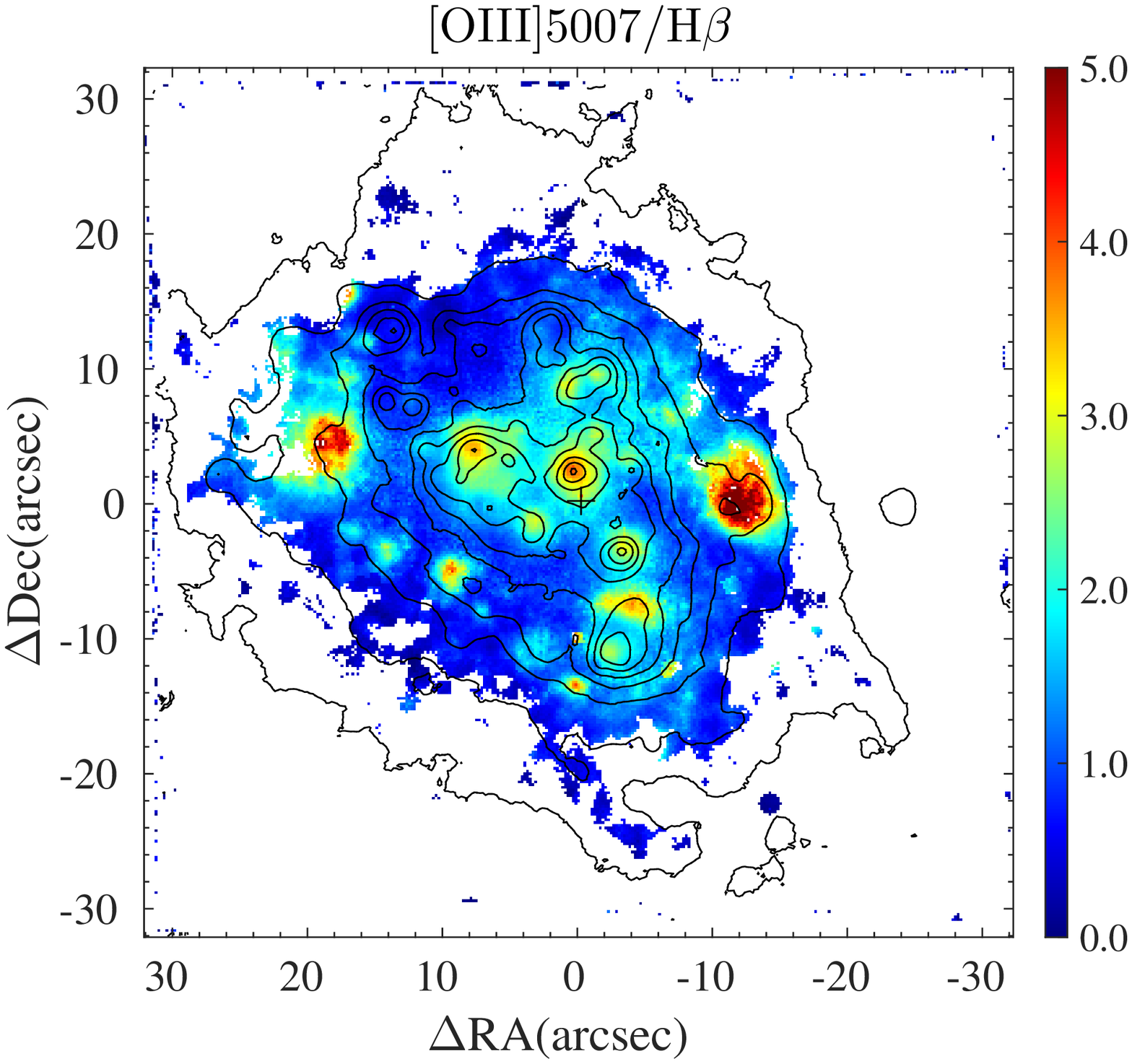}
\end{subfigure}
\begin{subfigure}{}
\includegraphics[width=8cm]{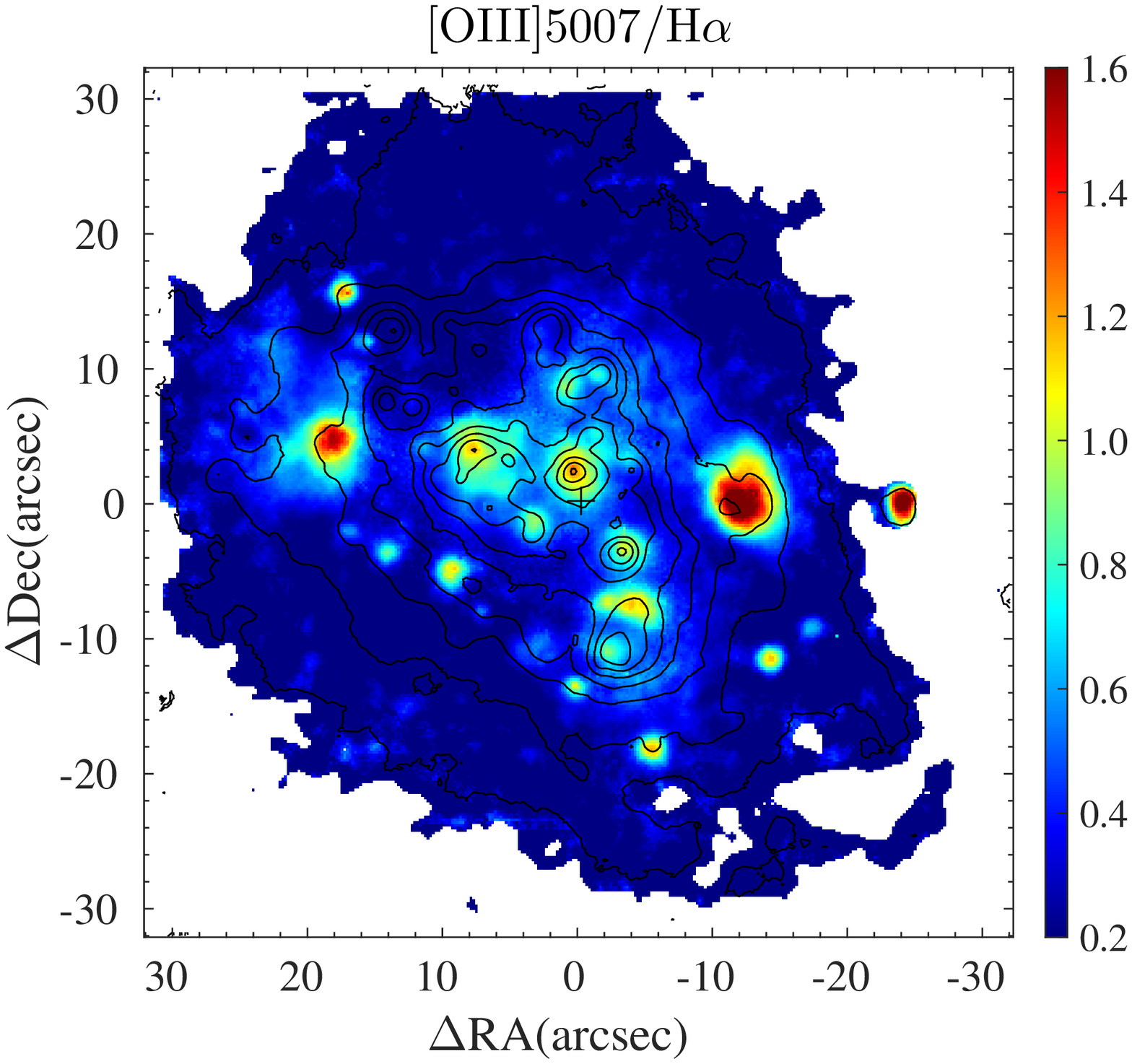}
\end{subfigure}
\caption{Emission line ratio maps of Haro~14. {\em Top left:} Balmer decrement, H$\alpha$/H$\beta$.
{\em Top right:} Ratio of the collisional excited sulfur lines, [\ion{S}{ii}]~$\lambda\lambda~6717,6731$. 
{\em Bottom left:} Excitation map, [\ion{O}{iii}]~$\lambda5007$/\Hb{}. 
{\em Bottom right:} Excitation map, [\ion{O}{iii}]~$\lambda5007$/\Ha{}.  The contour levels of the H$\alpha$ flux map are overplotted.
}
\label{Figure:diagnosticmaps} 
\end{figure*}

\begin{figure*}
\centering
\begin{subfigure}{}
\includegraphics[width=8cm]{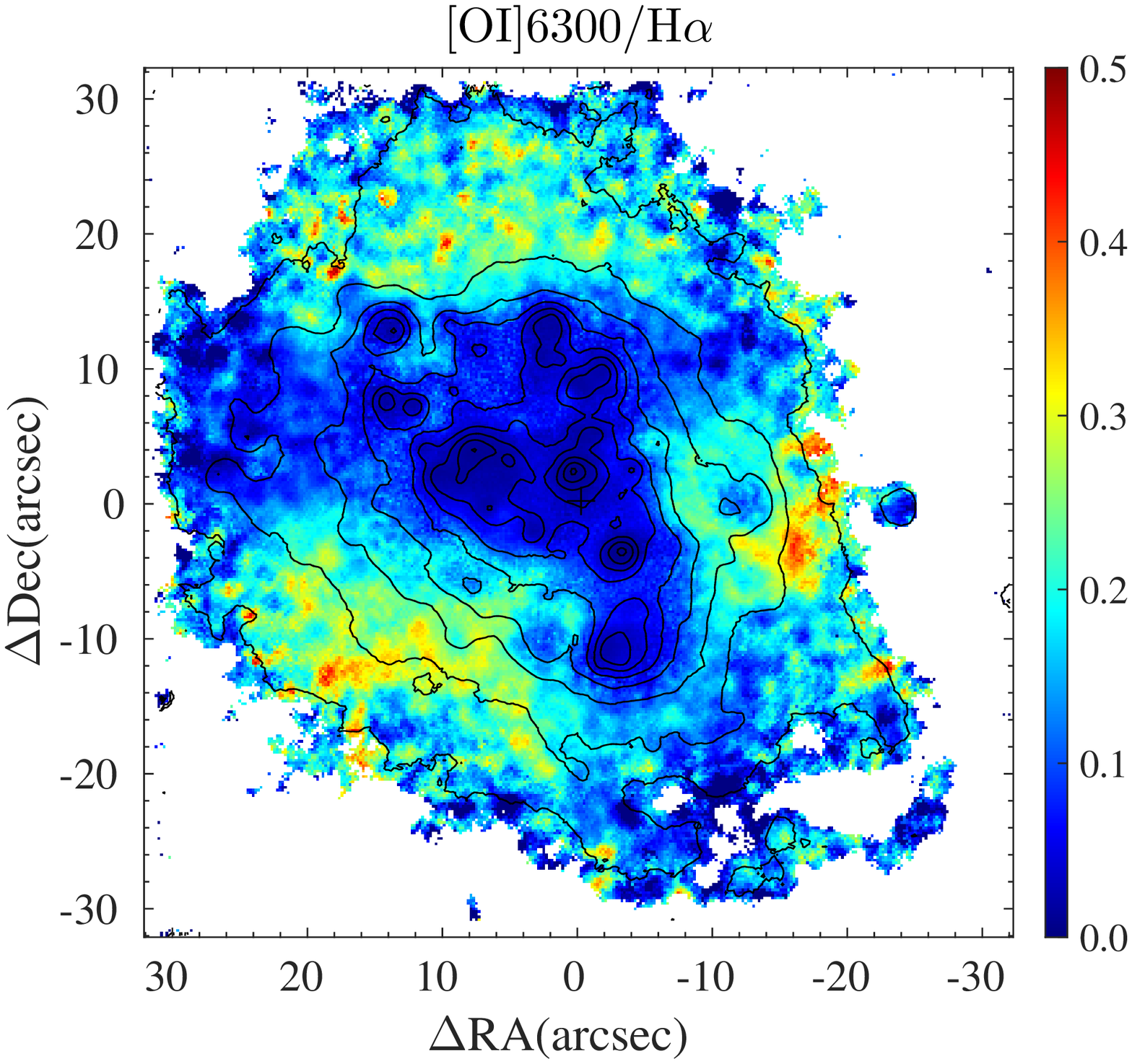}
\end{subfigure}
\begin{subfigure}{}
\includegraphics[width=8cm]{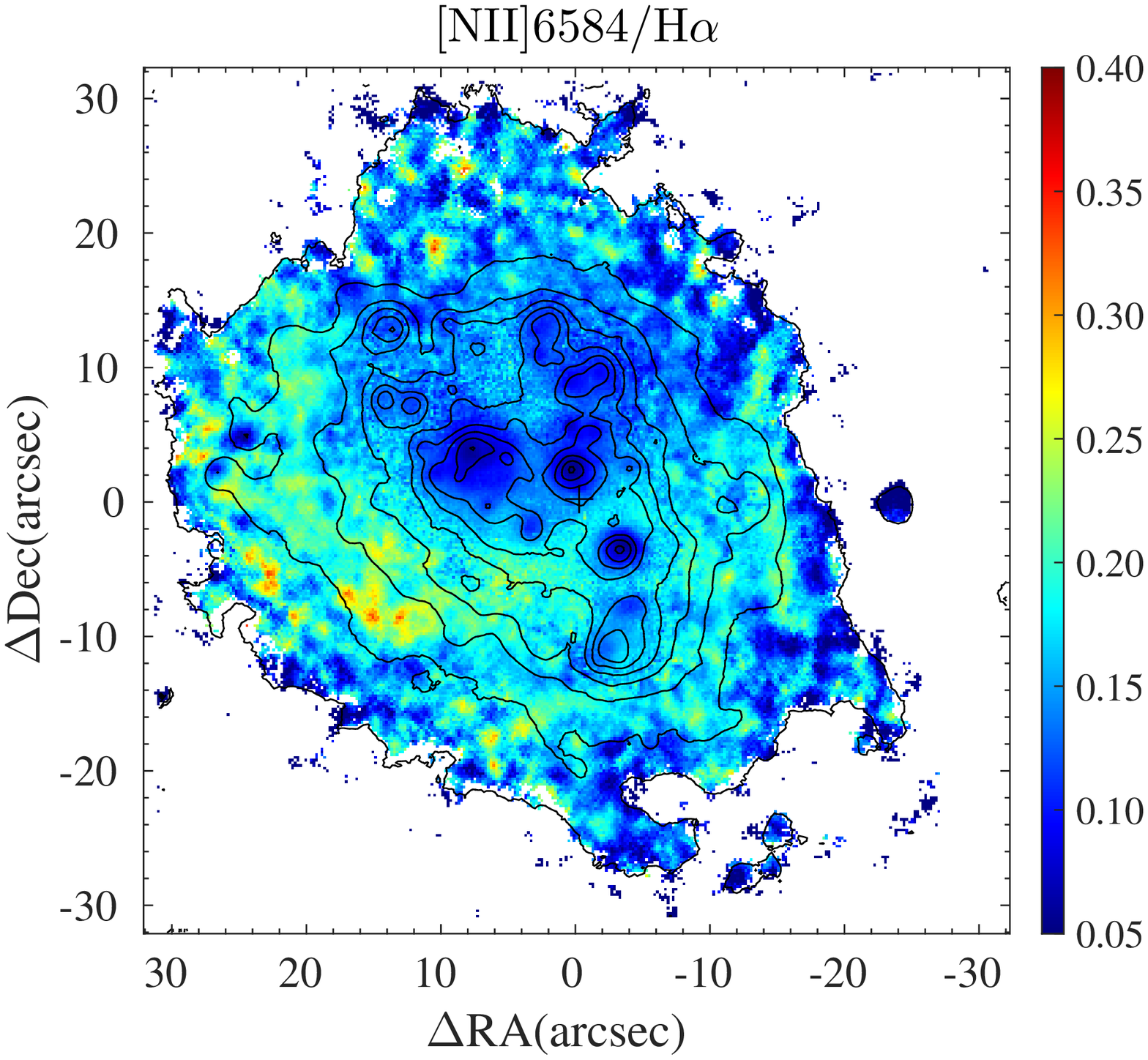}
\end{subfigure}
\begin{subfigure}{}
\includegraphics[width=8cm]{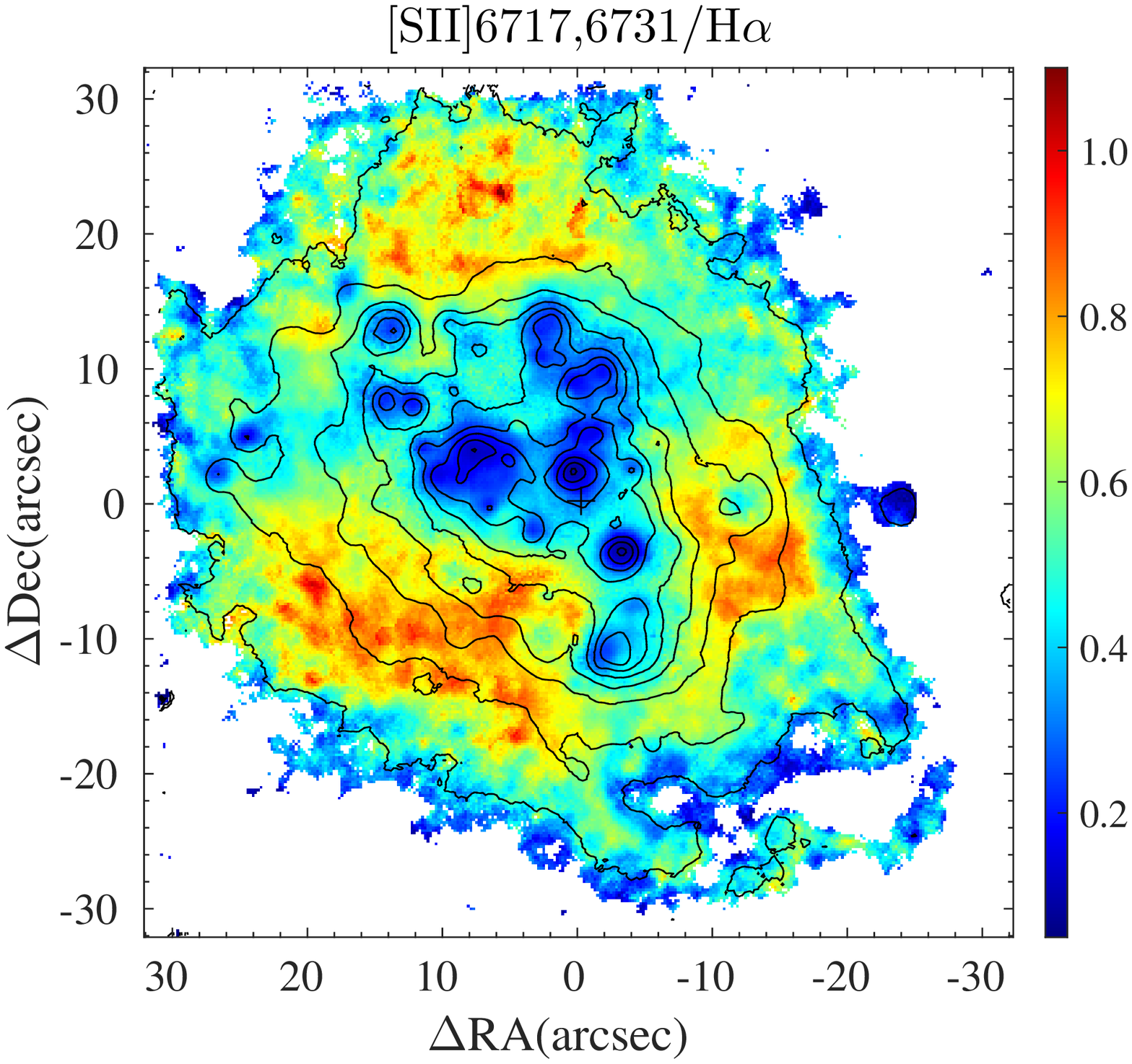}
\end{subfigure}
\begin{subfigure}{}
\includegraphics[width=8cm]{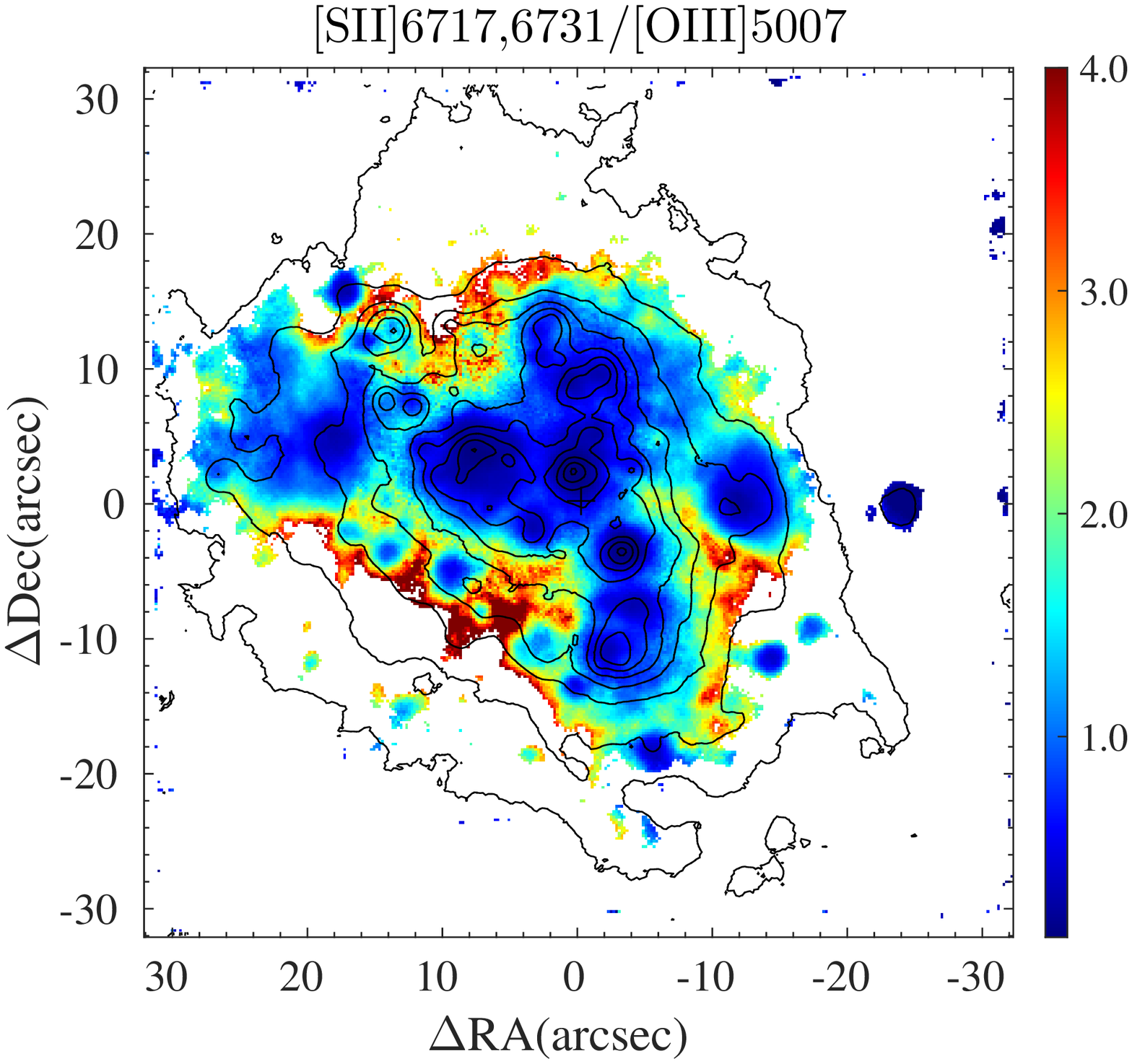}
\end{subfigure}
\caption{Diagnostic line ratio maps of Haro~14 involving low-ionization lines. 
{\em Top left:} [\ion{O}{i}]~$\lambda6300$/\Ha{}. 
{\em Top right:} [\ion{N}{ii}]~$\lambda6584$/\Ha{}. 
{\em Bottom left:}  [\ion{S}{ii}]~$\lambda\lambda6717,\,6731$/\Ha{}. 
{\em Bottom right:}  [\ion{S}{ii}]~$\lambda\lambda6717,\,6731$/[\ion{N}{ii}]~$\lambda6584$. 
The contour levels of the H$\alpha$ flux map are overplotted.}
\label{Figure:diagnosticmaps2} 
\end{figure*}

\subsection{Interstellar extinction}
\label{Section:extinctionmap}

The interstellar extinction in a nebula can be computed by comparing the observed Balmer
decrement values to the theoretical ones. The Balmer line ratios are
known from atomic theory; assuming that deviations from the predicted values are due
to absorption by dust, the interstellar extinction coefficient, C(H$_{\beta}$), 
is such that 
\begin{equation}
\frac{F_{\lambda}}{F(H_{\beta})}=\left[\frac{F_{\lambda}}{F(H_{\beta})}\right]_0\times10^{-C(H_{\beta})[f(\lambda)-f(H_{\beta})]}
\label{ext_eq}
,\end{equation}
\noindent where 
$F_{\lambda}/F(H_{\beta}$) is the observed ratio of Balmer emission-line intensities 
relative to H$_{\beta}$, $[F_{\lambda}/F(H_{\beta})]_0$ is the 
theoretical ratio, and
$f(\lambda$) is the adopted extinction law  \citep{Dopita2003,Osterbrock2006}. 
Equation~(\ref{ext_eq}) can be applied to 
every  spaxel on the frame to determine the spatial distribution of the 
dust when we work with integral field spectroscopy data.

\smallskip

We generated the  interstellar extinction map of Haro~14 from  the H$\alpha$ and H$\beta$ intensity maps. H$\alpha$ and H$\beta$ have a different PSF, and we have degraded the best-seeing map, convolving with a Gaussian to match both PSFs. We adopted a theoretical ratio of 2.87 (case~B recombination  
at a temperature of $10^4$~K; \citealp{Osterbrock2006}) and used the Galactic extinction law from
\cite{ODonnell1994}. 

\smallskip 

The derived extinction map exhibits a clear spatial pattern (see Figure~\ref{Figure:diagnosticmaps}, top-left). The lowest values of the extinction are found in the areas of intense SF, where $F(H\alpha)/F(H\beta) \sim3$ (this line ratio implies\footnote{With 
$A_{V}=2.1 \times C(H\beta)$ and $E(B-V)=0.697\times C(H\beta)$, following   \cite{Dopita2003}} magnitudes of extinction in $V$ of  $A_{V}=0.12$  and a color excess of $E(B-V)=0.04$). The extinction increases considerably 
when moving toward the periphery of the knots, reaching values of up to $F(H\alpha)/F(H\beta) \sim6$ (hence,
$A_{V}=2$ and $E(B-V)=0.64$). These results agree very well with our  previous VIMOS findings \citep{Cairos2017a}. The dust accumulates mostly in two zones ---located at the southeast and northwest---  that are somewhat symmetric with respect to the galaxy center. 
Such spatial distribution of the dust, with mostly dust-free SF areas and accumulation of dust in the periphery,  is consistent with  a scenario in which dust is destroyed or swept away by the  most massive stars.

\smallskip

The extinction map is spatially limited by the extent of the H$\beta$ emission, which is considerably fainter than H$\alpha$, [\ion{O}{iii}], or [\ion{S}{ii}]. This is clearly seen in the top-left panel of Figure~\ref{Figure:diagnosticmaps}, in which we have overplotted the H$\alpha$ contour maps (but see also Figures~\ref{Figure:emissionmaps1} and \ref{Figure:emissionmaps2}). For this reason, we have not corrected the maps presented in this paper for interstellar reddening: the extinction map presented here gives us an idea of the relative importance of the correction and its spatial variability; a detailed correction at the spaxel level would be of little value.

\smallskip 

 Large values of interstellar extinction considerably affect the fluxes, magnitudes, and luminosities, but the commonly used diagnostic line ratios ([\ion{O}{iii}]/\Hb{}, [\ion{S}{ii}]/\Ha{}, [\ion{N}{ii}]/\Ha,{} and [\ion{O}{i}]/\Ha{}) are almost insensitive to reddening (for an extinction in $V$ as large as $A_{V}=2$, these ratios would change by less than 10$\%$).

\subsection{Electron density distribution}
\label{Section:density}

Figure~\ref{Figure:diagnosticmaps} (top right) shows a map of the ratio of the collisionally excited sulfur lines,  [\ion{S}{ii}]~$\lambda\lambda6717,\,6731$, 
 which is a sensitive indicator of  density in the range $10^2$-$10^4$~cm$^{-3}$ \citep{Aller1984, Osterbrock2006}
.
\smallskip

In the central galaxy regions, where the SF is taking place, we find relatively high and homogeneous values of the [\ion{S}{ii}]~$\lambda$6717/[\ion{S}{ii}]~$\lambda$6731 ratio ($\geq$1.35),  indicative of low electron densities ($N_{\rm e}\leq 100$~cm$^{-3}$).
The ratio decreases slightly towards the outer regions, with values of between 1.2 and 0.9 (corresponding to densities 240~cm$^{-3}$ and 940~cm$^{-3}$, respectively, at $T_{\rm e}=10^4$~K).
Such a density pattern, that is,  low density in the areas of SF with increasing density towards the galaxy outskirts, is characteristic of a disturbed ISM, in which the higher densities are associated with filaments that form in the expanding fronts produced by stellar winds and SN explosions \citep{McCray1987,TenorioTagle1988}.

\begin{figure*}
\centering
\begin{subfigure}{}
\includegraphics[width=8cm]{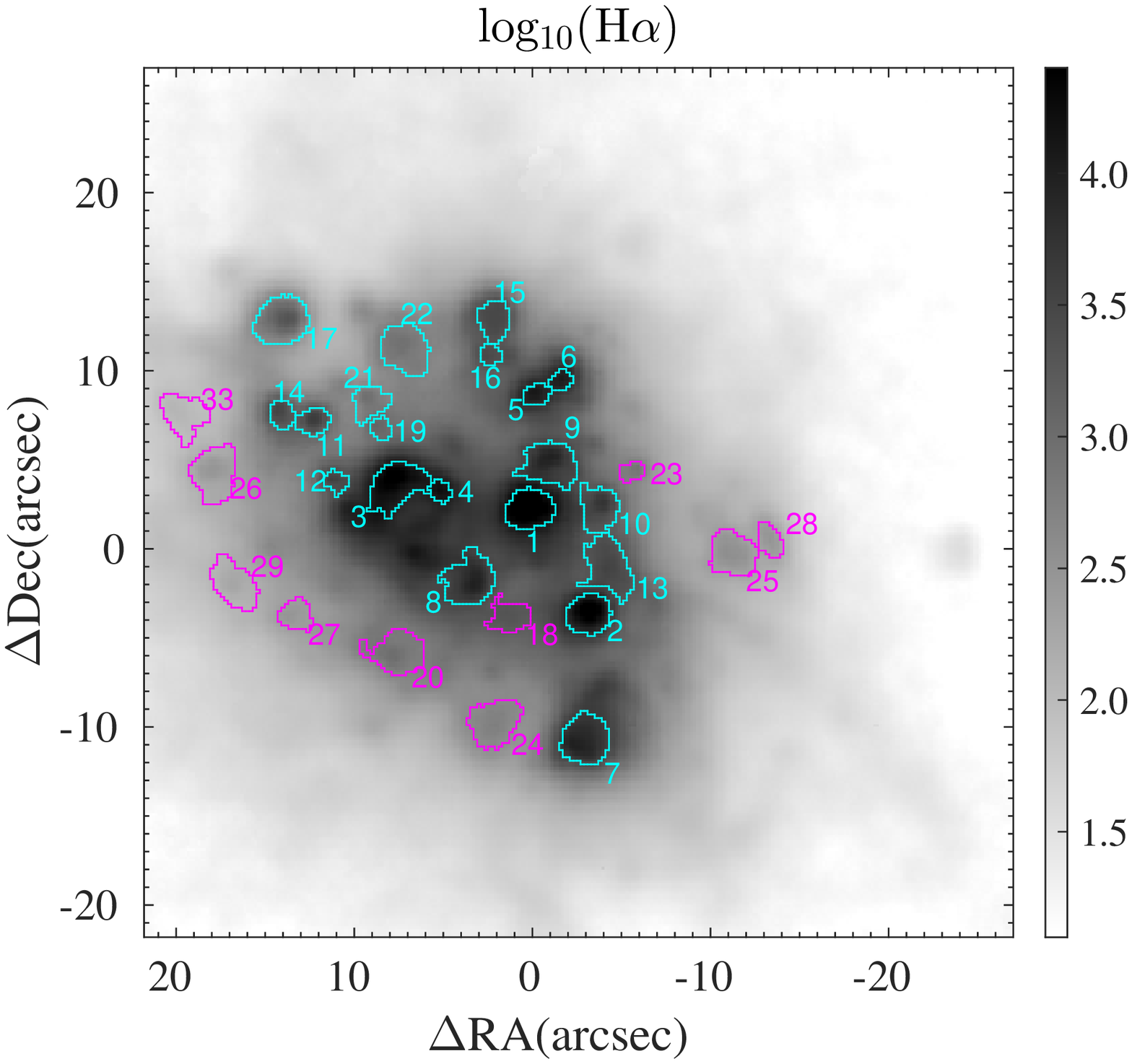}
\end{subfigure}
\begin{subfigure}{}
\includegraphics[width=8cm]{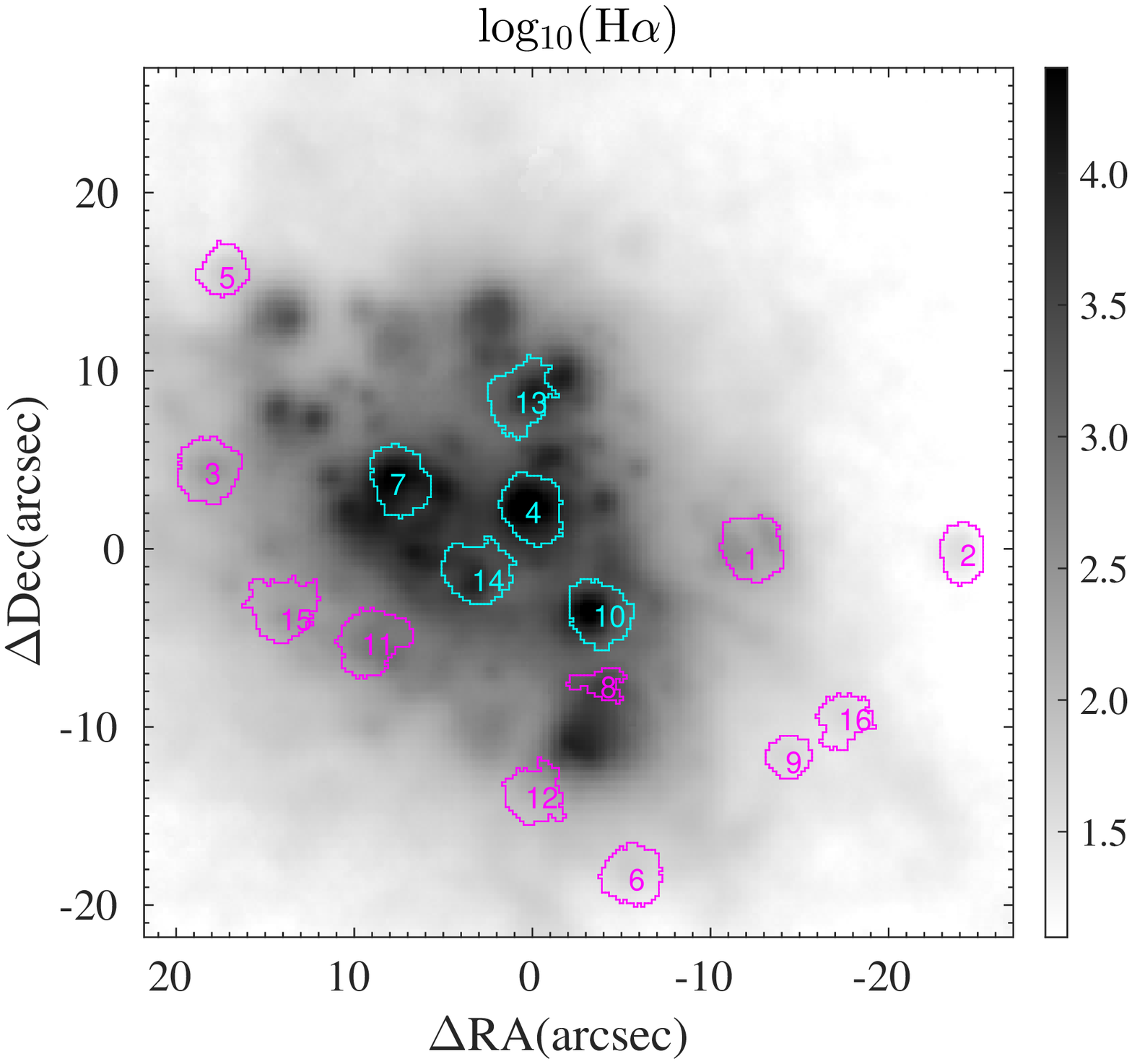}
\end{subfigure}
\caption{Individual regions of Haro~14 for which an integrated spectrum has been generated, which is overplotted 
on a H$\alpha$ flux map. The color of the knots indicates their position in the [\ion{O}{iii}]/\Hb{} versus [\ion{O}{i}]/\Ha{} diagnostic diagram: knots below the maximum starburst line are shown in blue, and knots above and to the right of the line are shown in magenta (see also Figure~\ref{Figure:diagnostic-brotes}). 
{\em Left panel:} Emission sources identified in the H$\alpha$ flux map. 
 Labels correspond to their number (L1, L2, ..., L33) in the catalog of Tables 3, A1, and A2.
{\em Right panel:} Regions of high excitation identified in the galaxy. 
 Labels correspond to their number (Ex1, Ex2, ..., Ex16) in the catalog of Tables 4 and 5. }
\label{Figure:emissionlinesources} 
\end{figure*}

\subsection{Diagnostic line ratio maps}
\label{Section:lineratiomaps}

A nebula can be excited and ionized by means of different mechanisms, the most common ones being  photoionization by OB stars, photoionization by a power-law continuum (AGN), and shock-wave heating \citep{Aller1984,Dopita2003,Osterbrock2006}. Because the characteristics of the emitted spectrum
strongly  depend on the power source, the strength and spatial distribution of emission lines can be used to constrain the excitation and ionizing mechanism(s). 
In particular, emission line intensity ratios involving strong lines, such as   [\ion{O}{iii}]/\Hb{},
[\ion{N}{ii}]/\Ha, [\ion{S}{ii}]/\Ha{}, and also
[\ion{O}{i}]/\Ha{},  are highly sensitive to the properties of the excitation source \citep{Baldwin1981,Veilleux1987}. The deep MUSE observations enable us to build line ratio maps that easily reach the galaxy periphery and allow us to probe the ionization structure of Haro~14 up to kiloparsec scales.

\smallskip

\begin{figure*}
\centering
\includegraphics[angle=0, width=\linewidth]{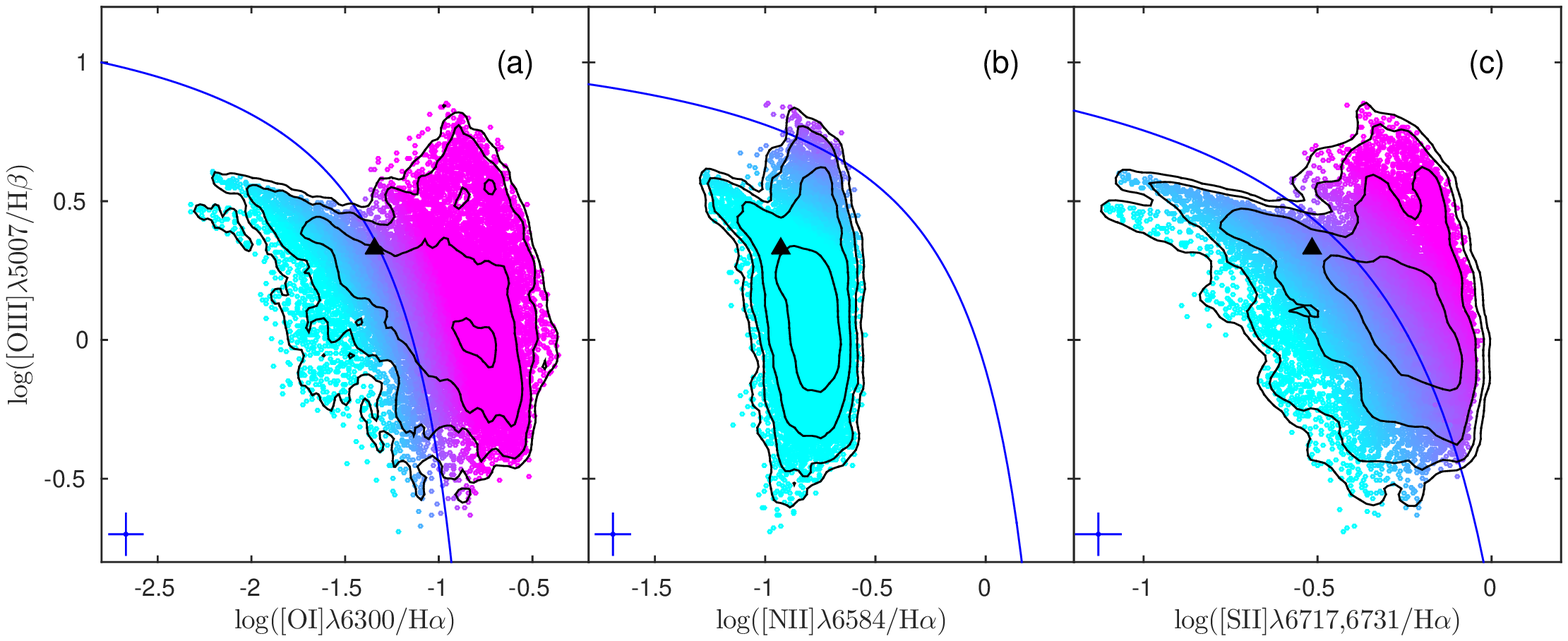}
\caption{Optical emission line  diagnostic
diagrams  for the individual spaxels in Haro~14.  
The solid blue line in the panels delineates the theoretical ``maximum starburst line''  
derived by \cite{Kewley2001b}; to better visualize the results, the points are
color-coded according to their distance to this line. 
In the three diagrams, the lower-left section of the plot is occupied by spaxels in which the dominant energy source is the radiation from hot stars (blue points in the figure); additional ionizing mechanisms
shift the spaxels to the top right and right part of the diagrams (from cyan to pink). 
The contours show the density of spaxels; each contour level corresponds to a density four times higher than the  next outer one. 
The cross in each diagram gives an estimate of the typical uncertainties on the ratios.
The black triangles indicate the diagnostic ratios for the integrated spectrum of the whole galaxy. 
}
\label{Figure:diagnostic-spaxel} 
\end{figure*}

\begin{figure*}
\centering
\begin{subfigure}{}
\includegraphics[width=8cm]{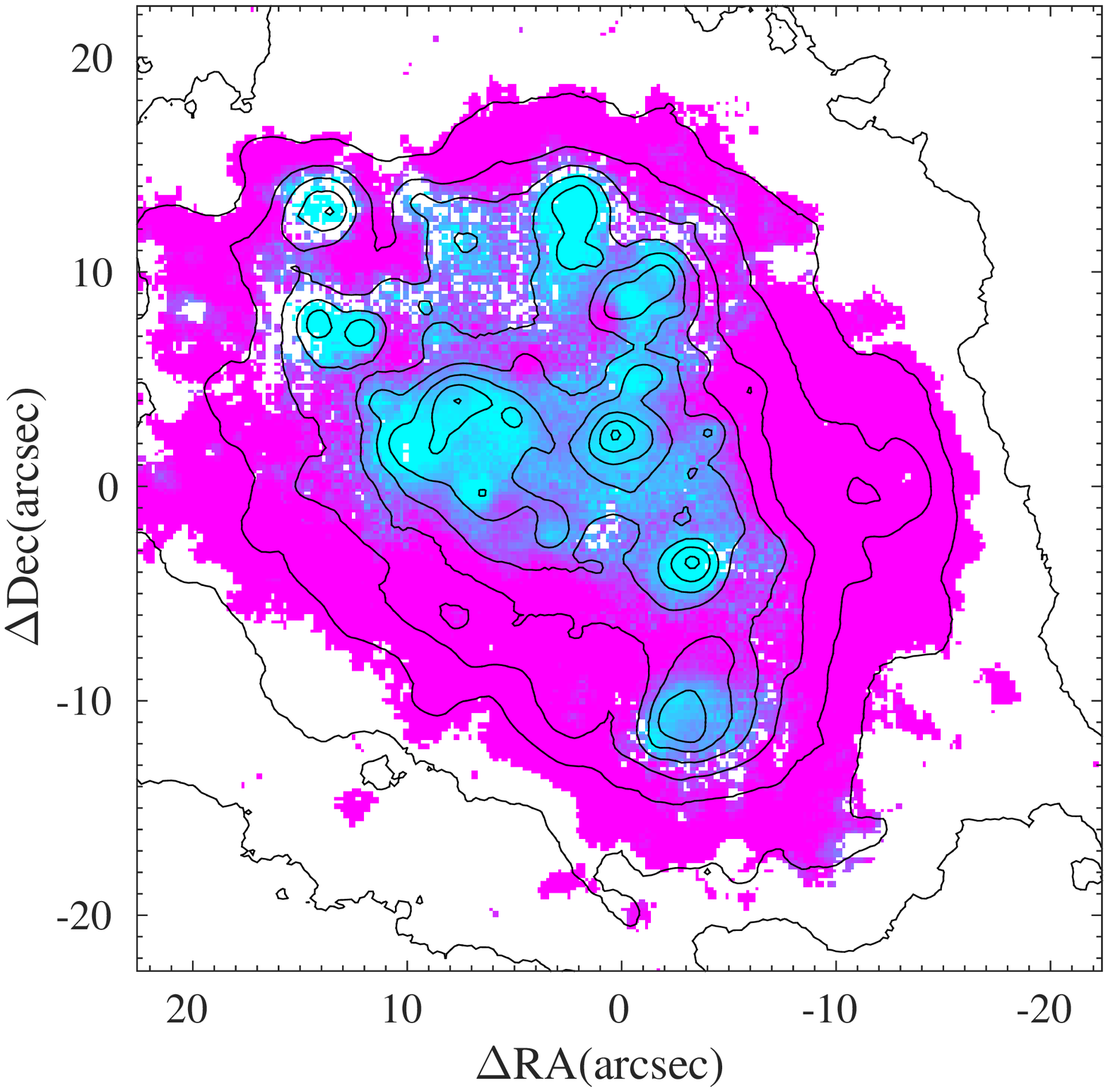}
\end{subfigure}
\begin{subfigure}{}
\includegraphics[width=8cm]{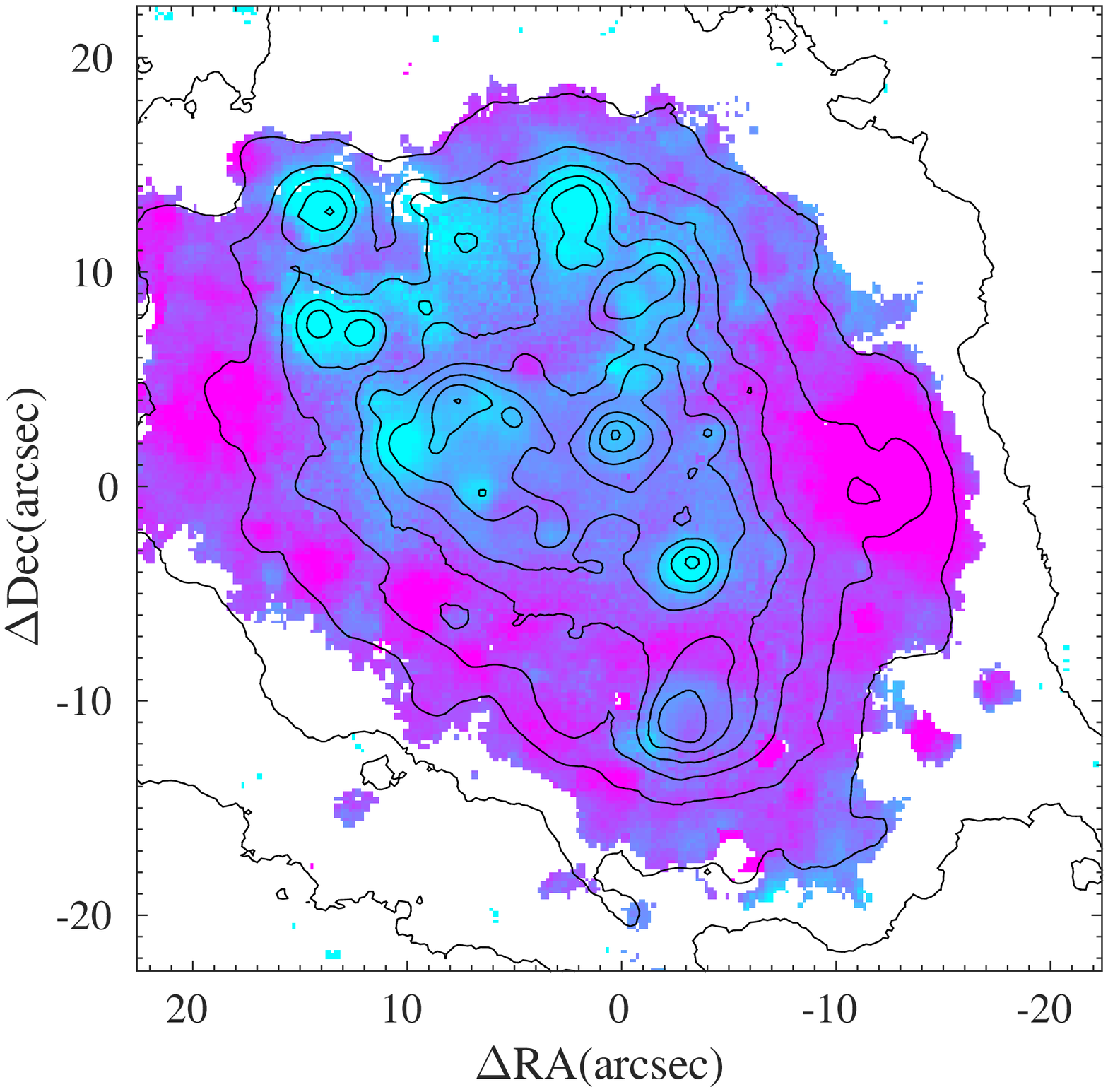}
\end{subfigure}
\caption{ Spaxels represented in the diagnostic diagrams of Haro~14  localized on the galaxy, 
following  the color code in Figure~\ref{Figure:diagnostic-spaxel}. Contours in H$\alpha$ are overplotted. {\em Left:} Spaxels in the [\ion{O}{i}] diagram: the largest fraction of the area is being ionized by a mechanism different from SF. {\em Right:} Spaxels in the [\ion{S}{ii}] diagram. 
}
\label{Figure:nonphotoregions} 
\end{figure*}

Emission line ratio maps of Haro~14, generated as described in Section~\ref{Section:themaps}, are displayed in Figures~\ref{Figure:diagnosticmaps} and ~\ref{Figure:diagnosticmaps2}. The correction for interstellar extinction is  negligible for the classical diagnostic line ratios ([\ion{O}{iii}]/\Hb{},
[\ion{N}{ii}]/\Ha, [\ion{S}{ii}]/\Ha{}, and 
[\ion{O}{i}]/\Ha{}) as the lines of each  pair  are very close in wavelength, but it does play a role in the  [\ion{O}{iii}]/\Ha{} map (see below). 

\smallskip

The [\ion{O}{iii}]/\Hb{} map (bottom left panel in Figure~\ref{Figure:diagnosticmaps}) exhibits an unusual and quite intriguing pattern, 
with the  excitation fluctuating strongly across the galaxy, and showing values ranging from about 0.8 to 5.5.
The inner area in the map reproduces our VIMOS findings \citep{Cairos2017a}: 
 the  excitation in the central knots  that define the linear  and horseshoe-like structures   is typical of regions photoionized by hot stars ([\ion{O}{iii}]/\Hb{}$\sim$3-3.5), but similarly high values are also found in several areas not associated with bright SF regions.
The MUSE excitation map reaches much larger galactocentric distances and clearly demonstrates that 
[\ion{O}{iii}]/\Hb{} does not peak in the central, brightest  \ion{H}{ii}-regions but in several clumps located in the galaxy periphery.  The  excitation reaches its maximum  ([\ion{O}{iii}]/\Hb{}$\sim$5.5)  in a large blob located  $\sim$12~arcsec (=750~pc) west from the continuum peak  (knot~Ex1 in Figure~\ref{Figure:emissionlinesources}). Another high-excitation blob ([\ion{O}{iii}]/\Hb{}$\sim$4.5) is situated $\sim$19~arcsec (1200~pc) east from the center (knot~Ex3 in Figure~\ref{Figure:emissionlinesources}). An interesting excitation peak, elongated and forming an arc,  is found in the central bar-like structure (knot~Ex8 in Figure~\ref{Figure:emissionlinesources}).  Various other secondary excitation peaks are  distinguishable in the southeast region,  close to the borders of the mapped area. 

\smallskip

The spatial  extent of [\ion{O}{iii}]/\Hb{} is considerably  limited because of
the relatively weak emission in H$\beta$. 
To better visualize the structure of the excitation in the galaxy outskirts, we also built the [\ion{O}{iii}]/\Ha{} map (Figure~\ref{Figure:diagnosticmaps}, bottom right). 
The [\ion{O}{iii}]/\Ha{} ratio is not as commonly used as a diagnostic because it is affected by interstellar extinction,  [\ion{O}{iii}] and \Ha\ being rather far apart in wavelength. 
Because the effect of extinction would be to lower the [\ion{O}{iii}] emission with respect to \Ha{}, [\ion{O}{iii}]/\Ha{} must be understood as  a lower bound to the excitation. In the case of Haro~14, the structure of  [\ion{O}{iii}]/\Ha{}  closely follows 
[\ion{O}{iii}]/\Hb{} in the areas where they overlap; 
as the ratio [\ion{O}{iii}]/\Hb{} is extinction independent,  dust does not seem to affect the spatial structure of the map in these regions.

\smallskip

The [\ion{O}{iii}]/\Ha{} map traces areas of high excitation in the galaxy outskirts with high fidelity. 
In this map, we clearly resolve a new blob situated about 24\arcsec (1.5~kpc) west from the peak in continuum  (knot~Ex2 in Figure~\ref{Figure:emissionlinesources}); interestingly,
it aligns with the other very high-excitation blobs  in a direction roughly perpendicular to the central SF bar. 
Several other excitation maxima are detected, most of them lying along two chains in the southeast and the southwest galaxy regions; they delimit a border
beyond which the excitation remains uniformly low. 
As expected, the high -excitation regions correspond to clumps that are brighter and much more clearly traced in the high-ionization [\ion{O}{iii}] line  than in H$\alpha$. Most of these high-excitation sources appear compact and roundish   in [\ion{O}{iii}], which suggests that they are associated with SF regions rather than with extended shocked areas. There is one clear exception: the elongated excitation peak  (knot~Ex8 in Figure~\ref{Figure:emissionlinesources}). 

\smallskip

The  [\ion{O}{i}]/\Ha{} map (Figure~\ref{Figure:diagnosticmaps2}, top-left)  displays a similar spatial pattern to the excitation, but here the clumps and/or peaks are not as well delineated. 
Very low [\ion{O}{i}]/\Ha{} ($\leq$0.05) values,  characteristic of \ion{H}{ii}-regions, appear in a band crossing the galaxy east to west in the northern region ($0"\leq\Delta~$Dec$\leq+18"$) and in the whole southwest area. The zone of low  [\ion{O}{i}]/\Ha{}  at the southwest coincides with the interior of the cavity delineated by the large filaments.
The galaxy outskirts present mostly high [\ion{O}{i}]/\Ha{}, with values of between 0.2 and 0.5---the largest ratios are reached in the  areas of  enhanced emission in the [\ion{O}{i}] intensity map (see Figure~\ref{Figure:emissionmaps2}, bottom-left).


\smallskip 

The [\ion{N}{ii}]/\Ha{} and [\ion{S}{ii}]/\Ha{} maps (Figure~\ref{Figure:diagnosticmaps2}, top-right and bottom-left) display a similar morphology (also roughly similar to  [\ion{O}{i}]/\Ha{}), but the outer galaxy regions are better traced  in the 
more extended [\ion{S}{ii}]/\Ha{} map. In both diagnostic maps,  the central clumps  appear well delineated and show values characteristic of \ion{H}{ii} photoionization  ([\ion{N}{ii}]/\Ha{} $\leq$0.15 and [\ion{S}{ii}]/\Ha{} $\leq$0.5) and the line ratios increase when we move outwards. The  ratio [\ion{S}{ii}]/\Ha{} rises up to 1 in the regions of the periphery that also shows enhanced values of  [\ion{O}{i}]/\Ha{}.


Photoionization models alone cannot accurately predict [\ion{O}{i}]/\Ha{} $\geq$0.2 and [\ion{S}{ii}]/\Ha{}  $\geq$0.8 (see, e.g.,  
\citealp{Mathis1986,Domgorgen1994,Hoopes2003}); larger values of these diagnostic line ratios have previously been explained in terms of  shock heating  \citep{Shull1979,Dopita1995,Allen1999,Rich2010,Rich2011}.

\subsection{Spatially resolved diagnostic diagrams}
\label{Section:diagnosticdiagram}

Line ratio diagnostic diagrams are simple but effective  tools to discriminate emission-line galaxies according to their dominant excitation and ionization  mechanism \citep{Baldwin1981,Veilleux1987,Kewley2001b}. This technique was originally developed for single-aperture spectroscopy and  mostly applied to integrated and nuclear spectra of galaxies \citep{Kewley2001a,Kauffmann2003,Kewley2006,Yuan2010}. However, it has gained a new, more profound diagnostic potential 
with the advent of integral field spectroscopy, in particular in combination with
the ability of mapping: tracing individual regions on the diagnostic diagram back to places on the object. Applying the diagnostic diagram approach to integral field data allows the excitation mechanism to be constrained over large galaxy areas and  the various power sources operating on the same object to be effectively located and identified \citep{Sharp2010,Rich2010,Rich2011,Rich2014,Rich2015,Leslie2014,Davies2014a,Davies2017,Cairos2017a,Cairos2017b,Cairos2020}.

\smallskip 

Figure~\ref{Figure:diagnostic-spaxel} presents the [\ion{O}{iii}]/\Hb{} versus [\ion{O}{i}]/\Ha{} (hereafter, [\ion{O}{i}] diagram), [\ion{O}{iii}]/\Hb{} versus [\ion{N}{ii}]/\Ha{} (hereafter, [\ion{N}{ii}] diagram), and [\ion{O}{iii}]/\Hb{} versus [\ion{S}{ii}]/\Ha{} (hereafter, [\ion{S}{ii}] diagram) diagnostic diagrams of Haro~14. We represent the line ratios for individual spaxels  in the plots,  in each case taking into account  only those with S/N$>$3 in the corresponding lines.   In each diagram, we also show the 
"maximum starburst line" derived by \cite{Kewley2001b}, which marks the theoretical limit 
for gas photoionized by young massive stars: 
the excitation of spaxels situated above and to the right of this line cannot be explained 
via SF alone and must have a significant contribution from alternative mechanisms (e.g., shocks or AGNs). 
The points in the figures are color-coded according to their distance to this maximum starburst line. We note that we have not made use of any rebinning strategy to increase the S/N
(one point corresponds to one spaxel); 
therefore, the density of points in a given region of the diagram
faithfully represents  the area of the galaxy with those properties.

\smallskip

The first result that emerges from Figure~\ref{Figure:diagnostic-spaxel} is the large discrepancy  between the  [\ion{N}{ii}]  (panel (b) in the figure) and the [\ion{O}{i}] and [\ion{S}{ii}] diagrams (panels (a) and (c)).
While the [\ion{O}{i}] and [\ion{S}{ii}] diagrams exhibit a similar shape, with the individual spaxels widely spread in both axes, the points appear  concentrated in a vertical strip  in the 
[\ion{N}{ii}] plot; the same behavior was found by  \cite{Cairos2017a}, building the diagrams for the smaller FoV of VIMOS. This is expected, because the  [\ion{N}{ii}] diagram  
is degenerated at low metallicities (0.2~Z$_\odot$<Z<0.4~Z$_\odot$), where the areas ionized by shocks  and hot stars  overlap \citep{Allen2008,Hong2013}. At the metallicity of Haro~14  (Z$\sim$Z$_\odot$/3; \citealp{Cairos2017a}), this diagram has a diminished diagnostic power. Therefore,  in the following we focus on the findings from the other two plots.

\smallskip

A striking feature of the diagnostic diagrams 
is the large fraction of spaxels that falls outside the zone corresponding to photoionization by hot stars. 
The spaxels above the starburst line in the  [\ion{O}{i}] and 
[\ion{S}{ii}] diagrams
represent $\sim$75\% and $\sim$50\% of the area of Haro~14, respectively. This is an important result that is in remarkable contrasts with the findings of previous works: in the spatially diagnostic diagrams built for BCGs so far,  the majority of datapoints appear situated below the maximum starburst line, from which it has often been concluded that hot massive stars are the dominant power source in these galaxies
\citep{Calzetti2004,James2010,James2016,Kehrig2008,Kehrig2016,Cresci2017,Kumari2017,Kumari2019,Oparin2020}. 

\smallskip 

The reason for the discrepancy between our findings and those of earlier studies is fundamentally of an observational nature. 
The dataset analyzed here represents a major advancement with respect to previous works, because
of the wider FoV and higher sensitivity  provided by MUSE.  Regarding IFS-based investigations, the main drawback until now has 
been their limited FoV, barely covering the 
starburst regions  (e.g., \citealp{ Kehrig2008,James2010,Kumari2017,Kumari2019}); but even working with a larger FoV, typical 2-4m telescopes do not allow accurate measurements of the outer regions of BCGs to be obtained within reasonable observing times (e.g., \citealp{Kehrig2016}). Alternatively, 
the ionized gas in a sample of nearby starbursts, including several BCGs,  was investigated by 
\cite{Calzetti2004} and \cite{Hong2013} by means of narrow-band imaging from the Hubble Space Telescope (HST). 
Although the mapped FoV was also restricted mostly to the starburst due to the proximity of the objects, the higher angular resolution of the HST allowed to identify shock-excited regions around the bright star forming clusters. These works reported a  small though discernible contribution of shocks.
Our VIMOS observations, which included the whole starburst and a significant fraction of the extended emission in Haro~14, Tololo~1937-423, and Mrk~900 also found a significant fraction of spaxels ionized by shocks \citep{Cairos2017a,Cairos2017b,Cairos2020}. 

\smallskip 

Therefore, the spatially resolved diagnostic diagrams previously built for BCGs did not allow the power sources operating in the galaxies to be probed, because they simply did not contain information on the whole system. It is not surprising to find that the dominant ionizing mechanism is OB stars, when the observations trace primarily the HSB regions of the galaxy, but this finding should not be extrapolated to the whole galaxy: conclusions regarding the ionizing mechanism acting in the faint galaxy regions should be considered only tentatively. The observations presented here, reaching  a  surface brightness of about 10$^{-18}$~erg~s$^{-1}$~cm$^{-2}$~arcsec$^{-2}$ 
in all diagnostic lines (including the faint [\ion{O}{i}]$\lambda$6300), allowed us, for the first time, to accurately investigate the excitation mechanism operating in the outer regions of a typical BCG.

\smallskip

We display in Figure~\ref{Figure:nonphotoregions} the spatial location on the galaxy of the spaxels following the same color code that we employed in the diagnostic diagrams. Here, it is clearly shown that 
spaxels within the photoionized zone in the diagrams can be traced back to the central SF regions,  
while the spaxels laying beyond the maximum starburst line  correspond    
mostly to fainter regions in the galaxy outskirts. 
The spaxels located outside the \ion{H}{ii}-region ionization zone 
in the [\ion{O}{i}] and [\ion{S}{ii}] diagrams 
account for $\sim$20\% and $\sim$13\% of the H$\alpha$ luminosity, respectively. This is a significant fraction, particularly if we take into account the fact that the maximum starburst line represents a rather conservative approach and that  spaxels lying below this line can also have a non-negligible  contribution from a harder ionizing source  \citep{Kauffmann2003,Kewley2006}. For comparison, for NGC~4214, NGC~3077, and NGC~5253, 
\cite{Calzetti2004} reported  a fraction of H$\alpha$ emission contributed by mechanisms other than OB-star photoionization  of around 3\%--4\%.

\subsection{Integrated spectroscopy}
\label{Integrated}

\begin{figure*}
\centering
\includegraphics[angle=0, width=\linewidth]{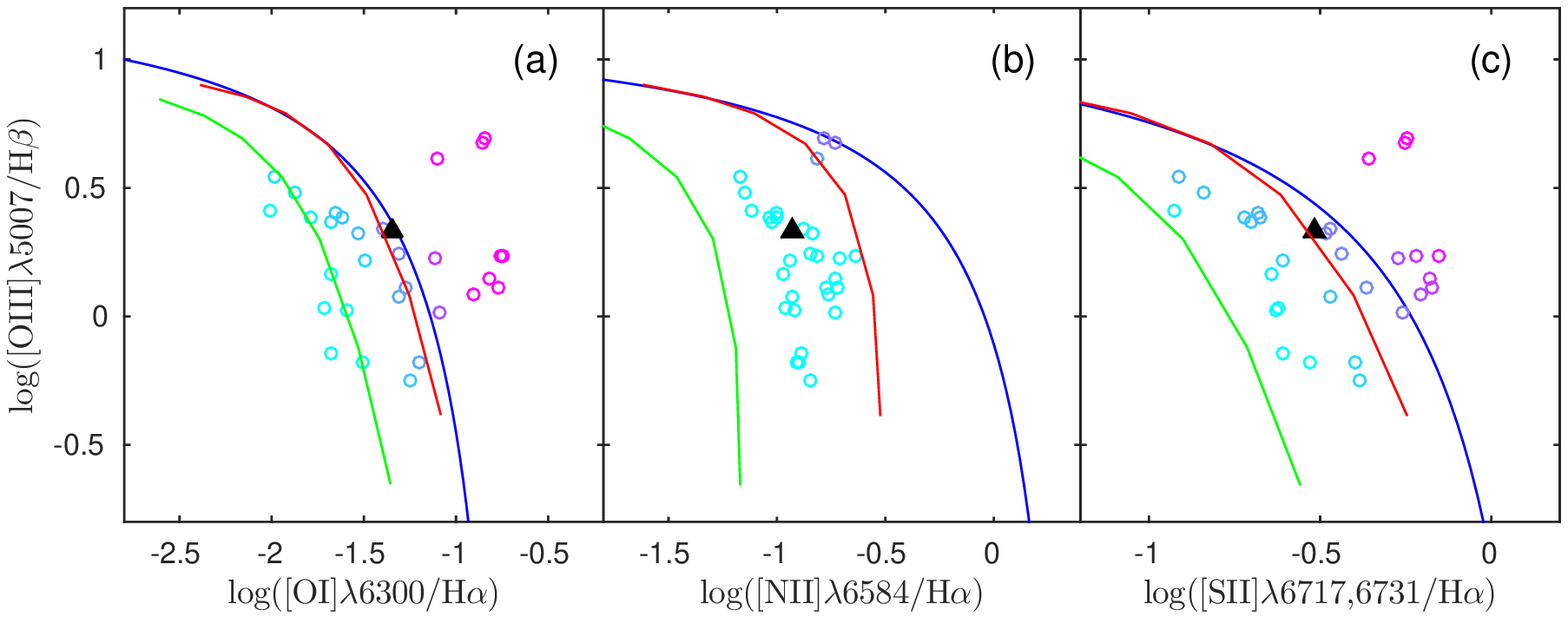}
\caption{As in Figure~\ref{Figure:diagnostic-spaxel},  but for the integrated spectra of the individual sources detected in H$\alpha$. 
The black triangles indicate the diagnostic ratios for the integrated spectrum of the whole galaxy. The red and green lines are the photoionization models  from \cite{Kewley2001b} for metallicities 0.5~Z$_\odot$  and 0.2~Z$_\odot$. }
\label{Figure:diagnostic-brotes} 
\end{figure*}

\begin{figure*}
\centering
\includegraphics[angle=0, width=\linewidth]{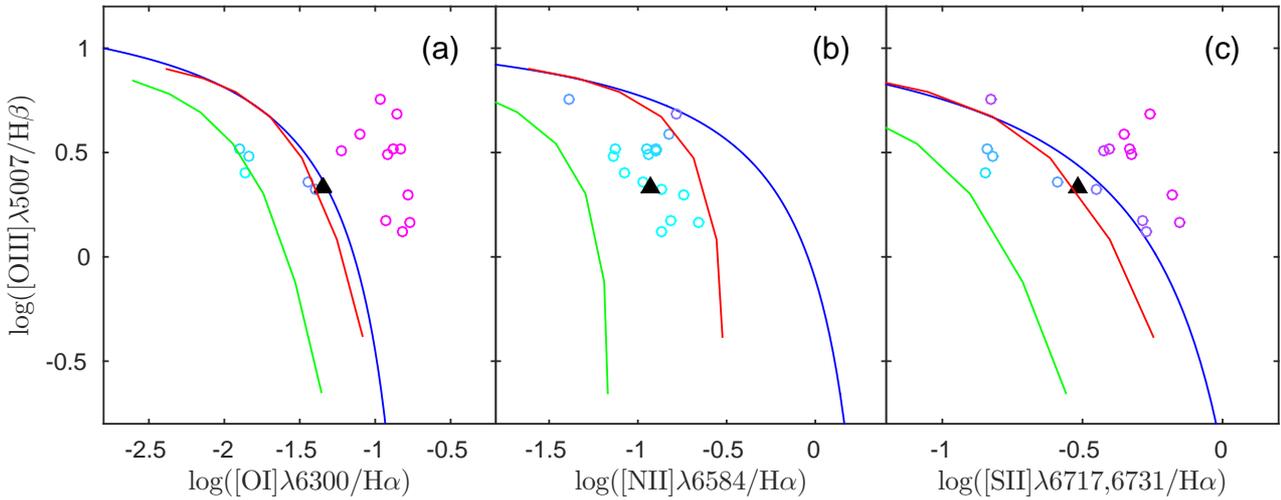}
\caption{As in Figure~\ref{Figure:diagnostic-spaxel}, but for the integrated spectra of the high-excitation blobs. The solid blue line in the panels delineates the theoretical ``maximum starburst line''  
derived by \cite{Kewley2001b}.
The black triangles indicate the diagnostic ratios for the integrated spectrum of the whole galaxy. The red and green lines are the photoionization models  from \cite{Kewley2001b} for metallicities 0.5~Z$_\odot$  and 0.2~Z$_\odot$. }
\label{Figure:diagnostic-excitation} 
\end{figure*}

The maps of Haro~14 enable us to identify interesting structures in the galaxy, such as areas of enhanced emission in a particular line, dust patches, and high-excitation blobs. As the next step in our analysis, we produced the integrated spectrum for Haro~14 and for the most distinctive structures, 
which include the brightest H$\alpha$ emitters and the regions of highest excitation. 
\subsubsection{The integrated spectrum of Haro~14}
\label{Integratedspectra}

The integrated spectrum of Haro~14 is shown in 
Figure~\ref{Figure:Haro14_spec}, and was generated by summing over all spaxels in the observed field. 
This is 
the spectrum of a composite stellar population ---with bright  emission lines 
(e.g., H$\alpha$, H$\beta$, [\ion{O}{iii}], [\ion{N}{ii}], [\ion{S}{ii}]) 
characteristic of \ion{H}{ii}-regions and starbursts--- placed on top of a blue and relatively high stellar continuum with evident absorption features, such as the pronounced wings of H$\beta$ and H$\alpha$ or the \ion{Ca}{ii} triplet in the near-infrared. 

\smallskip

The reddening-corrected emission line intensities (relative to H$\beta$)  for the integrated spectrum are presented  in Table~\ref{tab:lineratiosSF1}.
Line intensities were calculated taking into account the contribution of the underlying stellar population as described in Section~\ref{Section:fittinglines}; the 
interstellar extinction was derived from the Balmer decrement following the standard approach (Section~\ref{Section:extinctionmap}). 

\smallskip

We obtained a total reddening-corrected flux  in H$\alpha$ of  4.6$\pm$0.1$\times10^{-13}$erg~s$^{-1}$cm$^{-2}$, which translates into a  total H$\alpha$ luminosity of 9.2$\pm$0.2$\times$10$^{39}$~erg~s$^{-1}$. From here we estimated the SFR adopting the expression derived in  
\cite{Hunter2010}:
\begin{equation}
$${\rm SFR}(M_{\odot}~{\rm yr}^{-1})=6.9\times10^{-42}\,L_{H\alpha}({\rm erg~s}^{-1})\approx 0.06.$$
\nonumber
\end{equation}

We note that the H$\alpha$ flux here is significantly lower than the 
F(H$\alpha$)=7.3$\pm$0.1$\times10^{-13}$erg~s$^{-1}$cm$^{-2}$ value 
we derived from the VIMOS data in a smaller FoV\footnote{With VIMOS we covered a FoV of $27\arcsec\times27\arcsec$ in the sky.}  \citep{Cairos2017a}. 
This discrepancy likely results from differences in the interstellar extinction correction applied: the magnitude increments derived are  A$_{V}$=0.292$\pm$0.027 (MUSE) and A$_{V}$=0.819$\pm$0.007 (VIMOS). 
The observed fluxes (without  extinction correction) restricted to the smaller FoV of VIMOS 
are $3.3\pm 0.1\times10^{-13}$erg~s$^{-1}$~cm$^{-2}$ and $3.5\pm 0.1\times10^{-13}$erg~s$^{-1}$~cm$^{-2}$ from the  MUSE and VIMOS datasets,  respectively, which are also in 
reasonable good agreement. This confirms that the differences in the corrected values are due to the  distinct extinction coefficients adopted.

\smallskip 

Large discrepancies in the interstellar extinction coefficients are  a consequence of the large uncertainties associated with the determination of the H$\beta$ and H$\alpha$ fluxes in emission. Obtaining accurate fluxes for  the hydrogen Balmer series in emission  is  a complicated task,  because the  strengths of these lines can be severely affected by the absorption of the underlying population of stars (Section~\ref{Section:fittinglines}). Hence, the  values of the emission line fluxes are strongly dependent on the method used to correct for absorption. Here, we take into account the contribution of the underlying stellar component by performing a modeling of the SED, which provides more reliable values. 

\smallskip 

Table~\ref{tab:lineratiosSF1} also shows the diagnostic line ratios, electron density ($N_{\rm e}$), and oxygen abundance (12+log(O/H)) for the integrated spectrum.  $N_{e}$ was computed  from the  [\ion{S}{ii}]~$\lambda$6717/[\ion{S}{ii}]~$\lambda$6731 ratio following \cite{Osterbrock2006}. We estimated the oxygen abundances by adopting the empirical method introduced
by \cite{PilyuginGrebel2016}, which utilizes the intensities of 
the strong lines [\ion{O}{iii}]~$\lambda$$\lambda$4957,5007, [\ion{N}{ii}]~$\lambda$$\lambda$6548, 6584, and [\ion{S}{ii}]~$\lambda$$\lambda$6717,6731. The
relative accuracy of the abundance derived using this method is 0.1~dex. The oxygen abundance, 12+log(O/H)=8.20, is in excellent agreement with the value found from the VIMOS data (12+log(O/H)=8.25). 

\smallskip

The position of the line ratios computed from the integrated spectrum is shown in the diagnostic diagrams of Haro~14  (Figure~\ref{Figure:diagnostic-spaxel}). Interestingly, in the [\ion{O}{i}] diagram, the galaxy falls almost exactly on top of the maximum starburst line.

\subsubsection{Integrated spectroscopy of H$\alpha$ emission line sources}
\label{IntegratedspectraHa}

The most evident structures in the emission line maps of Haro~14 are the compact clumps visible up to large galactocentric distances; although the largest and brightest ones appear in the central regions of the galaxy. 
In Paper~I, we developed a routine to automatically detect and spatially delimit these sources (for a complete description of the process, see Section~3.6.2  in Paper~I). By running this 
routine in the H$\alpha$ flux map, we 
identified a total of 
40~sources and generated a catalog with their main properties, including position, size, and fluxes in the brightest lines (see Section 3.6, Figure~5, and Table~4 in Paper~I). 

\smallskip 
Here, we performed a more detailed study of these individual clumps. 
Because we aim to obtain reliable interstellar reddening coefficients and diagnostic line ratios, we restrict 
ourselves to sources with a S/N greater than 5 in H$\beta$ . In this way, we ended up with  the 30 emission line knots that appear identified in Figure~ \ref{Figure:emissionlinesources} (we note that we keep the number scheme introduced in Paper~I, and refer to these knots as L1, L2, etc. throughout the text).  Table~\ref{tab:lineratiosSF1} shows the reddening-corrected emission line fluxes and diagnostic line ratios for the first ten sources, which include the central  brightest ones; results for the remaining sources are provided in Appendix~1.

\smallskip

Diagnostic diagrams built from the line ratio of these  H$\alpha$ sources are shown in 
Figure~\ref{Figure:diagnostic-brotes}. The majority of the clumps are  located below the maximum starburst line, but still a large fraction (10 and 9 knots in the  [\ion{O}{i}] and  [\ion{S}{ii}] diagrams, respectively) 
fall outside of the star forming area. 
The bright sources that constitute the central bar-like and bubble structure  are classified, as expected, as \ion{H}{ii}-region  complexes  (knots in cyan in Figure~\ref{Figure:diagnostic-brotes}). By contrast, most of the H$\alpha$ sources  located in the galaxy periphery lie well above or to the right of the maximum starburst line (shown in magenta in the figure).

\smallskip 

The \ion{H}{ii} regions present a distribution of excitation and ionization similar to the  spatially resolved diagnostic diagram.
The highest excitation appears in the three brightest SF regions (L1, L2, L3), 
closest to the center of the galaxy, whereas  the lowest excitation is found in three sources (L17, L21, and L22) located between 0.6 and 1~kpc northeast. 
The distribution of the \ion{H}{ii}-regions in the diagnostic diagrams ---roughly along the maximum starburst line---  closely follows the predictions of  theoretical photoionization models for different ionization parameters ($q$) at a fixed metallicity \citep{Dopita2000,Kewley2001b}. 
The variation in $q$ would be compatible with a broad range in age for the ionizing stellar clusters.

\smallskip

The H$\alpha$ sources located above the maximum starburst line 
form two groups:  Three sources (L25, L26 and L28)  have very high excitation ( [\ion{O}{iii}]/\Hb{}$\geq$4); they are indeed the excitation maxima of the galaxy. The remaining sources present excitations of $\sim$1.5.
Five  out of the seven low-excitation sources (L20, L24, L27, L29, and L33) form a chain along the southeastern border of the bright ionization region of the galaxy (see Figure~\ref{Figure:emissionlinesources}).
The nature of these sources is difficult to assess. 
Different mechanisms might be responsible for the ionization of the gas here, and singling out one of them is not obvious.

\smallskip 

 Table~\ref{tab:lineratiosSF1} also presents the densities and oxygen abundances of the H$\alpha$ clumps. All clumps except L33 show densities in the low range, typical of regions of SF; the abundances are homogeneous considering the uncertainties, but there is a tendency to lower abundances in the outer galaxy regions.


\begin{landscape}
\begin{table}
\small
\caption{Reddening-corrected line intensity ratios, interstellar extinction, diagnostic line ratios, densities, and oxygen abundances for the emission line sources and integrated spectra of Haro~14.\label{tab:lineratiosSF1}}
\begin{center}
\begin{tabular}{lccccccccccc}
\hline 
\hline 
Ion      & L1 & L2 & L3 & L4 & L5 & L6 & L7 & L8 & L9 & L10 & Integrated     \\ 
\hline 
4861 H$\beta$ & 1.000 & 1.000 & 1.000 & 1.000 & 1.000 & 1.000 & 1.000 & 1.000 & 1.000 & 1.000 & 1.000 \\ 
4959 [OIII]    & 1.152$\pm$0.023 & 0.850$\pm$0.015 & 1.001$\pm$0.019 & 0.795$\pm$0.016 & 0.750$\pm$0.013 & 0.825$\pm$0.016 & 0.705$\pm$0.011 & 0.723$\pm$0.012 & 0.795$\pm$0.012 & 0.572$\pm$0.011 & 0.711$\pm$0.012  \\ 
5007 [OIII]   & 3.432$\pm$0.058 & 2.523$\pm$0.032 & 2.986$\pm$0.043 & 2.377$\pm$0.033 & 2.263$\pm$0.028 & 2.442$\pm$0.032 & 2.096$\pm$0.021 & 2.160$\pm$0.025 & 2.364$\pm$0.024 & 1.699$\pm$0.022 & 2.109$\pm$0.024  \\ 
6300 [OI]      & 0.030$\pm$0.005 & 0.028$\pm$0.004 & 0.038$\pm$0.004 & 0.047$\pm$0.004 & 0.062$\pm$0.004 & 0.066$\pm$0.004 & 0.086$\pm$0.004 & 0.117$\pm$0.005 & 0.071$\pm$0.004 & 0.142$\pm$0.005 & 0.131$\pm$0.005  \\ 
6363 [OI]      & 0.009$\pm$0.004 & 0.009$\pm$0.003 & 0.012$\pm$0.003 & 0.016$\pm$0.003 & 0.020$\pm$0.003 & 0.022$\pm$0.003 & 0.027$\pm$0.003 & 0.040$\pm$0.003 & 0.023$\pm$0.003 & 0.049$\pm$0.004 & 0.045$\pm$0.004  \\ 
6548 [NII]     & 0.062$\pm$0.004 & 0.071$\pm$0.006 & 0.066$\pm$0.003 & 0.086$\pm$0.004 & 0.088$\pm$0.004 & 0.090$\pm$0.004 & 0.137$\pm$0.005 & 0.121$\pm$0.008 & 0.091$\pm$0.008 & 0.132$\pm$0.008 & 0.108$\pm$0.009  \\ 
6563 H$\alpha$ & 2.870$\pm$0.067 & 2.870$\pm$0.056 & 2.870$\pm$0.052 & 2.870$\pm$0.048 & 2.870$\pm$0.048 & 2.870$\pm$0.049 & 2.870$\pm$0.036 & 2.870$\pm$0.049 & 2.870$\pm$0.049 & 2.870$\pm$0.050 & 2.870$\pm$0.044  \\ 
6584 [NII]     & 0.194$\pm$0.007 & 0.220$\pm$0.009 & 0.205$\pm$0.006 & 0.263$\pm$0.007 & 0.272$\pm$0.007 & 0.282$\pm$0.008 & 0.422$\pm$0.008 & 0.380$\pm$0.012 & 0.286$\pm$0.012 & 0.405$\pm$0.012 & 0.336$\pm$0.011  \\ 
6678 HeI       & 0.028$\pm$0.002 & 0.028$\pm$0.001 & 0.028$\pm$0.001 & 0.025$\pm$0.001 & 0.027$\pm$0.001 & 0.025$\pm$0.001 & 0.027$\pm$0.001 & 0.024$\pm$0.002 & 0.028$\pm$0.002 & 0.025$\pm$0.003 & 0.027$\pm$0.003  \\ 
6717 [SII]     & 0.202$\pm$0.005 & 0.199$\pm$0.004 & 0.243$\pm$0.005 & 0.320$\pm$0.006 & 0.330$\pm$0.007 & 0.345$\pm$0.007 & 0.552$\pm$0.007 & 0.570$\pm$0.011 & 0.351$\pm$0.006 & 0.610$\pm$0.012 & 0.512$\pm$0.009  \\ 
6731 [SII]     & 0.146$\pm$0.004 & 0.140$\pm$0.003 & 0.168$\pm$0.004 & 0.225$\pm$0.005 & 0.232$\pm$0.005 & 0.241$\pm$0.005 & 0.391$\pm$0.005 & 0.398$\pm$0.008 & 0.250$\pm$0.005 & 0.429$\pm$0.009 & 0.359$\pm$0.007  \\ 
7065 HeI       & 0.021$\pm$0.002 & 0.018$\pm$0.001 & 0.020$\pm$0.002 & 0.016$\pm$0.001 & 0.017$\pm$0.001 & 0.018$\pm$0.001 & 0.017$\pm$0.001 & 0.015$\pm$0.002 & 0.019$\pm$0.002 & 0.013$\pm$0.003 & 0.016$\pm$0.003  \\ 
7136 [ArIII]   & 0.089$\pm$0.004 & 0.085$\pm$0.005 & 0.077$\pm$0.003 & 0.071$\pm$0.003 & 0.079$\pm$0.003 & 0.077$\pm$0.003 & 0.078$\pm$0.004 & 0.078$\pm$0.012 & 0.083$\pm$0.007 & 0.084$\pm$0.015 & 0.081$\pm$0.014  \\ 
9068 [SIII]    & 0.213$\pm$0.011 & 0.213$\pm$0.010 & 0.184$\pm$0.008 & 0.181$\pm$0.009 & 0.190$\pm$0.007 & 0.185$\pm$0.007 & 0.169$\pm$0.007 & 0.157$\pm$0.016 & 0.170$\pm$0.011 & 0.153$\pm$0.019 & 0.158$\pm$0.018  \\ 
\hline 
F$_{H\beta}$  & 182.5$\pm$6.6 & 75.6$\pm$2.1 & 117.6$\pm$3.2 & 18.2$\pm$0.5 & 20.0$\pm$0.7 & 16.6$\pm$0.7 & 41.7$\pm$0.8 & 35.4$\pm$0.8 & 41.4$\pm$1.3 & 17.3$\pm$0.6 & 1589.5$\pm$37.3  \\ 
C$_{H\beta}$  & 0.146$\pm$0.019 & 0.116$\pm$0.016 & 0.123$\pm$0.015 & 0.148$\pm$0.014 & 0.261$\pm$0.014 & 0.331$\pm$0.014 & 0.130$\pm$0.010 & 0.094$\pm$0.014 & 0.231$\pm$0.014 & 0.214$\pm$0.015 & 0.135$\pm$0.013 \\ 
$A_{V}$        & 0.315$\pm$0.042 & 0.251$\pm$0.035 & 0.266$\pm$0.033 & 0.321$\pm$0.030 & 0.565$\pm$0.030 & 0.715$\pm$0.030 & 0.282$\pm$0.022 & 0.204$\pm$0.031 & 0.499$\pm$0.031 & 0.463$\pm$0.031 & 0.292$\pm$0.027 \\ 
E(B-V)         & 0.101$\pm$0.013 & 0.081$\pm$0.011 & 0.086$\pm$0.011 & 0.103$\pm$0.010 & 0.182$\pm$0.010 & 0.231$\pm$0.010 & 0.091$\pm$0.007 & 0.066$\pm$0.010 & 0.161$\pm$0.010 & 0.149$\pm$0.010 & 0.094$\pm$0.009 \\ 
\hline 
\hline 
$[\ion{O}{iii}]/\Hb$                 & 3.432$\pm$0.170 & 2.523$\pm$0.097 & 2.986$\pm$0.111 & 2.377$\pm$0.088 & 2.263$\pm$0.107 & 2.442$\pm$0.140 & 2.096$\pm$0.054 & 2.160$\pm$0.070 & 2.364$\pm$0.104 & 1.699$\pm$0.075 & 2.109$\pm$0.067 \\ 
$[\ion{O}{i}]/\Ha$                   & 0.010$\pm$0.002 & 0.010$\pm$0.001 & 0.013$\pm$0.001 & 0.016$\pm$0.001 & 0.021$\pm$0.002 & 0.023$\pm$0.002 & 0.030$\pm$0.002 & 0.041$\pm$0.002 & 0.025$\pm$0.001 & 0.050$\pm$0.002 & 0.046$\pm$0.002 \\ 
$[\ion{N}{ii}]/\Ha$                  & 0.068$\pm$0.003 & 0.077$\pm$0.003 & 0.072$\pm$0.002 & 0.092$\pm$0.003 & 0.095$\pm$0.003 & 0.098$\pm$0.004 & 0.147$\pm$0.003 & 0.132$\pm$0.005 & 0.100$\pm$0.005 & 0.141$\pm$0.005 & 0.117$\pm$0.004 \\ 
$[\ion{S}{ii}]/\Ha$     & 0.121$\pm$0.004 & 0.118$\pm$0.003 & 0.143$\pm$0.003 & 0.190$\pm$0.004 & 0.196$\pm$0.005 & 0.204$\pm$0.006 & 0.329$\pm$0.005 & 0.337$\pm$0.007 & 0.209$\pm$0.005 & 0.362$\pm$0.009 & 0.304$\pm$0.006 \\ 
\hline 
N$_{e}$   (cm$^{-3}$)           & < 100                         & < 100          & < 100                  & < 100                                & < 100                          & < 100                         & < 100                         & < 100                 & < 100           & < 100                 & < 100 \\ 
12+log(O/H)$^{2}$               & 8.22                  & 8.26           & 8.20                   & 8.22                         & 8.22                           & 8.22                   & 8.26                         & 8.22          & 8.22            & 8.22          & 8.20\\ 
\hline 
\hline 
\end{tabular}
\end{center}
Notes.- Line intensity ratios are normalized to H$\beta$; the reddening-corrected H$\beta$ flux is in units of 10$^{-16}$~erg~s$^{-1}$~cm$^{-2}$. 
\end{table}
\end{landscape}
\normalsize

\subsubsection{Integrated spectroscopy of the high-excitation blobs}
\label{IntegratedspectraEx}

The excitation maps of Haro~14 reveal a  large number of maxima distributed irregularly across the galaxy; interestingly, the zones of highest excitation do not spatially coincide with the brightest SF complexes, but are situated in the periphery of the HSB  galaxy regions (section~\ref{Section:lineratiomaps}). The large number of high-excitation clumps and, in particular, the fact that the peaks are located outside of the SF area are puzzling. The global properties of these regions, derived from their integrated spectra, will help to shed some light on the physical processes acting there.

\smallskip 

We detected and spatially delimited these high-excitation regions  using the same procedure as for the H$\alpha$ sources (section~\ref{IntegratedspectraEx} and Paper~I). In this case, we ran the routine in the [\ion{O}{iii}]/\Ha{} map and detected 16 clumps.
Tables~\ref{tab:lineratiosE1} and \ref{tab:lineratiosE2} present the reddening-corrected intensity ratios, the interstellar extinction coefficient, the diagnostic line ratios and  the oxygen abundances for each knot. 

\smallskip 

Figure~\ref{Figure:diagnostic-excitation} displays the diagnostic diagrams for the high-excitation blobs. Only five are situated within the hot star photoionization area of the diagram (in both [\ion{O}{i}] and [\ion{S}{ii}]), while most of them fall outside the starburst line. The knots under the starburst line roughly coincide with knots 1, 2, 3, 5, and 8 in H$\alpha$: they are bright \ion{H}{ii}-regions whose high excitation reflects the high temperature of the ionizing stars. However, the nature of the knots above the starburst line is  difficult to determine. They span a wide range of excitation, and although two groups are somehow suggested, they are not clearly defined. Their position in the diagram is consistent with different types of interstellar shocks, but their morphology, in many cases, suggests regions of SF.  Particularly interesting are the three knots aligned to the  north, which constitute the galaxy excitation peaks.

\begin{landscape}
\begin{table}
\small
\caption{Reddening-corrected line intensity ratios, interstellar extinction, diagnostic line ratios, densities, and oxygen abundances for the high-excitation blobs in Haro~14.\label{tab:lineratiosE1}}
\begin{center}
\begin{tabular}{lcccccccc}
\hline 
\hline 
Ion      & Ex1 & Ex2 & Ex3 & Ex4 & Ex5 & Ex6 & Ex7 & Ex8     \\ 
\hline 
4861 H$\beta$ & 1.000 & 1.000 & 1.000 & 1.000 & 1.000 & 1.000 & 1.000 & 1.000 \\ 
4959 [OIII]    & 1.593$\pm$0.048 & 1.798$\pm$0.561 & 1.291$\pm$0.050 & 1.102$\pm$0.018 & 0.966$\pm$0.175 & 1.055$\pm$0.208 & 1.000$\pm$0.017 & 1.080$\pm$0.016 \\ 
5007 [OIII]   & 4.799$\pm$0.123 & 5.371$\pm$1.610 & 3.787$\pm$0.122 & 3.281$\pm$0.045 & 3.147$\pm$0.444 & 3.104$\pm$0.513 & 2.981$\pm$0.039 & 3.211$\pm$0.038 \\ 
6300 [OI]      & 0.405$\pm$0.021 & 0.320$\pm$0.219 & 0.232$\pm$0.022 & 0.037$\pm$0.005 & 0.395$\pm$0.104 & 0.442$\pm$0.118 & 0.042$\pm$0.004 & 0.170$\pm$0.006 \\ 
6363 [OI]      & 0.145$\pm$0.013 & 0.120$\pm$0.150 & 0.058$\pm$0.015 & 0.012$\pm$0.004 & 0.152$\pm$0.067 & 0.091$\pm$0.056 & 0.013$\pm$0.003 & 0.057$\pm$0.004 \\ 
6548 [NII]     & 0.154$\pm$0.014 & 0.071$\pm$0.089 & 0.136$\pm$0.020 & 0.069$\pm$0.004 & 0.047$\pm$0.055 & 0.088$\pm$0.059 & 0.067$\pm$0.004 & 0.116$\pm$0.006 \\ 
6563 H$\alpha$ & 2.870$\pm$0.101 & 2.870$\pm$1.211 & 2.870$\pm$0.128 & 2.870$\pm$0.054 & 2.870$\pm$0.547 & 2.870$\pm$0.649 & 2.870$\pm$0.048 & 2.870$\pm$0.046 \\ 
6584 [NII]     & 0.473$\pm$0.023 & 0.116$\pm$0.101 & 0.432$\pm$0.030 & 0.216$\pm$0.006 & 0.318$\pm$0.092 & 0.365$\pm$0.108 & 0.211$\pm$0.006 & 0.356$\pm$0.009 \\ 
6678 HeI       & 0.028$\pm$0.010 & 0.043$\pm$0.065 & 0.043$\pm$0.016 & 0.028$\pm$0.002 & 0.135$\pm$0.065 & 0.049$\pm$0.030 & 0.028$\pm$0.001 & 0.028$\pm$0.002 \\ 
6717 [SII]     & 0.933$\pm$0.035 & 0.263$\pm$0.146 & 0.750$\pm$0.038 & 0.239$\pm$0.005 & 0.711$\pm$0.152 & 0.735$\pm$0.176 & 0.256$\pm$0.005 & 0.610$\pm$0.011   \\ 
6731 [SII]     & 0.644$\pm$0.025 & 0.156$\pm$0.105 & 0.512$\pm$0.028 & 0.171$\pm$0.004 & 0.396$\pm$0.095 & 0.559$\pm$0.137 & 0.177$\pm$0.004 & 0.427$\pm$0.008   \\ 
7065 HeI       & 0.023$\pm$0.008 & 0.167$\pm$0.113 & 0.030$\pm$0.015 & 0.021$\pm$0.002 & 0.072$\pm$0.041 & 0.067$\pm$0.060 & 0.019$\pm$0.002 & 0.017$\pm$0.003   \\ 
7136 [ArIII]   & 0.131$\pm$0.030 & 0.056$\pm$0.056 & 0.158$\pm$0.054 & 0.089$\pm$0.005 & 0.176$\pm$0.089 & 0.122$\pm$0.088 & 0.077$\pm$0.003 & 0.086$\pm$0.009   \\ 
9068 [SIII]    & 0.221$\pm$0.035 & 0.078$\pm$0.093 & 0.191$\pm$0.052 & 0.208$\pm$0.011 & 0.107$\pm$0.089 & 0.116$\pm$0.086 & 0.182$\pm$0.008 & 0.150$\pm$0.014   \\ 
\hline 
F$_{H\beta}$  &  2.5$\pm$0.1 &  0.4$\pm$0.6 &  1.9$\pm$0.1 & 225.2$\pm$6.1 &  0.7$\pm$0.5 &  1.1$\pm$1.5 & 129.6$\pm$3.1 &  9.0$\pm$0.2   \\ 
C$_{H\beta}$  & 0.104$\pm$0.029 & 0.497$\pm$0.351 & 0.138$\pm$0.037 & 0.113$\pm$0.016 & 0.479$\pm$0.159 & 0.650$\pm$0.188 & 0.101$\pm$0.014 & 0.114$\pm$0.013  \\ 
$A_{V}$        & 0.225$\pm$0.063 & 1.074$\pm$0.758 & 0.298$\pm$0.080 & 0.243$\pm$0.034 & 1.036$\pm$0.343 & 1.404$\pm$0.407 & 0.218$\pm$0.030 & 0.245$\pm$0.029  \\ 
E(B-V)         & 0.072$\pm$0.020 & 0.346$\pm$0.245 & 0.096$\pm$0.026 & 0.078$\pm$0.011 & 0.334$\pm$0.111 & 0.453$\pm$0.131 & 0.070$\pm$0.010 & 0.079$\pm$0.009  \\ 
\hline 
\hline 
$[\ion{O}{iii}] \lambda5007/\Hb$                 & 4.799$\pm$0.328 & 5.371$\pm$11.767 & 3.787$\pm$0.353 & 3.281$\pm$0.121 & 3.147$\pm$2.977 & 3.104$\pm$5.460 & 2.981$\pm$0.097 & 3.211$\pm$0.103  \\ 
$[\ion{O}{i}] \lambda6300/\Ha$                   & 0.050$\pm$0.005 & 0.042$\pm$0.066 & 0.020$\pm$0.005 & 0.004$\pm$0.001 & 0.053$\pm$0.032 & 0.032$\pm$0.030 & 0.005$\pm$0.001 & 0.020$\pm$0.001  \\ 
$[\ion{N}{ii}] \lambda6584/\Ha$                  & 0.165$\pm$0.009 & 0.040$\pm$0.051 & 0.151$\pm$0.012 & 0.075$\pm$0.003 & 0.111$\pm$0.055 & 0.127$\pm$0.096 & 0.073$\pm$0.002 & 0.124$\pm$0.003  \\ 
$[\ion{S}{ii}] \lambda\lambda6717,6731/\Ha$     & 0.549$\pm$0.020 & 0.146$\pm$0.135 & 0.440$\pm$0.023 & 0.143$\pm$0.003 & 0.386$\pm$0.152 & 0.451$\pm$0.281 & 0.151$\pm$0.003 & 0.361$\pm$0.007  \\ 
\hline 
N$_{e}$   (cm$^{-3}$) & < 100           & < 100           & < 100           & < 100           & < 100           & 110             & < 100           & < 100           \\ 
12+log(O/H)$^{1}$     & 8.22            & 8.20            & 8.22            & 8.22            & 8.09            & 8.13            & 8.20            & 8.19            \\ 
\hline 
\hline 
\end{tabular}
\end{center}
Notes.- Line intensity ratios are normalized to H$\beta$; the reddening-corrected H$\beta$ flux is in units of 10$^{-16}$~erg~s$^{-1}$~cm$^{-2}$. 
\end{table}
\end{landscape}

\begin{landscape}
\begin{table}
\small
\caption{Reddening-corrected line intensity ratios, interstellar extinction, diagnostic line ratios, densities, and oxygen abundances for the high excitation blobs in Haro~14.\label{tab:lineratiosE2}}
\begin{center}
\begin{tabular}{lcccccccc}
\hline 
\hline 
Ion      & Ex9 & Ex10 & Ex11 & Ex12 & Ex13 & Ex14 & Ex15 & Ex16   \\ 
\hline 
4861 H$\beta$ & 1.000 & 1.000 & 1.000 & 1.000 & 1.000 & 1.000 & 1.000 & 1.000 \\ 
4959 [OIII]    & 0.947$\pm$0.212 & 1.054$\pm$0.028 & 0.835$\pm$0.014 & 0.652$\pm$0.022 & 0.483$\pm$0.035 & 0.748$\pm$0.011 & 0.695$\pm$0.012 & 0.510$\pm$0.033 \\ 
5007 [OIII]   & 2.989$\pm$0.533 & 3.066$\pm$0.059 & 2.486$\pm$0.029 & 1.947$\pm$0.044 & 1.474$\pm$0.065 & 2.235$\pm$0.023 & 2.070$\pm$0.025 & 1.432$\pm$0.057 \\ 
6300 [OI]      & 0.363$\pm$0.122 & 0.216$\pm$0.011 & 0.040$\pm$0.004 & 0.479$\pm$0.017 & 0.336$\pm$0.029 & 0.104$\pm$0.004 & 0.117$\pm$0.005 & 0.488$\pm$0.028 \\ 
6363 [OI]      & 0.205$\pm$0.085 & 0.075$\pm$0.008 & 0.012$\pm$0.003 & 0.157$\pm$0.009 & 0.127$\pm$0.020 & 0.034$\pm$0.003 & 0.039$\pm$0.003 & 0.161$\pm$0.015 \\ 
6548 [NII]     & 0.135$\pm$0.066 & 0.138$\pm$0.009 & 0.077$\pm$0.006 & 0.176$\pm$0.014 & 0.139$\pm$0.019 & 0.099$\pm$0.007 & 0.124$\pm$0.009 & 0.216$\pm$0.019 \\ 
6563 H$\alpha$ & 2.870$\pm$0.695 & 2.870$\pm$0.072 & 2.870$\pm$0.050 & 2.870$\pm$0.082 & 2.870$\pm$0.153 & 2.870$\pm$0.044 & 2.870$\pm$0.052 & 2.870$\pm$0.132 \\ 
6584 [NII]     & 0.326$\pm$0.104 & 0.424$\pm$0.015 & 0.239$\pm$0.009 & 0.525$\pm$0.022 & 0.436$\pm$0.031 & 0.309$\pm$0.010 & 0.390$\pm$0.013 & 0.626$\pm$0.034 \\ 
6678 HeI       & 0.053$\pm$0.034 & 0.035$\pm$0.006 & 0.027$\pm$0.001 & 0.028$\pm$0.007 & 0.035$\pm$0.013 & 0.027$\pm$0.001 & 0.024$\pm$0.002 & 0.026$\pm$0.012 \\ 
6717 [SII]     & 0.731$\pm$0.195 & 0.781$\pm$0.022 & 0.238$\pm$0.005 & 1.104$\pm$0.034 & 0.873$\pm$0.050 & 0.429$\pm$0.009 & 0.592$\pm$0.012 & 1.169$\pm$0.057 \\ 
6731 [SII]     & 0.602$\pm$0.164 & 0.550$\pm$0.017 & 0.168$\pm$0.004 & 0.774$\pm$0.024 & 0.605$\pm$0.036 & 0.303$\pm$0.007 & 0.414$\pm$0.009 & 0.826$\pm$0.041 \\ 
7065 HeI       & 0.046$\pm$0.063 & 0.015$\pm$0.008 & 0.018$\pm$0.002 & 0.021$\pm$0.006 & 0.012$\pm$0.013 & 0.017$\pm$0.001 & 0.015$\pm$0.002 & 0.020$\pm$0.011 \\ 
7136 [ArIII]   & 0.075$\pm$0.051 & 0.096$\pm$0.017 & 0.086$\pm$0.007 & 0.113$\pm$0.041 & 0.109$\pm$0.052 & 0.080$\pm$0.005 & 0.081$\pm$0.015 & 0.116$\pm$0.046 \\ 
9068 [SIII]    & 0.125$\pm$0.102 & 0.152$\pm$0.025 & 0.206$\pm$0.012 & 0.115$\pm$0.039 & 0.118$\pm$0.063 & 0.178$\pm$0.008 & 0.149$\pm$0.019 & 0.123$\pm$0.051 \\ 
\hline 
F$_{H\beta}$  &  0.3$\pm$0.3 &  1.1$\pm$0.0 & 88.9$\pm$2.1 &  7.5$\pm$0.4 &  1.6$\pm$0.1 & 46.0$\pm$1.1 & 49.7$\pm$1.2 &  2.7$\pm$0.2 \\ 
C$_{H\beta}$  & 0.413$\pm$0.201 & 0.107$\pm$0.021 & 0.090$\pm$0.015 & 0.165$\pm$0.024 & 0.066$\pm$0.044 & 0.169$\pm$0.013 & 0.090$\pm$0.015 & 0.183$\pm$0.038 \\ 
$A_{V}$        & 0.892$\pm$0.435 & 0.231$\pm$0.045 & 0.195$\pm$0.032 & 0.356$\pm$0.052 & 0.143$\pm$0.096 & 0.365$\pm$0.027 & 0.195$\pm$0.033 & 0.396$\pm$0.083 \\ 
E(B-V)         & 0.288$\pm$0.140 & 0.074$\pm$0.015 & 0.063$\pm$0.010 & 0.115$\pm$0.017 & 0.046$\pm$0.031 & 0.118$\pm$0.009 & 0.063$\pm$0.011 & 0.128$\pm$0.027 \\ 
\hline 
\hline 
$[\ion{O}{iii}] \lambda5007/\Hb$                 & 2.989$\pm$3.023 & 3.066$\pm$0.151 & 2.486$\pm$0.081 & 1.947$\pm$0.126 & 1.474$\pm$0.143 & 2.235$\pm$0.076 & 2.070$\pm$0.070 & 1.432$\pm$0.157 \\ 
$[\ion{O}{i}] \lambda6300/\Ha$                   & 0.071$\pm$0.044 & 0.026$\pm$0.003 & 0.004$\pm$0.001 & 0.055$\pm$0.004 & 0.044$\pm$0.007 & 0.012$\pm$0.001 & 0.014$\pm$0.001 & 0.056$\pm$0.006 \\ 
$[\ion{N}{ii}] \lambda6584/\Ha$                  & 0.113$\pm$0.062 & 0.148$\pm$0.006 & 0.083$\pm$0.003 & 0.183$\pm$0.009 & 0.152$\pm$0.011 & 0.108$\pm$0.004 & 0.136$\pm$0.005 & 0.218$\pm$0.015 \\ 
$[\ion{S}{ii}] \lambda\lambda6717,6731/\Ha$     & 0.464$\pm$0.202 & 0.464$\pm$0.013 & 0.142$\pm$0.003 & 0.654$\pm$0.022 & 0.515$\pm$0.027 & 0.255$\pm$0.006 & 0.350$\pm$0.007 & 0.695$\pm$0.038 \\ 
\hline 
N$_{e}$   (cm$^{-3}$) & 220             & < 100           & < 100           & < 100           & < 100           & < 100           & < 100           & < 100           \\ 
12+log(O/H)$^{2}$     & 8.13            & 8.20            & 8.25            & 8.20            & 8.17            & 8.21            & 8.22            & 8.25            \\ 
\hline 
\hline 
\end{tabular}
\end{center}
Notes.- Line intensity ratios are normalized to H$\beta$; the reddening-corrected H$\beta$ flux is in units of 10$^{-16}$~erg~s$^{-1}$~cm$^{-2}$. 
\end{table}
\end{landscape}

\section{Discussion}
\label{discussion}

MUSE allowed us to observe not only the central HSB star forming regions of Haro~14, but also the ionized halo and faint structures that dominate the emission at large galactocentric distances, and  with unprecedented depth. This is an exceptional asset  with respect to previous BCGs works, as 
it is in these dimmer regions of the galaxy periphery that we expect to see the feedback mechanisms at work and where alternative ionization sources (e.g., shock heating, evolved stellar populations, or the dilute radiation from photoionizing sources) are expected to play a major role.

\subsection{The diverse ionizing  mechanisms operating in Haro~14}
\label{ionizationmechanism}

 Considering both the morphology and the surface brightness,
the emission line maps of Haro~14  present
two different components: a HSB region made of individual clumps, which occupies the central galaxy zones, and a LSB component, which extends up to kiloparsec scales and consists of widespread  diffuse emission,
clearly delineated curvilinear structures, and faint clumps\footnote{The gas outside classical \ion{H}{ii}-regions appears commonly referred to in the literature as {\em diffuse ionized gas} (DIG) or {\em warm ionized medium} (WIM). These terms comprise the three different constituents that we mention here: the smooth extended emission, the filamentary structures, and the faint blobs.}.

\smallskip 

The HSB area exhibits the same clumpy morphology in all lines, which is also apparent in the diagnostic maps, where the clumps are clearly delineated and spatially coincide with maxima in the excitation and minima 
in the other ratios. This is characteristic of areas photoionized by massive hot stars.

\smallskip 

By contrast,  the LSB component shows marked differences among the emission in different lines. For example, 
several clumps and filaments are visible in some lines but not in others.
All  diagnostic maps present variability on large spatial scales in the galaxy periphery. 
Most of the LSB area displays low values of [\ion{O}{iii}]/\Hb{} and high values of 
[\ion{O}{i}]/\Ha{}, [\ion{N}{ii}]/\Ha,{} and  [\ion{S}{ii}]/\Ha{}, an ionization pattern usually interpreted in terms of 
diluted radiation from OB stars or shock-heating  \citep{Mathis1986,Mathis2000,Dopita2003,Haffner2009}.
However,  [\ion{O}{i}]/\Ha{} and  [\ion{S}{ii}]/\Ha{} are extremely high in various zones (up to 0.5 and 1, respectively); these large values cannot be explained in terms of OB stars alone, as photoionization models do not predict  ratios significantly larger than
[\ion{O}{i}]/\Ha{}$\sim$0.2 and [\ion{S}{ii}]/\Ha{}$\sim$0.8 (e.g.,  
 \citealp{Domgorgen1994,Hoopes2003}). In starburst systems,  such enhanced [\ion{O}{i}]/\Ha{} and  [\ion{S}{ii}]/\Ha{} in the galaxy periphery have previously been associated with shock excitation \citep{Calzetti2004,Veilleux2002,MonrealIbero2006,MonrealIbero2010,Sharp2010,Rich2010,Rich2011}.
Interestingly,  in the galaxy outskirts, we also find regions of very high excitation and low values of the other ratios; indeed, the excitation peaks are located there, in two knots placed $\sim$750 and 1600~pc to the west of the continuum peak, and a third one located $\sim$1200~pc to the east. The ionization source operating here is not evident; we come back to this point in Section~\ref{alternativepower}.

\smallskip 

We make these arguments quantitative by building  
the spatially resolved diagnostic line ratio diagrams of Haro~14 (Figure~\ref{Figure:diagnostic-spaxel}). According to their position in these diagrams, most spaxels in the galaxy are being ionized by a mechanism other than SF:  adopting the maximum starburst line proposed in \cite{Kewley2001a} we find that 
in $\sim$75\% and $\sim$50\% of the galaxy area (in the  [\ion{O}{i}] and 
[\ion{S}{ii}] diagrams, respectively)  an alternative mechanism  is operating. 
 This corroborates the finding of the galaxy maps, which show that the central clumpy area is made of photoionized
\ion{H}{ii} regions whereas an additional mechanism of ionization (and most probably, more than one) is required to explain the values of the line ratios all over the LSB area (Figure~\ref{Figure:nonphotoregions}).

\smallskip 

In contrast to the spaxels falling below the starburst line, it is not straightforward to associate an excitation mechanism(s) to those above the line. The traditional approach distinguishes two sectors here: the AGN and the low-ionization narrow emission line region (LINER) domains \citep{Kewley2001b,Kewley2006,Kauffmann2003}. However, it is now well established that other mechanisms can generate line ratios mimicking those of AGNs and LINERs, such as   
interstellar shocks \citep{Sharp2010,Farage2010,Rich2010,Rich2011,Rich2015}, 
energetic photons leaking out from  \ion{H}{ii}-regions \citep{HidalgoGamez2005,HidalgoGamez2007,Weilbacher2018,Belfiore2022},
or hot low-mass evolved stars \citep[HOLMES; ][]{FloresFajardo2011,Zhang2017,Belfiore2016,Belfiore2022}.

\smallskip

From the diagnostic diagrams alone we cannot discriminate between these different sources,
but interpreting their results together with the morphology and ionization structure revealed by the maps will allow us to put additional constraints on the physical process(es) responsible for the ionization in the galaxy outskirts.

\subsubsection{Below the maximum starburst line}

The spaxels that fall below the maximum starburst line in the  diagnostic diagrams 
correspond spatially with the clumps that comprise the central bar-like and the horseshoe-shaped features  (Figures~\ref{Figure:emissionmaps1} and ~\ref{Figure:nonphotoregions}), confirming that the central clumps are  large complexes of \ion{H}{ii}-regions. 

\smallskip

This  \ion{H}{ii}-region photoionized area in the diagram
spans a broad range in excitation ($\sim$1.2 dex) and follows a sequence that runs almost parallel to the maximum starburst line. 
The distribution of the points along this sequence can be interpreted as variations in the ionization parameter   which, in turn, may be reflecting the large range of ages of the ionizing clusters, as is expected in an extended (kiloparsec scales) starburst such as the one observed in Haro~14.  This
is consistent with the [\ion{O}{iii}]/\Ha{} line ratio maps, in which there is an evident excitation gradient among the  central clumps, and is also further 
corroborated by the diagnostic diagram of the individual H$\alpha$ sources (Figure~\ref{Figure:emissionlinesources}), in which the SF knots define the same excitation sequence as that seen in the spaxel resolved diagrams. 

\smallskip

Stellar clusters with different ages naturally arise in different scenarios of SF; for example,
sequential SF resulting from the impact of  massive stellar evolution in the surrounding  ISM  \citep{Elmegreen1977,Whitworth1994} or 
continuous SF,  as in a merger or interaction event. In the case of Haro~14, the observations are consistent with both scenarios: massive clusters, whose evolution must have  produced a large number of SN events, are present in the galaxy, but also the morphology is suggestive of past interactions or mergers \citep{Cairos2021}; we note that neither scenario excludes the other. The presence of a marked age gradient among the  \ion{H}{ii}-regions in Haro~14 is an interesting phenomenon, and
will be addressed in more detail in a forthcoming publication where we will use synthesis population models to reproduce the observables of the individual clusters. 

\smallskip

Similar spaxel distributions below the maximum starburst line have been found for 
NGC~4214 \citep{Calzetti2004}, Mrk~900 \citep{Cairos2020}, and Tololo~1937-423 \citep{Cairos2017b}, objects that also present an extended area of ongoing SF made of multiple \ion{H}{ii}-region complexes. In Haro~14 (as well as in NGC~4214), the 
galaxy inclination (we see the galaxy almost face on) allows us to resolve  individual sources that in edge-on galaxies would appear integrated along the line of sight.

\subsubsection{Beyond the maximum starburst line}
\label{alternativepower}
 
 As we state above, it is not straightforward to identify the mechanism responsible for the ionization of the spaxels above the maximum starburst line. 
In the case of Haro~14, an AGN can be naturally discarded, but to discriminate between shocks and other mechanisms on the basis of line ratios alone is rather difficult.

\smallskip

 Some information can be extracted from the distribution of the points in the diagnostic diagrams.
The individual spaxels span a huge range in excitation, with [\ion{O}{iii}]/\Hb{} up to 10;  however, the 
individual  H$\alpha$ and high-excitation sources do not extend over the whole excitation range, but appear split into two groups,   with the H$\alpha$ sources being much more clearly defined (Figures~\ref{Figure:diagnostic-brotes}  and 
\ref{Figure:diagnostic-excitation}). The large variation in excitation together with the fact that the individual sources appear gathered in groups  point to  more than one power source acting in the galaxy outskirts. Also, that the spaxels cover a broad and densely populated area (Figure~\ref{Figure:diagnostic-spaxel}) means that the ionization conditions  vary in 
a continuous manner; as expected, for instance, when the ionizing photons are diluted radiation.

\smallskip

Additional information can be derived from the morphology and ionization pattern associated with these spaxels.
The spaxels beyond the maximum starburst line  encompass the whole LSB component  (Figure~\ref{Figure:nonphotoregions}); that is, they include 
 the diffuse extended halo, coherent and likely expanding structures (such as filaments and shells), and several faint clumps  (see Figures~\ref{Figure:emissionmaps1}, ~\ref{Figure:emissionmaps2},  and ~\ref{Figure:nonphotoregions}). It seems a plausible hypothesis that these different morphologies  arise from different physical processes; it would be indeed difficult to identify a unique mechanism able to originate the three of them. 

\smallskip 

The morphology of the diffuse constituent is consistent with the ionization being produced by the dilute radiation from OB stars. In this scenario, the ionization structure changes gradually as we move away from the central heating source. Because the most energetic photons are absorbed close to the exciting stars,  the hardness of the radiation (and the excitation) decreases with the distance while the line ratios of the low-ionization species increases \citep{Miller1993,Dove1994,Domgorgen1994,Haffner2009}. This produces a smooth distribution of ionized gas that can extend to large distances, in which the excitation anti-correlates with the H$\alpha$ surface brightness \citep{Ferguson1996,Zurita2000,Zurita2002}. This is the behaviour that we observe in the largest fraction of the LSB  ionized gas in  Haro~14 (see the bottom panels of Figure~\ref{Figure:diagnosticmaps}).

\smallskip

The most striking  features in the LSB regions of Haro~14 are the numerous curvilinear 
features (filaments, loops, and shells) departing from the central HSB areas and extending up to kiloparsec scales.  The morphology suggests shock excitation, with the ionized filaments and shells being the edges of  giant expanding bubbles  associated with supernova-driven galactic winds \citep{Marlowe1995,Lehnert1996,Martin1998,Strickland2004a,Strickland2004b,Book2008,Egorov2014,Egorov2017,Egorov2018}. Particularly eye-catching are the two filaments extending up to 2 and 2.3 kpc southwest (Figure~\ref{Figure:emissionmapssat}).
Similar filaments have been observed in other BCGs and local starbursts (e.g., NGC~1569 \citep{Heckman1995,Martin1998,Buckalew2006,Westmoquette2008}, NGC~3077 \citep{Martin1998,Ott2003,Oparin2020}, and Henize~2-10 \citep{Mendez1999,Johnson2000,Cresci2017}), and  were interpreted as signatures of bipolar flows connected with nuclear  starburst activity. 

\smallskip

Finally, the mechanism responsible for the ionization of the LSB  clumps in the galaxy periphery is more difficult to determine unequivocally. The fact that these sources split into two groups  in the diagnostic diagrams (Figures~\ref{Figure:diagnostic-brotes} and \ref{Figure:diagnostic-excitation})  points to different kinds of objects; otherwise we would expect a smooth transition. The morphological pattern of the knots with lower ionization suggests they were generated in shocked areas, as most of them are distributed in a chain depicting the  border between the high- and low-surface-brightness  gas components (Figure~\ref{Figure:emissionlinesources}, left panel); in this scenario the clumps may be the debris of fragmented shells and/or filaments. 
The nature of the high ionization knots  is puzzling; in particular, the three clumps that coincide with peaks in the excitation maps (1, 2, and 3 in Figure~\ref{Figure:emissionlinesources} and Table~\ref{tab:lineratiosE1}). These three knots, which fall in the AGN zone of the diagnostic diagram mostly due to their high excitation, appear aligned in a  direction roughly orthogonal to the central SF bar. Their morphology suggests large SF complexes, but their diagnostic ratios imply  different physical conditions from those found 
in typical \ion{H}{ii}-regions; they are also very different from the SF regions in the central areas. Several mechanisms can contribute to alter the values of the excitation in  \ion{H}{ii}-regions; for example, the presence of dust, variations in the metallicity, or the fact that these are density-limited regions. In Haro~14, neither the dust contribution nor the oxygen abundances seem to play a significant role: the ionized clumps with high excitation present interstellar extinction values and oxygen abundances in the same range as those of the central  \ion{H}{ii}-regions (see Table~\ref{tab:lineratiosSF3} in the Appendix).

\smallskip

A more complete study that includes the analysis of the kinematical results is required to  further investigate and  disentangle the nature of the distinct power sources acting in the galaxy periphery.  This will be addressed in a forthcoming publication.

\subsection{Some remarks on the  spatially resolved diagnostic diagrams of Haro~14}
\label{discussdiagram}

Spatially resolved diagnostic diagrams are fundamentally different tools from
 the classic diagrams  introduced to classify galaxies based on their integrated spectra. 
Rather than from a single point in a graph,  a galaxy is now characterized by a distribution of points giving
the relationship between the diagnostic line ratios throughout the observed area.  
This allows us  to identify and probe different power sources acting in the same object and, in combination
with the spectral maps, to determine their spatial position. In addition, 
the characteristics of the cloud of points (e.g., its  extent, shape, or density) 
may provide additional information that
helps to reconstruct a consistent picture of the physical processes taking place in the galaxy and  their impact on the surrounding ISM.

\smallskip 

 Some features on the spatially resolved diagnostic diagrams of Haro~14 (Figure~\ref{Figure:diagnostic-spaxel}) merit further discussion. In particular, because these diagrams show notable quantitative and qualitative differences with respect to the BCG diagnostic diagrams presented in previous works (e.g., \citealp{Calzetti2004,Kehrig2008,Kehrig2016,Kehrig2018,Kehrig2020,James2010,James2016,Lagos2012,Lagos2014,Hong2013,Cresci2017,Cairos2017a,Cairos2017b,Cairos2020}). We focused here on the [\ion{O}{i}] and [\ion{S}{ii}] diagrams of Haro~14, because
the [\ion{N}{ii}] diagram loses its diagnostic power  at low metallicities  (see Section~\ref{Section:diagnosticdiagram}). 

\smallskip

The most salient features of the diagrams of Haro~14 are the following:
\begin{enumerate}
\item They appear quite crowded, with the data points densely covering a well-defined and quite extended area
spanning more than 1.5~dex in excitation and [\ion{O}{I}]/H$\alpha$, and 1~dex in [\ion{S}{II}]/H$\alpha$. This is a consequence of the larger FoV, the higher spatial resolution, and the higher sensitivity of the observations.

\smallskip

\item The data points in the [\ion{O}{i}] and the [\ion{S}{ii}] diagrams spread 
symmetrically over a broad strip of up to 1.5 dex in width, 
and roughly along the maximum starburst line. 

\smallskip

\item There is a  well-defined, curved exclusion zone at large excitation 
([\ion{O}{iii}]/\Hb{}~$\geq$3) that segregates the points into two  symmetrical populations, one  either side of the maximum starburst line.

\smallskip

\item At low excitation ([\ion{O}{iii}]/\Hb{}~$\leq$0.3), there is a gradual
concentration around the maximum starburst line  
([\ion{O}{i}]/\Ha{}$\sim$0.1 and [\ion{S}{ii}]/\Ha{}$\sim$0.6).  
\end{enumerate}

\smallskip

These features confer to the diagnostic diagrams the distinctive shape of a
butterfly, a 
general arrangement that could already be seen, though much less clearly, 
in the diagrams of Haro~14 built from VIMOS data \citep{Cairos2017a}.  Interestingly, this characteristic shape is also visible in the diagrams of the BCGs Mrk~900 \citep{Cairos2020} and Tololo~1937-423 \citep{Cairos2017b}. It would be interesting to explore whether or not this butterfly shape also appears in the diagnostic diagrams of other BCGs and, if so, how often and for which conditions. Such an analysis, which requires observations reaching  similar surface brightness levels in the galaxy outskirts for a number of objects, is out of the scope of this paper. We address this question in a forthcoming publication of this series, currently under preparation.

\section{Conclusions}
\label{conclusions}

This is the second paper in a series reporting the results of a MUSE/VLT based study of the BCG Haro~14. 
Here, we performed an exhaustive analysis of the properties of the warm ionized gas emission in the galaxy. 
Capitalizing on the large FoV  (3.8$\times$3.8~kpc$^{2}$ at the adopted distance) of MUSE
and its unprecedented sensitivity, we investigated the morphology, dust distribution, physical properties, and ionization structure of the ionized gas, not only in the central HSB galaxy regions but also in the LSB areas of the periphery; this is an important achievement with respect to previous integral field spectroscopy of BCGs.

\smallskip

An additional advantage of our {deep observational strategy} is that MUSE made possible
to obtain reliable measurements of the faint [\ion{O}{i}]$\lambda$6300 line  up to kiloparsec scales. Maps of this line at large galactocentric distances are {essential} if we aim to explore  the contribution of shocks to the global ionization \citep{Dopita1976,Dopita1996}, as is the case here. 
However, the detection of  [\ion{O}{i}]$\lambda$6300 in galaxies has
so far been quite challenging: this line is not only extremely faint, but in the Milky-Way and nearby objects it becomes strongly contaminated  by  the bright ($\sim$100 times brighter than the interstellar line) [\ion{O}{i}]$\lambda$6300 airglow line \citep{Reynolds1998,Voges2006}.

\smallskip

From our analysis of the extremely rich MUSE dataset, we highlight the following results:

\begin{itemize}

\item  The ongoing SF episode in Haro~14 extends up to kiloparsec scales, with the brightest  \ion{H}{ii}-region complexes located close to the galaxy center.  These  \ion{H}{ii}-regions show large variations in the excitation, which likely reflect a range of ages in the ionizing clusters.

\smallskip

\item    An extended and faint halo  of ionized gas encircles the \ion{H}{ii}-regions complexes.  This  halo is not smooth, but here we resolved well-defined curvilinear structures, such as arcs, shells, filaments, and compact clumps. Particularly noteworthy are the two curvilinear filaments extending up to 2 and 2.3 kpc southwest, and a shell with a diameter about 800~pc to the north.

\smallskip

\item We find evidence of a photoionization mechanism other than young massive stars operating  in Haro~14. The central HSB component is mostly made of \ion{H}{ii}-region complexes, but an additional heating source is required to explain the line ratios and morphological pattern in the whole LSB component. Indeed, the largest fraction of the galaxy area is being ionized by one (or several) alternative power source(s): spaxels above the maximum starburst line represent  $\sim$75\% and $\sim$50\% of the area in the [\ion{O}{i}] and [\ion{S}{ii}] diagrams respectively and  account for $\sim$20\% and $\sim$13\% of the H$\alpha$ luminosity, respectively.

\smallskip 

This result contrasts with previous BCG investigations, with the reason for this discrepancy lying fundamentally in the observations: our data, reaching surface brightnesses of about 10$^{-18}$~erg~s$^{-1}$~cm$^{-2}$~arcsec$^{-2}$ 
in all diagnostic lines (including the faint [\ion{O}{i}]$\lambda$6300), allowed us, for the first time, to come to robust conclusions as to the excitation mechanism operating in the outer regions of a typical BCG.

\smallskip

\item Both the morphology and 
the distribution of the spaxels in the diagnostic diagrams point to different mechanisms acting in the galaxy periphery. The three distinct morphological constituents of the LSB ionized gas, that is, the widespread diffuse halo, the filamentary structures, and the individual clumps, are likely ionized by different mechanisms. 
We suggest that the extended halo is being ionized by diluted radiation escaping from the central \ion{H}{ii}-regions, and that the large filaments and shells are shocked areas in the walls of expanding bubbles associated with outflows. The mechanism responsible for the ionization of the individual clumps is more difficult to asses. The power sources acting in the galaxy outskirts will be investigated in a forthcoming publication based on kinematical results.

\end{itemize}

In conclusion, this work  stresses the important role played by wide-field integral field instruments such as MUSE in BCG research and, in particular, in the study of their warm ionized gas. At the typical distances of BCGs (between 10 and 40~Mpc), MUSE allow us to map the whole system, 
making it possible to characterize  the ionized gas at large galactocentric distances. 
Constraining the ionizing source, not just in the central galaxy regions but also in the galaxy outskirts, enables us to  separate areas of the galaxy ionized by hot stars from areas where other mechanisms are responsible for the ionization. This makes it possible to derive accurate ages, abundances,
and  excitation patterns, leading to a  much better characterization of the SF event. However, investigations of the outskirts of BCGs are not only important in the context of dwarf galaxy research and nearby starbursts, but are also essential in cosmological studies, as it is in the galaxy outskirts where primordial gas may still be accreting today.

\begin{acknowledgements} LMC acknowledges support from the Deutsche Forschungsgemeinschaft
(CA~1243/1-1 and CA~1234/1-2). PMW received support from BMBF Verbundforschung (project VLT-BlueMUSE, grant 05A20BAB). 
Based on observations collected at the European Southern Observatory under ESO programme ID 60.A-9186(A). 
This research has made use of the NASA/IPAC Extragalactic
Database (NED), which is operated by the Jet Propulsion Laboratory, Caltech,
under contract with the National Aeronautics and Space Administration. We acknowledge the usage of the HyperLeda database ({\tt http://leda.univ-lyon1.fr}). This work made use of v2.2.1 of the Binary Population and Spectral Synthesis (BPASS) models as described in \cite{Eldridge2017} and \cite{Stanway2018}.  We used IRAF package, which are distributed by the National Optical Astronomy Observatory, which is operated by the Association of Universities for Research in Astronomy, Inc., under contract with the National Science Foundation.
\end{acknowledgements}

\bibliographystyle{aa}
\bibliography{musegas}

\appendix
\section{Supplementary tables}

\begin{landscape}
\begin{table}
\small
\caption{Reddening-corrected line intensity ratios, interstellar extinction, diagnostic line ratios, densities, and oxygen abundances for the emission line sources of Haro~14.\label{tab:lineratiosSF2}}
\begin{center}
\begin{tabular}{lcccccccccc}
\hline 
\hline 
Ion      & L11 & L12 & L13 & L14 & L15 & L16 & L17 & L18 & L19 & L20    \\ 
\hline 
4861 H$\beta$ & 1.000 & 1.000 & 1.000 & 1.000 & 1.000 & 1.000 & 1.000 & 1.000 & 1.000 & 1.000 \\ 
4959 [OIII]    & 0.354$\pm$0.007 & 0.539$\pm$0.010 & 0.433$\pm$0.014 & 0.230$\pm$0.006 & 0.361$\pm$0.006 & 0.490$\pm$0.012 & 0.216$\pm$0.006 & 0.355$\pm$0.017 & 0.382$\pm$0.015 & 0.455$\pm$0.020 \\ 
5007 [OIII]   & 1.033$\pm$0.012 & 1.621$\pm$0.019 & 1.271$\pm$0.025 & 0.701$\pm$0.010 & 1.070$\pm$0.011 & 1.450$\pm$0.021 & 0.647$\pm$0.009 & 1.014$\pm$0.026 & 1.184$\pm$0.028 & 1.374$\pm$0.038  \\ 
6300 [OI]      & 0.075$\pm$0.003 & 0.092$\pm$0.004 & 0.153$\pm$0.006 & 0.062$\pm$0.004 & 0.056$\pm$0.003 & 0.061$\pm$0.005 & 0.090$\pm$0.005 & 0.235$\pm$0.010 & 0.142$\pm$0.011 & 0.448$\pm$0.018 \\ 
6363 [OI]      & 0.026$\pm$0.002 & 0.031$\pm$0.003 & 0.053$\pm$0.004 & 0.021$\pm$0.003 & 0.016$\pm$0.002 & 0.019$\pm$0.003 & 0.029$\pm$0.003 & 0.081$\pm$0.007 & 0.041$\pm$0.007 & 0.147$\pm$0.010 \\ 
6548 [NII]     & 0.110$\pm$0.006 & 0.107$\pm$0.005 & 0.157$\pm$0.012 & 0.123$\pm$0.005 & 0.105$\pm$0.004 & 0.100$\pm$0.006 & 0.120$\pm$0.005 & 0.170$\pm$0.014 & 0.111$\pm$0.010 & 0.174$\pm$0.017 \\ 
6563 H$\alpha$ & 2.870$\pm$0.041 & 2.870$\pm$0.041 & 2.870$\pm$0.068 & 2.870$\pm$0.040 & 2.870$\pm$0.032 & 2.870$\pm$0.047 & 2.870$\pm$0.035 & 2.870$\pm$0.079 & 2.870$\pm$0.080 & 2.870$\pm$0.097 \\ 
6584 [NII]     & 0.342$\pm$0.009 & 0.329$\pm$0.008 & 0.486$\pm$0.018 & 0.374$\pm$0.008 & 0.317$\pm$0.006 & 0.306$\pm$0.008 & 0.358$\pm$0.007 & 0.532$\pm$0.022 & 0.336$\pm$0.015 & 0.533$\pm$0.026 \\ 
6678 HeI       & 0.024$\pm$0.002 & 0.028$\pm$0.002 & 0.025$\pm$0.003 & 0.021$\pm$0.002 & 0.026$\pm$0.002 & 0.028$\pm$0.002 & 0.021$\pm$0.003 & 0.030$\pm$0.005 & 0.019$\pm$0.005 & 0.031$\pm$0.006  \\ 
6717 [SII]     & 0.391$\pm$0.007 & 0.414$\pm$0.009 & 0.725$\pm$0.019 & 0.412$\pm$0.007 & 0.405$\pm$0.006 & 0.388$\pm$0.008 & 0.499$\pm$0.008 & 0.921$\pm$0.027 & 0.566$\pm$0.018 & 1.095$\pm$0.039 \\ 
6731 [SII]     & 0.276$\pm$0.005 & 0.284$\pm$0.007 & 0.509$\pm$0.014 & 0.287$\pm$0.005 & 0.278$\pm$0.005 & 0.267$\pm$0.006 & 0.345$\pm$0.006 & 0.643$\pm$0.020 & 0.405$\pm$0.014 & 0.770$\pm$0.028 \\ 
7065 HeI       & 0.015$\pm$0.002 & 0.018$\pm$0.002 & 0.015$\pm$0.004 & 0.014$\pm$0.002 & 0.014$\pm$0.001 & 0.014$\pm$0.002 & 0.014$\pm$0.003 & 0.016$\pm$0.005 & 0.015$\pm$0.006 & 0.023$\pm$0.006 \\ 
7136 [ArIII]   & 0.058$\pm$0.004 & 0.068$\pm$0.005 & 0.094$\pm$0.028 & 0.052$\pm$0.004 & 0.057$\pm$0.003 & 0.072$\pm$0.005 & 0.042$\pm$0.006 & 0.098$\pm$0.034 & 0.056$\pm$0.012 & 0.092$\pm$0.038 \\ 
9068 [SIII]    & 0.162$\pm$0.007 & 0.180$\pm$0.008 & 0.123$\pm$0.029 & 0.160$\pm$0.007 & 0.155$\pm$0.006 & 0.177$\pm$0.008 & 0.127$\pm$0.009 & 0.100$\pm$0.032 & 0.124$\pm$0.016 & 0.095$\pm$0.036 \\ 
\hline 
F$_{H\beta}$  &  7.5$\pm$0.2 &  4.0$\pm$0.1 & 19.2$\pm$0.7 &  5.2$\pm$0.1 &  9.4$\pm$0.2 &  3.2$\pm$0.1 &  7.2$\pm$0.1 &  7.0$\pm$0.3 &  1.5$\pm$0.1 &  5.7$\pm$0.4  \\ 
C$_{H\beta}$  & 0.246$\pm$0.012 & 0.105$\pm$0.012 & 0.126$\pm$0.020 & 0.199$\pm$0.012 & 0.121$\pm$0.009 & 0.126$\pm$0.014 & 0.180$\pm$0.010 & 0.151$\pm$0.023 & 0.131$\pm$0.023 & 0.225$\pm$0.028 \\ 
$A_{V}$        & 0.531$\pm$0.026 & 0.228$\pm$0.026 & 0.272$\pm$0.043 & 0.430$\pm$0.025 & 0.260$\pm$0.020 & 0.272$\pm$0.029 & 0.388$\pm$0.022 & 0.327$\pm$0.050 & 0.283$\pm$0.050 & 0.487$\pm$0.061 \\ 
E(B-V)         & 0.171$\pm$0.008 & 0.073$\pm$0.008 & 0.088$\pm$0.014 & 0.139$\pm$0.008 & 0.084$\pm$0.007 & 0.088$\pm$0.009 & 0.125$\pm$0.007 & 0.105$\pm$0.016 & 0.091$\pm$0.016 & 0.157$\pm$0.020 \\ 
\hline 
\hline 
$[\ion{O}{iii}] \lambda5007/\Hb$                 & 1.033$\pm$0.041 & 1.621$\pm$0.046 & 1.271$\pm$0.063 & 0.701$\pm$0.025 & 1.070$\pm$0.025 & 1.450$\pm$0.050 & 0.647$\pm$0.019 & 1.014$\pm$0.063 & 1.184$\pm$0.069 & 1.374$\pm$0.121 \\ 
$[\ion{O}{i}] \lambda6300/\Ha$                   & 0.026$\pm$0.001 & 0.032$\pm$0.002 & 0.053$\pm$0.002 & 0.021$\pm$0.001 & 0.019$\pm$0.001 & 0.021$\pm$0.002 & 0.031$\pm$0.002 & 0.082$\pm$0.004 & 0.049$\pm$0.004 & 0.156$\pm$0.009 \\ 
$[\ion{N}{ii}] \lambda6584/\Ha$                  & 0.119$\pm$0.004 & 0.114$\pm$0.003 & 0.169$\pm$0.007 & 0.130$\pm$0.003 & 0.110$\pm$0.002 & 0.106$\pm$0.003 & 0.125$\pm$0.003 & 0.185$\pm$0.009 & 0.117$\pm$0.006 & 0.186$\pm$0.012 \\ 
$[\ion{S}{ii}] \lambda\lambda6717,6731/\Ha$     & 0.232$\pm$0.005 & 0.243$\pm$0.005 & 0.430$\pm$0.012 & 0.244$\pm$0.005 & 0.238$\pm$0.003 & 0.228$\pm$0.005 & 0.294$\pm$0.005 & 0.545$\pm$0.018 & 0.338$\pm$0.011 & 0.650$\pm$0.029 \\ 
\hline 
N$_{e}$   (cm$^{-3}$) & < 100                   & < 100                         & < 100                   & < 100                 & < 100                  & < 100                                 & < 100          & < 100                                & < 100 & < 100 \\ 
12+log(O/H)$^{2}$       & 8.26                   & 8.24                  & 8.25                   & 8.27                  & 8.23                  & 8.23                           & 8.22          & 8.22                          & 8.17 & 8.20\\ 
\hline 
\hline 
\end{tabular}
\end{center}
Notes.- Line intensity ratios are normalized to H$\beta$; the reddening-corrected H$\beta$ flux is in units of 10$^{-16}$~erg~s$^{-1}$~cm$^{-2}$. 
\end{table}
\end{landscape}

\begin{landscape}
\begin{table}
\small
\caption{Reddening-corrected line intensity ratios, interstellar extinction, diagnostic line ratios, densities, and oxygen abundances for the emission line sources of Haro~14.\label{tab:lineratiosSF3}}
\begin{center}
\begin{tabular}{lcccccccccc}
\hline 
\hline 
Ion      & L21 & L22 & L23 & L24 & L25 & L26 & L27 & L28 & L29 & L33    \\ 
\hline 
4861 H$\beta$ & 1.000 & 1.000 & 1.000 & 1.000 & 1.000 & 1.000 & 1.000 & 1.000 & 1.000 & 1.000 \\ 
4959 [OIII]    & 0.218$\pm$0.013 & 0.180$\pm$0.009 & 0.536$\pm$0.034 & 0.411$\pm$0.026 & 1.613$\pm$0.053 & 1.360$\pm$0.049 & 0.582$\pm$0.035 & 1.543$\pm$0.074 & 0.430$\pm$0.054 & 0.575$\pm$0.073 \\ 
5007 [OIII]   & 0.654$\pm$0.018 & 0.555$\pm$0.013 & 1.635$\pm$0.070 & 1.187$\pm$0.041 & 4.820$\pm$0.133 & 4.020$\pm$0.119 & 1.697$\pm$0.071 & 4.625$\pm$0.196 & 1.271$\pm$0.090 & 1.632$\pm$0.123  \\ 
6300 [OI]      & 0.181$\pm$0.010 & 0.162$\pm$0.008 & 0.531$\pm$0.033 & 0.365$\pm$0.017 & 0.422$\pm$0.022 & 0.232$\pm$0.022 & 0.503$\pm$0.032 & 0.403$\pm$0.033 & 0.501$\pm$0.047 & 0.222$\pm$0.041 \\ 
6363 [OI]      & 0.068$\pm$0.007 & 0.058$\pm$0.006 & 0.150$\pm$0.016 & 0.129$\pm$0.011 & 0.139$\pm$0.013 & 0.052$\pm$0.015 & 0.169$\pm$0.017 & 0.140$\pm$0.020 & 0.149$\pm$0.024 & 0.094$\pm$0.031 \\ 
6548 [NII]     & 0.110$\pm$0.011 & 0.134$\pm$0.010 & 0.144$\pm$0.027 & 0.160$\pm$0.012 & 0.151$\pm$0.015 & 0.132$\pm$0.018 & 0.222$\pm$0.022 & 0.162$\pm$0.020 & 0.173$\pm$0.029 & 0.168$\pm$0.039 \\ 
6563 H$\alpha$ & 2.870$\pm$0.068 & 2.870$\pm$0.050 & 2.870$\pm$0.165 & 2.870$\pm$0.104 & 2.870$\pm$0.110 & 2.870$\pm$0.118 & 2.870$\pm$0.152 & 2.870$\pm$0.168 & 2.870$\pm$0.229 & 2.870$\pm$0.253 \\ 
6584 [NII]     & 0.359$\pm$0.015 & 0.407$\pm$0.013 & 0.437$\pm$0.040 & 0.497$\pm$0.023 & 0.470$\pm$0.024 & 0.438$\pm$0.028 & 0.660$\pm$0.041 & 0.533$\pm$0.038 & 0.548$\pm$0.053 & 0.553$\pm$0.066 \\ 
6678 HeI       & 0.017$\pm$0.005 & 0.023$\pm$0.004 & 0.026$\pm$0.011 & 0.034$\pm$0.011 & 0.026$\pm$0.009 & 0.046$\pm$0.015 & 0.018$\pm$0.017 & 0.036$\pm$0.019 & 0.021$\pm$0.022 & 0.054$\pm$0.029  \\ 
6717 [SII]     & 0.669$\pm$0.018 & 0.702$\pm$0.013 & 0.999$\pm$0.062 & 1.052$\pm$0.041 & 0.956$\pm$0.040 & 0.745$\pm$0.035 & 1.176$\pm$0.066 & 0.954$\pm$0.059 & 1.110$\pm$0.096 & 0.857$\pm$0.082 \\ 
6731 [SII]     & 0.468$\pm$0.014 & 0.484$\pm$0.010 & 0.697$\pm$0.045 & 0.728$\pm$0.030 & 0.667$\pm$0.029 & 0.513$\pm$0.026 & 0.835$\pm$0.049 & 0.646$\pm$0.041 & 0.784$\pm$0.070 & 0.664$\pm$0.065 \\ 
7065 HeI       & 0.014$\pm$0.006 & 0.015$\pm$0.004 & 0.005$\pm$0.008 & 0.035$\pm$0.014 & 0.034$\pm$0.012 & 0.032$\pm$0.016 & 0.013$\pm$0.021 & 0.022$\pm$0.013 & 0.038$\pm$0.021 & 0.061$\pm$0.040 \\ 
7136 [ArIII]   & 0.062$\pm$0.019 & 0.054$\pm$0.010 & 0.070$\pm$0.031 & 0.108$\pm$0.047 & 0.126$\pm$0.034 & 0.148$\pm$0.047 & 0.141$\pm$0.057 & 0.139$\pm$0.027 & 0.099$\pm$0.067 & 0.191$\pm$0.075 \\ 
9068 [SIII]    & 0.107$\pm$0.020 & 0.115$\pm$0.014 & 0.085$\pm$0.031 & 0.104$\pm$0.049 & 0.207$\pm$0.040 & 0.183$\pm$0.045 & 0.128$\pm$0.046 & 0.215$\pm$0.035 & 0.095$\pm$0.063 & 0.170$\pm$0.091 \\ 
\hline 
F$_{H\beta}$  &  2.0$\pm$0.1 &  4.2$\pm$0.1 &  1.5$\pm$0.3 &  3.4$\pm$0.2 &  1.8$\pm$0.1 &  1.5$\pm$0.1 &  0.8$\pm$0.1 &  0.6$\pm$0.1 &  1.1$\pm$0.2 &  0.7$\pm$0.1  \\ 
C$_{H\beta}$  & 0.090$\pm$0.020 & 0.081$\pm$0.014 & 0.398$\pm$0.048 & 0.171$\pm$0.030 & 0.171$\pm$0.032 & 0.164$\pm$0.034 & 0.150$\pm$0.044 & 0.179$\pm$0.049 & 0.244$\pm$0.066 & 0.217$\pm$0.073 \\ 
$A_{V}$        & 0.195$\pm$0.042 & 0.176$\pm$0.031 & 0.859$\pm$0.103 & 0.370$\pm$0.065 & 0.369$\pm$0.069 & 0.354$\pm$0.074 & 0.323$\pm$0.095 & 0.386$\pm$0.105 & 0.527$\pm$0.143 & 0.469$\pm$0.159 \\ 
E(B-V)         & 0.063$\pm$0.014 & 0.057$\pm$0.010 & 0.277$\pm$0.033 & 0.119$\pm$0.021 & 0.119$\pm$0.022 & 0.114$\pm$0.024 & 0.104$\pm$0.031 & 0.124$\pm$0.034 & 0.170$\pm$0.046 & 0.151$\pm$0.051 \\ 
\hline 
\hline 
$[\ion{O}{iii}] \lambda5007/\Hb$                 & 0.654$\pm$0.032 & 0.555$\pm$0.021 & 1.635$\pm$0.377 & 1.187$\pm$0.101 & 4.820$\pm$0.417 & 4.020$\pm$0.367 & 1.697$\pm$0.196 & 4.625$\pm$0.626 & 1.271$\pm$0.280 & 1.632$\pm$0.371 \\ 
$[\ion{O}{i}] \lambda6300/\Ha$                   & 0.063$\pm$0.004 & 0.056$\pm$0.003 & 0.185$\pm$0.023 & 0.127$\pm$0.008 & 0.147$\pm$0.009 & 0.081$\pm$0.008 & 0.175$\pm$0.014 & 0.141$\pm$0.014 & 0.175$\pm$0.024 & 0.077$\pm$0.016 \\ 
$[\ion{N}{ii}] \lambda6584/\Ha$                  & 0.125$\pm$0.006 & 0.142$\pm$0.005 & 0.152$\pm$0.021 & 0.173$\pm$0.010 & 0.164$\pm$0.010 & 0.153$\pm$0.011 & 0.230$\pm$0.017 & 0.186$\pm$0.016 & 0.191$\pm$0.025 & 0.193$\pm$0.029 \\ 
$[\ion{S}{ii}] \lambda\lambda6717,6731/\Ha$     & 0.396$\pm$0.010 & 0.413$\pm$0.008 & 0.591$\pm$0.060 & 0.620$\pm$0.027 & 0.565$\pm$0.026 & 0.438$\pm$0.022 & 0.701$\pm$0.042 & 0.557$\pm$0.039 & 0.660$\pm$0.070 & 0.530$\pm$0.060 \\ 
\hline 
N$_{e}$   (cm$^{-3}$)  & < 100                  & < 100                  & < 100                  & < 100                & < 100                 & < 100 & < 100 & < 100& < 100 & $\sim$100 \\ 
12+log(O/H)$^{2}$       & 8.14                  & 8.18                   & 8.15                   & 8.17         & 8.21           & 8.22          & 8.27 & 8.26 & 8.20 & 8.13 \\ 
\hline 
\hline 
\end{tabular}
\end{center}
Notes.- Line intensity ratios are normalized to H$\beta$; the reddening-corrected H$\beta$ flux is in units of 10$^{-16}$~erg~s$^{-1}$~cm$^{-2}$. 
\end{table}
\end{landscape}

\end{document}